# 3D observations discover a new paradigm in rubber elasticity

Z. Wang, S. Das[*], A. Joshi[*], A.J.D. Shaikeea[§] and V.S. Deshpande[§]

**Author Affiliations:**
*Department of Engineering, University of Cambridge, Cambridge CB2 1PZ, UK*

[*] Authors contributed equally

[§] **Corresponding authors:** A.J.D. Shaikeea: ajds3@cam.ac.uk
                                   V.S. Deshpande: vsd@eng.cam.ac.uk

**Summary paragraph**
The mechanical response of rubbers has been ubiquitously assumed to be only a function of the imposed strain. Using innovative X-ray measurements capturing the three-dimensional spatial volumetric strain fields, we demonstrate that rubbers and indeed many common engineering polymers, undergo significant local volume changes. But remarkably the overall specimen volume remains constant regardless of the imposed loading. This strange behaviour which also leads to apparent negative local bulk moduli is due to the presence of a mobile phase within these materials. Combining X-ray tomographic observations with high-speed radiography to track the motion of the mobile phase we have revised classical thermodynamic frameworks of rubber elasticity. The work opens new avenues to understand not only the mechanical behaviour of rubbers but a large class of widely used engineering polymers.

Rubber elasticity is the generic term describing the behaviour of polymeric solids made up of flexible chains which are joined together to form a three-dimensional network. These solids have the capacity to sustain very large deformations followed by complete recovery. It is one of the most established branches of the mechanics of solids with detailed molecular theories [1-7] supporting macroscopic measurements [7-11] of the deformation behaviour and more recent multiscale insights [12] improving properties such fatigue resistance. The ubiquitous feature of all elastomeric models for rubbers is that the Helmholtz free-energy is written purely as a function of temperature and deformation, and no other state variable. In fact, this is the widely accepted definition of a rubber or elastomer. Here using detailed three-dimensional (3D) X-ray computed tomography (XCT) strain measurements and high-speed radiography, supported by laser measurements and two-dimensional digital image correlation (DIC), we demonstrate that the understanding of rubber elasticity needs revision. Our findings show that rubbers/elastomers, and synthetic polymers, including thermoplastics require new state variables to describe their deformation behaviour. This has wide-ranging and significant implications for understanding the mechanical responses of polymer composites.

We have developed a laboratory-based XCT method, dubbed "Flux Enhanced Tomography for Correlation (FETC)", to non-intrusively measure full field 3D strains in nominally homogeneous solids ranging from polymers to metals. FETC (Extended Data Fig. 1 and Supplementary S1) captures subtle variations in density or heterogeneities such as differences in crosslinking densities in polymers by separating out the grayscale distinctions in the X-ray attenuation spectra (Fig. 1a). This provides a natural speckle pattern in the material, which we leverage to perform digital volume correlation (DVC) [13-15] and an in-situ multi-stage loading protocol (Fig. 1a, Methods). We then derive volumetric spatial distributions of displacement fields, enabling precise measurement of all 9 components of the deformation gradient $F_{ij}$. An inherent advantage of FETC is that it does not use artificial tracer particles to provide the speckle for the DVC and therefore the measurements directly reflect the innate material response.

**Large local volume changes but specimens remain incompressible**
Silicone rubber is considered an exemplar rubber and extensively characterised by combining micro-mechanical models of rubber elasticity with measurements [7, 16]. We first consider cylindrical Silicone rubber specimens (Methods) of height $H$ and diameter $D$ which are glued to platens (Fig. 1b) and tested in tension or compression by applying an axial displacement rate $\dot{U}$ and measuring the conjugate axial force $P$. The glued ends imply that the deformation field is spatially heterogenous with the specimen under tensile loading thinning at mid-section but constrained at the platens (Fig. 1c). The nominal tensile stress $\sigma^\infty \equiv 4P/(\pi D^2)$ versus strain $\varepsilon^\infty \equiv U/H$ responses (Fig. 1d) for an imposed quasi-static rate $\dot{\varepsilon}^\infty = 10^{-3}$ s$^{-1}$ are in line with expectations of an elastomer, viz. the behaviour is reversible, shows some non-linearity at larger deformations and the stiffness increases with decreasing aspect ratio $H/D$. FETC provides a rich dataset on the deformation modes with the evolution of the spatial distributions of $F_{ij}$ with $\varepsilon^\infty$ (Extended Data Fig. 2). The first interesting observation presents itself when we examine the evolution of the spatial distribution of the volumetric strain $\Delta V/V_0 \equiv \det(F_{ij}) - 1$ within the $H/D = 1$ specimen (Fig. 1c and Supplementary Movie S1). Clearly, the specimen does not deform in an incompressible manner as widely assumed for many elastomers but rather undergoes local volumetric strains of $\sim \pm 10\%$ (laser measurements independent of FETC further confirm these results; see Extended Data Fig. 3). That in itself may not be remarkable as compressible rubber elasticity has also been proposed [9, 17] although without a molecular basis but more often in the context of porous rubbers or foams [18]. The real surprise here is that under tensile loading we observe a mixture of compressive and dilatory volumetric strains with the central core of the specimen losing volume and an outer cylindrical shell gaining volume (Fig. 1c). The puzzle becomes more perplexing when we observe that the overall specimen volume remains constant over all loading stages, i.e., $\langle V/V_0 \rangle = \langle \det(F_{ij}) \rangle \approx 1$; see Fig. 1c and Extended Data Fig. 2a (here $\langle \cdot \rangle$ denotes the volume-average over the whole specimen). The overall incompressibility of the specimen is also confirmed from tomographic reconstruction of the specimen outline (Extended Data Fig. 3a). These conclusions that deformation is accompanied by local volume changes of $\pm 10\%$ but the specimen remains incompressible also extend to compressive loading (Extended Data Fig. 4).

To further investigate whether the specimen level volume constraint is a material property and not restricted to the stress-state generated within the $H/D = 1$ specimen, we performed tensile tests on specimens with a wide range of aspect ratios (the stress triaxiality reduces and becomes more uniaxial with increasing $H/D$). These measurements (Fig. 1e) clearly illustrate that while there are large local volume ($\sim \pm 10\%$) changes, the specimen volume is always conserved. However, with increasing specimen aspect ratio, the central portion of the specimen becomes increasingly spatially uniform and deforms in a nearly incompressible manner with regions of volumetric deformation restricted to the regions near the glued loading platens where gradients of the strain are present. The immediate question then arises is whether existing elastomeric models can capture these observations? We employed the finite element model updating (FEMU) [19-21] method to infer elastomeric constitutive model parameters for three common elastomeric models from the spatial data of $F_{ij}$. No model was able to capture this mixture of volumetric dilation and compaction observed in the FETC measurements (Supplementary S2).

**Deformation gradients drive the motion of a mobile phase**
The current hyperelastic models cannot capture the observations is likely related to the fact that the volumetric deformations occur when strain/stress gradients are large (recall effect of $H/D$ in Fig. 1e). We confirmed that the apparent effect of strain gradients does not result in a specimen size effect (Supplementary S3). This motivated us to investigate the effect of strain gradients in a more controlled setting. Four-point bending of a beam (Fig. 2a) imposes a pure bending moment over the section of length $s_1$ between the central rollers. The measured moment $M$ versus curvature $\kappa$ relation (Fig. 2b) for a beam with a square cross-section $h \times h$ and subjected to a loading rate $\dot{\kappa}h$ is approximately linear with negligible loading/unloading hysteresis. Further the response is independent of loading rates over 3 orders of magnitude.

This apparent macroscopic elastic behaviour is also confirmed by relaxation tests (Fig. 2c) where we linearly ramped up the curvature over a time period $0 \leq t \leq t_R$ and then held the curvature constant (inset of Fig. 2c): the moment increases linearly during the ramp and then remains constant with no relaxation. While these are behaviours expected of a rubber our next major finding occurs while examining the distributions of the volumetric strains $\Delta V/V_0$ calculated from $F_{ij}$ (Fig. 2d and Supplementary Movie S2). We observe that the rubber gains volume on the compressive side of the beam and loses volume on the tensile side with the strains again on the order of $\pm 10\%$ (see Extended Data Fig. 3 for independent validation experiments using 3D DIC for the surface strain components). These unexpected volumetric strains occur even though the central section of the beam is under pure bending; see Fig. 2e for selected components of the Green-Lagrange strain. The as-cast Silicone rubber is isotropic and homogeneous (Supplementary S3), and it is immediately clear that no existing theory for rubbers/elastomers can capture this response as it will require a negative bulk modulus. In fact, thermodynamic consistency arguments therefore suggest that state variables in addition to deformation (and temperature) are required to formulate the Helmholtz free-energy of Silicone rubber.

The observation that there are local volume changes accompanied by overall incompressibility of the specimen drives the hypothesis that there exists an incompressible mobile phase within the rubber. The transport of this phase then induces local volume changes but with global volume conservation. The micro-mechanical rationale for a mobile phase is that synthetic rubbers such as Silicone rubber are produced by a polymerisation reaction [22, 23] which might leave a certain fraction of the polymer un-crosslinked. The rubber is therefore comprised of a crosslinked polymer network and an un-crosslinked component that is mobile within the open structure crosslinked network (Supplementary S4). To investigate the presence of such a mobile phase, we manufactured a "composite" Silicone rubber beam where half of the beam contained 5% wt. $ZnI_2$ dissolved into the catalyst (which serves as an X-ray tracer) and the other half is pure Silicone rubber (Fig. 3a and Methods). This composite beam was then subjected to 4-point bending with the neutral axis of bending delineating the halves of the beam with and without the tracer. Tomographic reconstructions of the beam subjected to curvatures $\kappa h = \pm 0.24$ are included in Fig. 3b, i.e., the beam was successively bent with the tracer portion on the tensile and compressive side of the beam, respectively. The dissolved tracers spread out from the compressive side onto the tensile side but there is little transport observed when the tracers are on the tensile side (the motion of the tracers is reversible with the XCT reconstructions of the undeformed beam prior to and after loading/unloading indistinguishable; see Supplementary Movie S3). Given that adhesion between $ZnI_2$ and Silicone rubber is excellent (Methods) these observations definitively show that there exists a mobile phase within the rubber. The dissolved $ZnI_2$ is attached proportionally to the mobile un-crosslinked polymer and the immobile crosslinked polymer: the observed motion of the tracers is associated with the transport of the mobile phase.

It is now apparent that the concentration of the mobile phase is a key additional state variable that is required to describe the mechanical behaviour of the Silicone rubber. But an additional anomalous feature of the rubber is that the material can volumetrically expand under compressive hydrostatic stress even though observations (Fig. 3b) show that material egresses out of the compressive region. Expansion in the presence of a compressive hydrostatic stress implies a negative contribution from the hydrostatic component of the mechanical work. Thus, unlike in hydrogels [24-26] and theories of other materials [27-29] involving a mobile phase, the driving force for the egress of un-crosslinked phase cannot be pressure. Rather we hypothesize that the un-crosslinked phase comprises non spherical molecules such as dimers/trimers. The shape of these molecules implies that this phase does not induce pure volumetric straining but a combination of volumetric and deviatoric straining (Supplementary S4). Then positive deviatoric work associated with the un-crosslinked phase provides the driving force for the egress/ingress of the un-crosslinked phase. We have developed thermodynamically consistent model based on this hypothesis (Supplementary S4). Under isothermal conditions the Helmholtz free-energy is a function of not only the deformation parameterised by $F_{ij}$ but also the occupancy $\theta$ of the un-crosslinked phase in the crosslinked network ($\theta$ is directly related to the concentration of the mobile phase). A key feature of the free-energy is the non-isotropic straining due to the mobile un-crosslinked phase which also implies that deformation induces reversible anisotropy of the Silicone rubber. Predictions of the equilibrium (i.e., for loading at the limiting rate $\dot{\kappa} \to 0$) moment $M$ versus normalised curvature $\kappa h$ response (Fig. 2b) and the spatial variation of the volumetric strain (Fig. 3c) across the beam thickness at $\kappa h = 0.24$ are in excellent agreement with measurements. The corresponding predictions of the normalised concentration $\bar{\theta}$ of the mobile phase across the beam thickness (Fig. 3c) also confirm that that phase egresses out of the compressive region in line with observations in Fig. 3b ($\bar{\theta} = 1$ is the normalised concentration in the undeformed state so that $\bar{\theta} < 1$ and $\bar{\theta} > 1$ indicate a reduction and increase, respectively in the concentration of the mobile phase compared to the undeformed state). While the continuum model presented here demonstrates that a thermodynamically consistent theory can be constructed to describe the strange behaviours reported here, a deeper molecular level understanding remains to be established: qualitative understandings of molecular mechanisms are discussed in Supplementary S4.

**Direct observations of the motion of the mobile phase**
An immediate implication of our finding of a mobile phase within the rubber is that we expect the rubber to have a rate dependent response that is dictated by the mobility of the mobile phase. Such a behaviour would then, even in terms of macroscopic load versus displacement measurements, be inconsistent with hyperelastic models usually used to describe the deformation of rubbers. However, in line with behaviours of rubbers reported in the literature, our measurements too show little or no rate dependence of the response over more than 3 orders of magnitude in $\dot{\kappa} h$ (Fig. 2b) and similarly no relaxation (Fig. 2c). These measurements are seemingly incompatible with the notion of a mobile phase: a sufficiently high mobility such that the spatio-temporal distribution of the mobile phase is independent of $\dot{\kappa} h$ but only dependent on the level of loading $\kappa h$ could rationalise this apparent inconsistency. To visualise this spatio-temporal distribution we repeated the step loading 4-point test with the composite Silicone rubber beam (Fig. 3a) but now obtaining radiographs at an interframe times of 0.5 s to visualise the real time motion of the dissolved $ZnI_2$ tracers (Methods). A montage of the radiographs (Fig. 3d and Supplementary Movie S4) for the $\dot{\kappa} h = 0.12 \text{ s}^{-1}$ test loaded to $\kappa^\infty h = 0.48$ ($\kappa^\infty \equiv \dot{\kappa} t_R$) shows that the tracers move from the compressive side to the tensile side of the beam as anticipated. But remarkably the transport is rapid: the position $X_f$ of the $ZnI_2$ front is plotted in Fig. 3e (Supplementary S3) and we observe that the front travels a distance $\ell \approx 2.7$ mm over the duration $t_R = 4$ s of the step loading and reaches its final steady-state concurrently at the end of the step at $t = t_R$. Thus, the transport of the mobile phase is so rapid that even at strain rates on the order of $\dot{\kappa} h = 0.12 \text{ s}^{-1}$ the beam remains approximately in equilibrium resulting in no observed rate dependence of the response.

Given this understanding of the kinetics, we extended our model to include the mobility of the mobile phase (Supplementary S4). The fact that there is no observed rate dependence of the response even at $\dot{\kappa} h = 0.12 \text{ s}^{-1}$ suggests an estimate for the effective diffusion co-efficient of the mobile phase to be $D \approx \ell^2 \dot{\kappa}/\kappa^\infty = 1.8 \times 10^{-6}$ m$^2$s$^{-1}$. This is an unrealistically high linear diffusion co-efficient (typical diffusion co-efficient for water in hydrogels are $\sim 10^{-8}$ m$^2$s$^{-1}$ – $10^{-9}$ m$^2$s$^{-1}$ [30, 31]). We rationalise the fast transport in our model by arguing that transport of the mobile phase differs somewhat from conventional Darcy flow [32, 33] of a Newtonian fluid through a porous network (the crosslinked network in this case). Given that the mobile phase is a polymer we hypothesize that it behaves as a shear thinning fluid [34] where the resistance to flow (or viscosity) reduces with increasing strain rate. Consequently, a linear diffusion co-efficient cannot be defined for the mobile phase [35]. Using shear thinning parameter values that makes the dissipation rate of the mobile phase to be nearly independent of flux (Supplementary S4) provides excellent agreement with the measurements. These predictions are included in Figs. 2b and 2c for the rate dependence of the moment curvature and relaxation responses, respectively. The model also predicts the temporal evolution of the position of the front of the tracer particles (Supplementary S4) for the case shown in Fig. 3d and two additional cases: (i) a slower rate of $\dot{\kappa}h = 0.012$ s$^{-1}$ to the same final curvature $\kappa^\infty h = 0.48$ and (ii) the same rate of $\dot{\kappa}h = 0.12$ s$^{-1}$ loaded to a lower curvature of $\kappa^\infty h = 0.24$. The predictions and the corresponding measurements (Fig. 3e) are in excellent agreement confirming that the transport of the mobile phase is sufficiently rapid that the tracer front position is only dependent on the current value of the imposed curvature over the loading rates investigated here.

**Outlook**

We have reported numerous counterintuitive behaviours of a Silicone rubber that are contrary to the current understanding of rubber elasticity. These include: (i) the existence of a mobile phase which results in large local volume changes ($\pm 10\%$) while the specimen volume remains unchanged; (ii) dilation under compressive hydrostatic pressures concurrent with the egress of the mobile phase and (iii) extremely high mobility within the rubber. Taken together it is now clear why this behaviour had to-date eluded researchers: (a) traditional techniques only measured global or specimen level volume changes and (b) the very high mobility implies that material rate dependence does not manifest itself at the usual loading rates used to investigate properties of rubbers. These findings while indeed intriguing raise the question whether they are rather niche and without broader implications. The key requirement for the strange behaviours reported here is the existence of a mobile phase which is the outcome of an incomplete polymerisation reaction. Many synthetic engineering polymers are produced by such reactions, and they include widely used thermoplastics like Nylon and high-density polyethylene (HDPE). Thus, it is conceivable that similar effects are not restricted to rubbers but also present in engineering plastics. To test this, we performed three-point bend experiments (Fig. 4a) on Silicone rubber, Nylon6 (Supplementary Movie S5) and HDPE beams of identical geometries (Methods). The measured imposed load $P$ versus central roller displacement $U$ response is shown in Fig. 4b illustrating the very different mechanical responses of these three materials. The rubber has a recoverable response while the Nylon6 and HDPE are not only significantly stiffer but also undergo plastic or permanent deformation upon unloading. The measured spatial distributions of the volumetric strains $\Delta V/V_0$ at $U = 4$ mm included in Fig. 4c clearly show large local volumetric deformations. But again, when averaged over the specimen $\langle V/V_0 \rangle = \langle \det(F_{ij}) \rangle \approx 1$ for all these materials. Moreover, consistent with the 4-point bending results on the Silicone rubber we observe local dilation on the compressive side of the 3-point bend specimen for all these materials. Together these findings suggest that the phenomena we have uncovered are rather general and apply to a wide range of synthetic polymers.

The FETC method we have developed has broad applications to the emerging field of data-driven mechanics. To-date most studies [36-38] developing Machine Learning (ML) methods for constitutive law development have employed only synthetic data. This is due to lack of measurement techniques that provide 3D data: FETC fills this gap. Moreover, measured data using FETC opens doors for new discoveries such as those reported here with wide ranging implications for mechanical constitutive models of engineering rubbers and polymers. For example, the plastic response of polymers is often thought to be incompressible but with a pressure dependence on the yield [39-41]. This has motivated a range of non-associated flow plastic models for polymers which often result in thermodynamic inconsistencies. Here we suggest that there exists a mechanism that motivates revisiting mechanical models for rubbers and polymers including an understanding of the effect of strain/stress gradients within polymers when these materials are used as matrices in composite materials.

**References**
[1] James, H.M., Guth, E. (1943). Theory of the Elastic properties of rubber, *Journal of Chemical Physics*, 11, 455-481.
[2] James, H.M., Guth, E. (1944). Theory of the elasticity of rubber. *Journal of Applied Physics*, 15, 294-303.
[3] Flory, P.J. (1944). Network structure and the elastic properties of vulcanized rubber. *Chemical Reviews*, 35, 51-75.
[4] Flory, P.J., Rehner, J. (1943). Statistical mechanics of cross-linked polymer networks. I. *The Journal of Chemical Physics*, 11, 512-520.
[5] Treloar, L.R.G. (1943). The elasticity of a network of long-chain molecules. I. *Transactions of the Faraday Society*, 39, 36-41.
[6] Treloar, L.R.G. (1943). The elasticity of a network of long-chain molecules. II. *Transactions of the Faraday Society*, 39, 241-246.
[7] Arruda, E.M., Boyce, M.C. (1993). A three-dimensional model for the large stretch behavior of rubber elastic materials. *Journal of the Mechanics and Physics of Solids*, 41, 389-412.
[8] Treloar, L.R.G. (1944). Stress-strain data for vulcanized rubber under various types of deformation. *Transactions of the Faraday Society,* 42, 59-70.
[9] Rivlin, R.S. (1948). Large elastic deformations of isotropic materials. I. Fundamental concepts. *Philosophical Transactions of the Royal Society of London. Series A*, 240, 459-490.
[10] Rivlin, R.S., Saunders, D.W. (1951). Large elastic deformations of isotropic materials, VII. Experiments on the deformation of rubber. *Philosophical Transactions of the Royal Society of London. Series A*, 243, 251-288.
[11] Mooney, M. (1940). A theory of large elastic deformation. *Journal of Applied Physics*, 11, 582–592.
[12] Steck, J., Kim, J., Kutsovsky, Y., Suo, Z. (2023) Multiscale stress deconcentration amplifies fatigue resistance of rubber. *Nature*, 624, 303-308
[13] Gillard, F., Boardman, R., Mavrogordato, M., Hollis, D., Sinclair, I., Pierron, F., Browne, M. (2014). The application of digital volume correlation (DVC) to study the microstructural behaviour of trabecular bone during compression, *Journal of the Mechanical Behavior of Biomedical Materials*, 29, 480-499.
[14] Dall'Ara, E., Tozzi G (2022), Digital volume correlation for the characterization of musculoskeletal tissues: Current challenges and future developments. *Frontiers in Bioengineering and Biotechnology*, 10, doi.org/10.3389/fbioe.2022.1010056.
[15] Buljac, A., Jailin, C., Mendoza, A., Neggers, J., Taillandier-Thomas, T., Bouterf, A., Smaniotto, B., Hild, F., Roux, S. (2020) Digital Volume Correlation: Review of Progress and

Challenges. In: Lin, M.T., et al. Advancements in Optical Methods & Digital Image Correlation in Experimental Mechanics, Volume 3. *Conference Proceedings of the Society for Experimental Mechanics Series*, Springer.

[16] Martins, P.A.L.S., Natal Jorge, R.M., Ferreira, A.J.M. (2006). A Comparative Study of Several Material Models for Prediction of Hyperelastic Properties: Application to Silicone-Rubber and Soft Tissues. *Strain*, 42, 135-147.

[17] Yao, Y., Chen, S., Huang, Z. (2022). A generalized Ogden model for the compressibility of rubber-like solids. *Philosophical Transactions of the Royal Society A*, 380, 20210320.

[18] Storåkers, B. (1986). On Material Representation and Constitutive Branching in Finite Compressible Elasticity. *Journal of the Mechanics and Physics of Solids*, 34, 125-145.

[19] Avril, S., Pierron, F. (2007). General framework for the identification of constitutive parameters from full-field measurements in linear elasticity. *International Journal of Solids and Structures*, 44, 4978-5002.

[20] Molimard, J., Riche, R.Le, Vautrin, A., Lee, J.R., (2005). Identification of the four orthotropic plate stiffnesses using a single open-hole tensile test. *Experimental Mechanics*, 45, 404-411.

[21] Ereiz, S., Duvnjak, I., Jiménez-Alonso, J.F. (2022). Review of finite element model updating methods for structural applications. *Structures,* 41,684-723.

[22] Harkous, A., Colomines, C., Leroy, E., Mousseau, P., Deterre, R. (2016). The kinetic behavior of Liquid Silicone Rubber: A comparison between thermal and rheological approaches based on gel point determination. *Reactive and Functional Polymers*, 101, 20-27.

[23] de Buyl, F., Hayez, V., Harkness, B., Kimberlain, J., Shephard, N. (2023). Advances in structural silicone adhesives, In Dillard, D.A. *Advances in Structural Adhesive Bonding (Second Edition)*, Woodhead Publishing, 179-219.

[24] Flory, P.J. (1942). Thermodynamics of high polymer solutions. *The Journal of Chemical Physics*, 10, 51-61.

[25] Huggins, M.L. (1941). Solutions of long chain compounds. *The Journal of Chemical Physics*, 9, 440-440.

[26] Cai, S., Suo, Z. (2011). Mechanics and chemical thermodynamics of phase transition in temperature-sensitive hydrogels. *Journal of the Mechanics and Physics of Solids*, 59, 2259-2278.

[27] Sofronis, P., McMeeking, R.M. (1989). Numerical analysis of hydrogen transport near a blunting crack tip. *Journal of the Mechanics and Physics of Solids*, 37, 317-350.

[28] Purkayastha, R.T., and McMeeking, R.M. (2012). A linearized model for lithium-ion batteries and maps for their performance and failure. *Journal of Applied Mechanics*, 79, 031021.

[29] Bohn, E., Eckl, T., Kamlah, M., and McMeeking, R.M. (2013). A model for lithium diffusion and stress generation in an intercalation storage particle with phase change. *Journal of the Electrochemical Society,* 160, A1638-A1652.

[30] Hu, Y., Chen, X., Whitesides, G.M., Vlassak, J.J., Suo, Z. (2011). Indentation of polydimethylsiloxane submerged in organic solvents. *Journal of Materials Research,* 26, 785-795.

[31] Hu, Y., Zhao, X., Vlassak, J.J., Suo, Z. (2010). Using indentation to characterize the poroelasticity of gels. *Applied Physics Letters,* 96, 121904.

[32] Biot, M.A. (1941). General Theory of Three-Dimensional Consolidation. *Journal of Applied Physics,* 12, 155-164.

[33] Rice, J.R., Cleary, M.P. (1976). Some Basic Stress-Diffusion Solutions for Fluid-Saturated Elastic Porous Media with Compressible Constituents. *Reviews of Geophysics and Space Physics*, 14, 227-241.

[34] Vergne, P. (2007) Super Low Traction under EHD & Mixed Lubrication Regimes. In Erdemir, A., Martin, J.M. *Superlubricity*, Elsevier Science, 427-443.

[35] Airiau, C., Bottaro, A. (2020). Flow of shear-thinning fluids through porous media. *Advances in Water Resources*, 143, 103658.

[36] Thakolkaran, P., Joshi, A., Zheng, Y., Flaschel, M., De Lorenzis, L., Kumar, S. (2022). NN-EUCLID: Deep-learning hyperelasticity without stress data. *Journal of the Mechanics and Physics of Solids*, Volume 169, 105076.

[37] Flaschel, M., Kumar, S., De Lorenzis, L. (2022). Discovering plasticity models without stress data. *npj computational materials,* 8, 91, doi.org/10.1038/s41524-022-00752-4.

[38] Mozaffar, M., Bostanabad, R., Chen, W., Bessa, M.A. (2019). Deep learning predicts path-dependent plasticity. *Proceedings of the National Academy of Sciences*, 116, 26414-26420.

[39] Silano, A.A., Pae, K.D., Sauer, J.A. (1977). Effects of hydrostatic pressure on shear deformation of polymers. *Journal of Applied Physics*, 48, 4076-4084.

[40] Spitzig, W.A., Richmond, O. (1979). Effect of hydrostatic pressure on the deformation behavior of polyethylene and polycarbonate in tension and in compression. *Polymer Engineering and Science*, 19, 1129-1139.

[41] Sanomura, Y. (2003). Constitutive Equation for Plastic Behavior of Hydrostatic-Pressure-Dependent Polymers. *Materials Science and Research International*, 9, 243-247.

**Methods**

*Fabrication of Silicone rubber specimens*

We employed the Polycraft GP3481-F RTV mould-making Silicone rubber. This is a two-part material comprising the Polycraft GP3481-F RTV base which when mixed with the Polycraft RTC-10F/VF catalyst results in a room temperature condensation curing reaction. The base and catalyst were mixed in ratio 20:1 (by weight) and poured into Perspex moulds to produce specimens of the required geometry. The cast rubber was then cured for 48 hrs at 4°C to slow the curing process. This allowed any trapped air bubbles to escape. The samples were de-moulded after a further 4 days of curing at room temperature. We confirmed the absence of porosity in the samples; see high resolution scanning electron micrographs and XCT images in Supplementary Figs. S1a and S1b. The HDPE and Nylon specimens were machined from 12 mm thick sheets of (HD-PE300 corresponding to 0.3 million g mol$^{-1}$) and Nylon6 (cast polyamide 6), respectively. All specimen geometries along with specimen dimensions are detailed in Supplementary Table S1.

The composite Silicone rubber beams with the $ZnI_2$ tracers were manufactured using the same protocol except that the beam was cast in two parts. A mixture comprising the base and 5% by wt. $ZnI_2$ dissolved into the catalyst and was first poured into the mould to cast the bottom half of the beam. Subsequently, we poured a mixture comprising just the base and catalyst to produce a beam with the tracers below the neutral axis (Fig. 3a). To confirm the adhesion of $ZnI_2$ to Silicone rubber we painted patterns of $ZnI_2$ on the surface of Silicone rubber specimens. It was clear that the $ZnI_2$ adhered to the rubber and this was further evidenced by deforming the specimens and observing that the $ZnI_2$ pattern deformed conformally with the surface of the rubber (Supplementary Fig. S2).

*In-situ X-ray tomography rigs*

To facilitate in-situ mechanical loading of specimens in conjunction with the high accuracy X-ray tomography of required for FETC, we developed bespoke loading rigs (Fig. 1a and Supplementary Fig. S3). The versatility of the testing stage allows for seamless adaptation to accommodate various fixtures, including tensile jaws, compression platens as well as 4-point and 3-point bending rigs (Supplementary Fig. S3). As is typical, our tomography acquisition and reconstruction processes were calibrated for cylindrical sample geometries. This is ideal for the cylindrical shapes of our compression/tension specimens. To ensure accuracy of the reconstructions for the bending specimens, we developed bending fixtures to allow the bend tests to be conducted such that the longitudinal direction of the beam was vertical. This implied a near axisymmetric scan geometry during the rotation of the stage. All jigs and fixtures were made from acrylic/Perspex, and the loading frame was built from 5 mm thick Perspex tubes (Supplementary Fig. S3). The selection of these materials was guided by their low X-ray attenuation properties which ensured minimal interference with X-rays, and a reduction in the X-ray scattering to mitigate against scan artifacts. For load measurement, a precision 100 N load cell was employed, while crosshead displacements were monitored using a vernier scale with a 1 µm resolution.

*X-ray tomography method for FETC*
X-ray tomographs were obtained using the Nikon XTH 225 ST system. X-rays were generated using a Tungsten source (target), and a 2000 pixel × 2000-pixel CCD detector with a resolution of 150 µm pixel$^{-1}$ captured the transmitted X-rays. Geometric magnification achieved via the location of the detector and sample stage relative to the X-ray source enhanced the resolution of the scans to 30 µm. Copper filters were introduced at the X-ray source to eliminate low-energy X-rays in the Bremsstrahlung spectrum, mitigating beam hardening effects at the specimen-air interface. These filters are also crucial for the FETC method to enable an increase in flux without oversaturating the detector (Supplementary S1). We employed an exposure time of 1 s to strike a balance between a reasonable scan time (~50 minutes for 3142 projections) while also maintaining a high signal-to-noise ratio (the photon count through the specimen is at least 1.5 times the background noise of the detector). To optimize the X-ray source, we initially adjusted the anode voltage to achieve a transmissivity of ~30% through the specimen, with the digital gain kept at a minimum. Subsequently, the current was increased until it reached a point of oversaturation on the detector. Various thicknesses of copper filters (ranging from 0.1 mm to 1 mm) were then employed to further increase the flux by progressively increasing the anode current until the spot size of the X-ray source at 30 W attained the desired scan resolution of 30 µm (Supplementary Table S1 lists the X-ray scan parameters for all measurements reported here). Tomographic reconstruction using scans of the 3142 projections was performed using a backpropagation algorithm and beam hardening corrections applied using polynomial curves to ensure the mean deviation in gray values between the centre and the edge of the specimen was less than 10%. The resulting reconstructed volumes were exported as uncompressed .raw files and imported into VGStudioMax 2022.4 for post-processing analysis. The in-situ loading, using the rigs described above, was interrupted periodically to acquire scans via this protocol. Examples of reconstructed images at various stages of the loading are shown in Supplementary Fig. S4. Detailed verifications of the FETC observations were performed using laser and 3D digital image correlation (3D DIC) measurements (Supplementary S3 and Extended Data Fig. 3). At-least 3 repeat tests were conducted for each measurement reported here to confirm reproducibility of the results.

*Fast radiographs for observation of the motion of the tracers*
The 50 mins required for each tomographic scan implied that we could not observe via tomography the anticipated rapid motion of the $ZnI_2$ tracers attached to the mobile phase. We therefore developed a protocol to acquire 2D radiographs at an interframe time of 500 ms during 4-point bend tests. In radiographs we need to minimise any parallax error associated with the X-ray cone being emitted. For this we made two modifications: (i) we switched from the Tungsten (W) target used for tomography to a Molybdenum (Mo) target which reduces the cone angle of the emitted X-rays from ~40º to ~20º and (ii) employed a collimation rectangular slit (Fig. 3a) of size 10 mm × 15 mm made from a 5 mm thick steel plate that was placed near the source. This created a near collimated X-ray beam over a 10 mm × 15 mm widow to visualise a portion of the beam undergoing deformation (Fig. 3a). Radiographs (Fig. 3d) were captured utilizing X-rays generated at 50 kVp (limited by the Mo target) and 10 W power. These radiographs were used to extract the tracer front position (Supplementary S3) as a function of time (Fig. 3e) and the reproducibility confirmed via at-least 3 repeat tests.

*Digital Volume Correlation (DVC) analysis*
The uncompressed ".raw" volume files were loaded into the commercial software VGStudioMax 2022.4 and organised sequentially in order of the loading stages. For example, for the tension results shown in Fig. 1 scans were conducted at displacement increments of $\Delta U = 2$ mm. In the initial DVC analysis, two stages corresponding to $U = 0$ mm and 2 mm were loaded into the analysis and designated as Volume A and Volume B, respectively. Before initiating the DVC analysis, the air within Volume A was eliminated using a region grower algorithm based on the gray value range within the Silicone rubber. An ~300 μm layer of air around the specimen was retained in Volume A to delineate the boundary. The Perspex platens were also incorporated in the cropped Volume A. This inclusion served as a reference for optimizing the global DVC analysis. Subsequently, the DVC analysis module was activated, and a correlation was performed between Volume A and Volume B. The correlation box was optimised during the analysis, and we typically achieved converged results for the displacement field with a correlation box comprising $10 \times 10 \times 10$ voxels (with a voxel size of ~ 30 μm). We performed numerous checks on the converged displacement field including (i) agreement with the platen displacements and (ii) correct prediction of the overall deformed shape of the specimen. The resultant displacements were then exported to a cubic mesh of the same size as the correlation window. This DVC analysis process was repeated for each consecutive loading stage and the total deformation gradients determined using a multi-stage analysis (Supplementary S1).

**Acknowledgements**
The authors acknowledge funding from the UKRI Frontier Research grant "Graph-based Learning and design of Advanced Mechanical Metamaterials" with award number EP/X02394X/1.

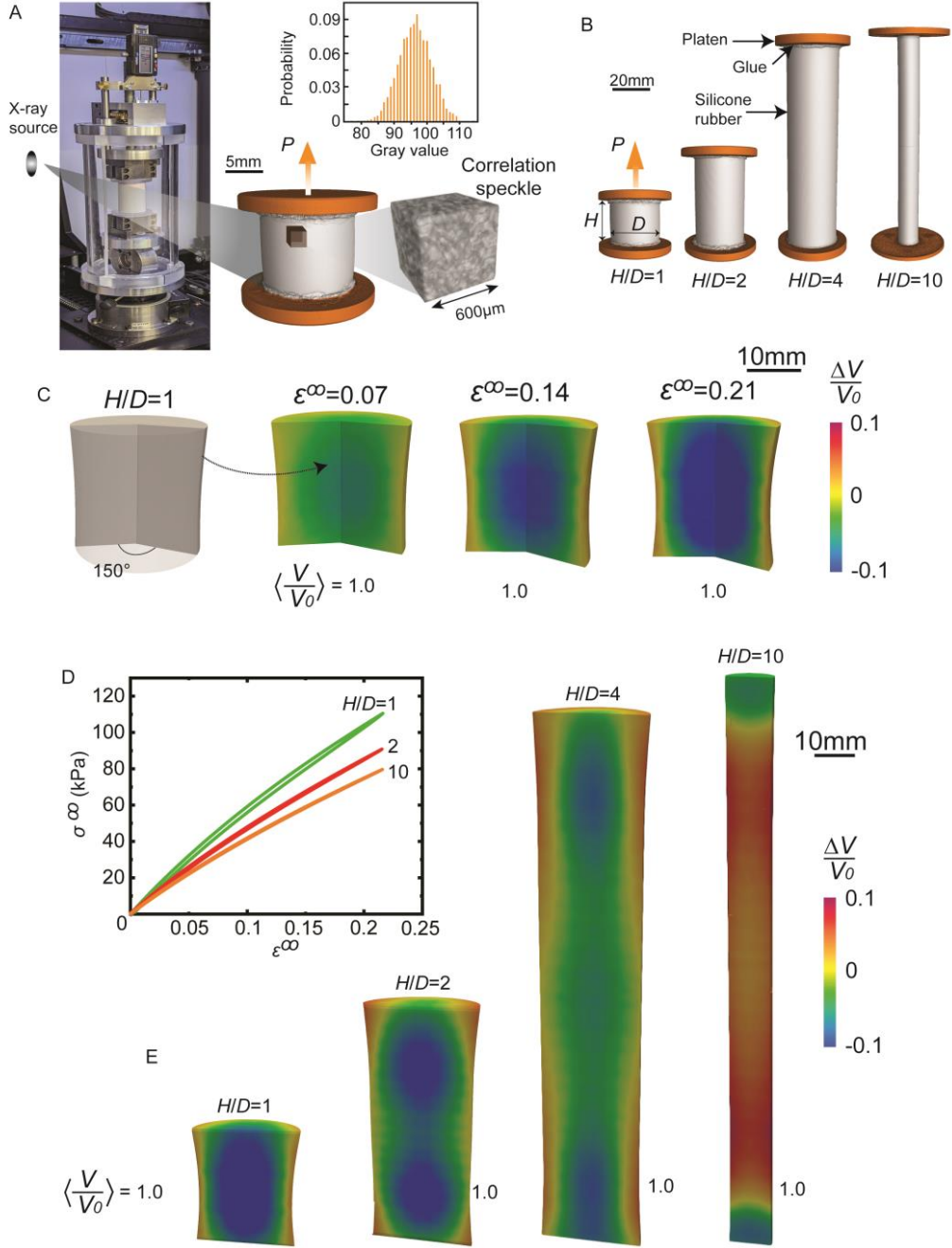

**Figure 1: Globally incompressible deformation of Silicone rubber accompanied by local volume changes.** (a) The loading setup, with in-situ XCT imaging. The computed grayscale tomograph of a 600 μm cube within a cylindrical Silicone rubber specimen obtained from FETC illustrates the inherent speckle within the material. (b) Cylindrical Silicone rubber specimens of varying aspect ratios $H/D$ glued to platens and subjected to uniaxial tension. (c) XCT image of the $H/D = 1$ specimen deformed to $\varepsilon^\infty = 0.21$ along with the evolution of the spatial distribution of volumetric strain $\Delta V/V_0$ on internal planes with increasing strain $\varepsilon^\infty$. (d) The nominal tensile stress $\sigma^\infty$ versus strain $\varepsilon^\infty$ curves for specimens with aspect ratios in the range $1 \leq H/D \leq 10$ and (e) comparison of the volumetric strains at $\varepsilon^\infty = 0.21$ for the range of specimen aspect ratios investigated here. In (c) and (e) the volumetric strains are plotted on the deformed configuration, and we include for each case the volumetric deformation $\langle V/V_0 \rangle$ averaged over the whole specimen.

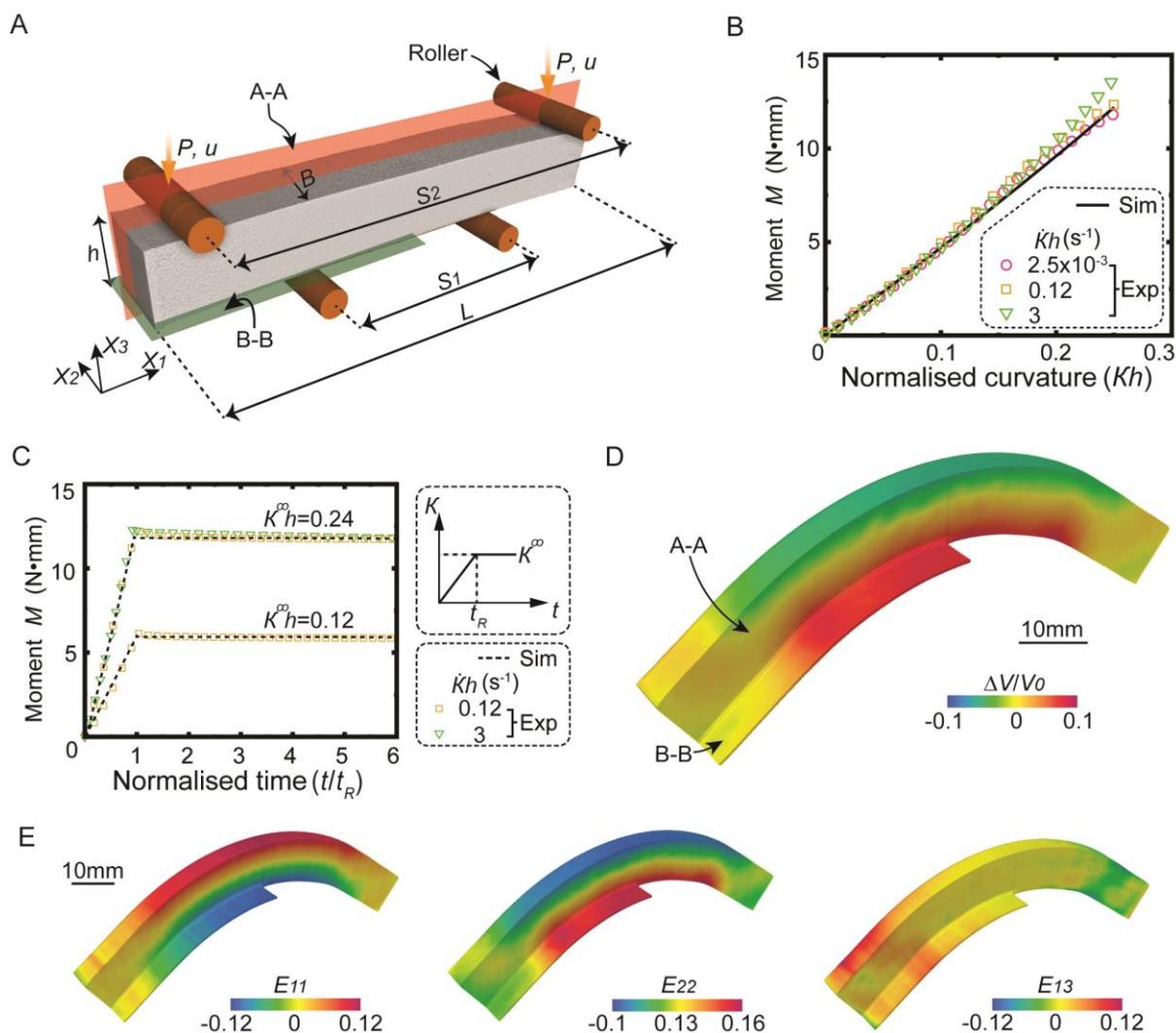

**Figure 2: Bending of the Silicone rubber provides insights into deformation mechanisms.**
(a) Sketch of the 4-point bending apparatus with leading dimensions labelled (dimensions listed in Supplementary Table S1). (b) Comparison of the measured and predicted moment $M$ versus normalised curvature $\kappa h$ response for selected loading rates $\dot{\kappa} h$. (c) Measurements and predictions of the moment relaxation response for step loading under 4-point bending. The inset shows a sketch of the loading history with the imposed curvature increasing linearly up to time $t = t_R$ such that the final curvature is $\kappa^\infty \equiv \dot{\kappa} t_R$. (d) Spatial distribution of volumetric strain $\Delta V/V_0$ within the beam bent to $\kappa^\infty h = 0.24$ and (e) corresponding distributions of selected components of the Green-Lagrange strain $E_{ij}$.

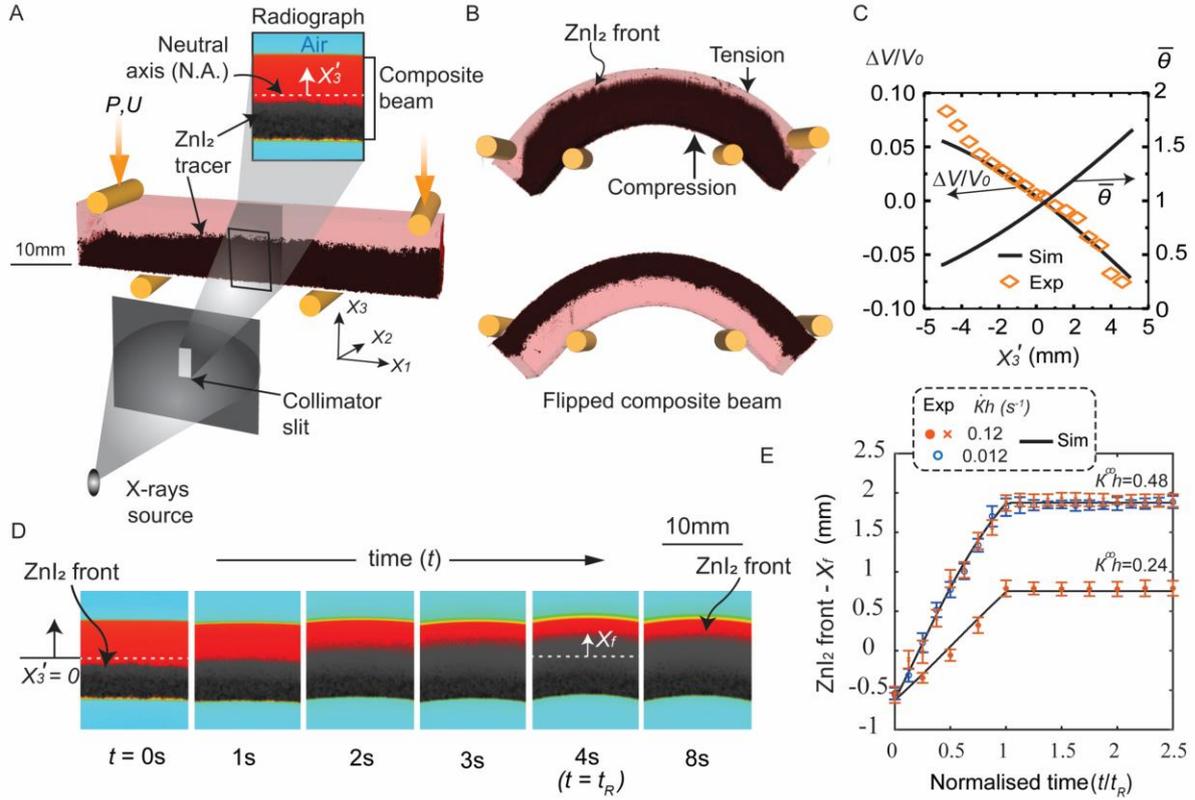

**Figure 3: Evidence of the presence of a mobile phase within Silicone rubber and measurements of its mobility.** (a) X-ray tomographic reconstruction of a composite Silicone rubber beam with dissolved $ZnI_2$ in the lower half serving as an X-ray tracer. The sketch shows the setup of the fast radiography 4-point bend setup with the collimation slit. The definition of the local co-ordinate system $x_3'$ fixed to the neutral axis of the beam is also indicated. (b) X-ray tomographic reconstructions of the composite beam subjected a normalised curvature $\kappa h = \pm 0.24$. The two curvatures correspond to bending of the beam with the tracers on tensile and compression side, respectively. (c) Predictions and measurements of the variation of the volumetric strain $\Delta V/V_0$ as well as predictions of the normalised concentration $\bar{\theta}$ of the mobile phase across the thickness of the beam subjected to $\kappa h = 0.24$. (d) Montage of radiographs showing a central section of the composite beam subjected to a step loading 4-point bend experiment with $\dot{\kappa} h = 0.12 \text{ s}^{-1}$ and $t_R = 4 \text{ s}$ (i.e. $\kappa^\infty h = 0.48$). (e) Comparison between measurements and predictions of the tracer front position $X_f$ as a function of normalised time $t/t_R$. The two cases with $\dot{\kappa} h = 0.12 \text{ s}^{-1}$ ($\kappa^\infty h = 0.48$ and $0.24$) correspond to $t_R = 4 \text{ s}$ and $2 \text{ s}$. Error bars associated with the measurements are included (Supplementary S3). In (d & e) time $t = 0$ corresponds to the instant that loading commenced.

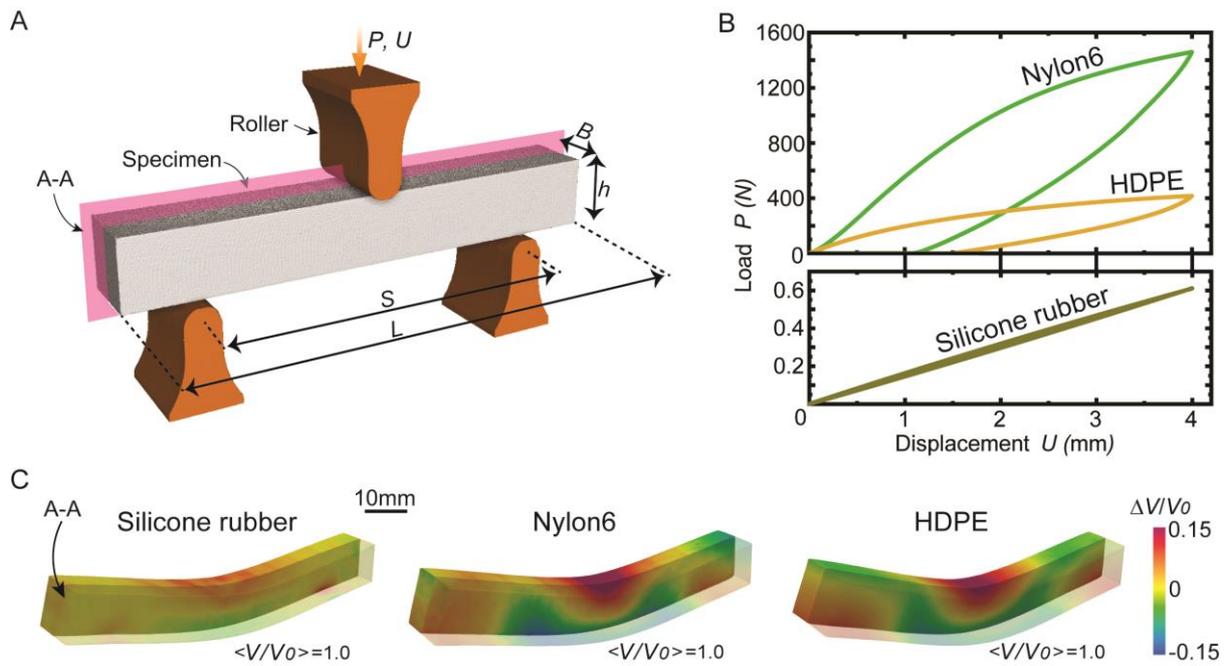

**Figure 4: Local compressibility but global incompressibility is a phenomenon that is observed for a range of synthetic polymers ranging from rubbers to engineering thermoplastics.** (a) X-ray tomographic reconstruction of the 3-point bending apparatus with leading dimensions labelled (dimensions listed in Supplementary Table S1). (b) The measured applied load $P$ versus displacement $U$ of central roller for the 3-point bending of geometrically identical Silicone rubber, Nylon6 and HDPE beams. (c) Distributions of the volumetric strains $\Delta V/V_0$ within the Silicone rubber, Nylon6 and HDPE beams deformed to $U = 4$ mm. The strains are plotted on the deformed configurations of the beam. In each case we include the volumetric deformation $\langle V/V_0 \rangle$ averaged over the whole specimen.

**Extended Data Figures**

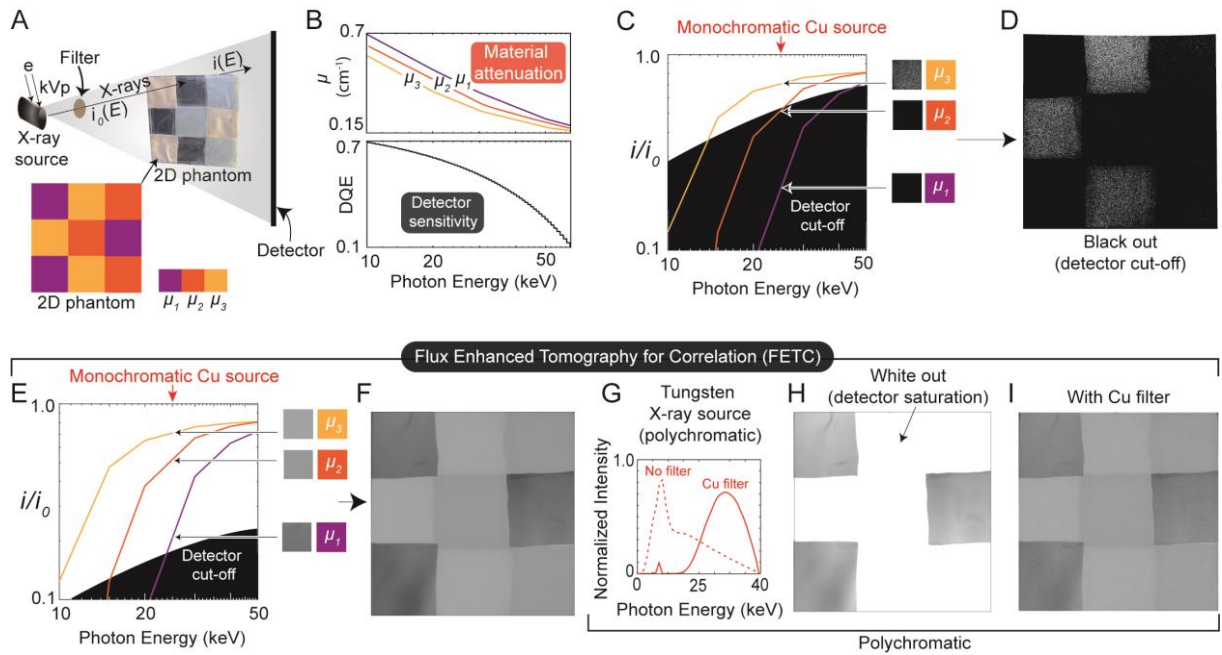

**Extended Data Figure 1: The principle of FETC.** (a) The experimental setup of a 2D phantom to illustrate the idea to magnify the phase contrast in nominally homogeneous materials. (b) The linear attenuation co-efficients $\mu_j$ of the 3 materials used in the 2D phantom versus X-ray photon energy and the corresponding detection quantum efficiency (DQE) of the 100 μm thick Silicon sensor array detector used in the measurements. (c) Predictions of the normalized transmitted intensities $\hat{\iota} \equiv i/i_0$ for the 3-phase of the phantom as a function of the photon energy of a monochromatic X-ray source. The normalized detector cut-off limit $i_{cr}/i_0$ based on the detector DQE from (b) is also indicated for an X-ray flux corresponding to a 50 μA electron current. (d) The measured radiograph of the 2D phantom with a Copper X-ray target using a 25 kVp source and a 50 μA electron current. (e) Effect of increasing electron current to 150 μA on the transmitted intensities and the detector cut-off. (f) Radiograph of the 2D phantom with a Copper X-ray target using a 25 kVp source and a 150 μA electron current. (g) The energy spectrum using a Tungsten target with 40 kVp source and current of 150 μA with and without a 0.25 mm thick copper filter. (h, i) Adding a 0.25 mm thick copper filter between the X-ray source and 2D phantom allows all phases of the phantom to be distinguished using a Tungsten target with 40 kVp source and current of 150 μA.

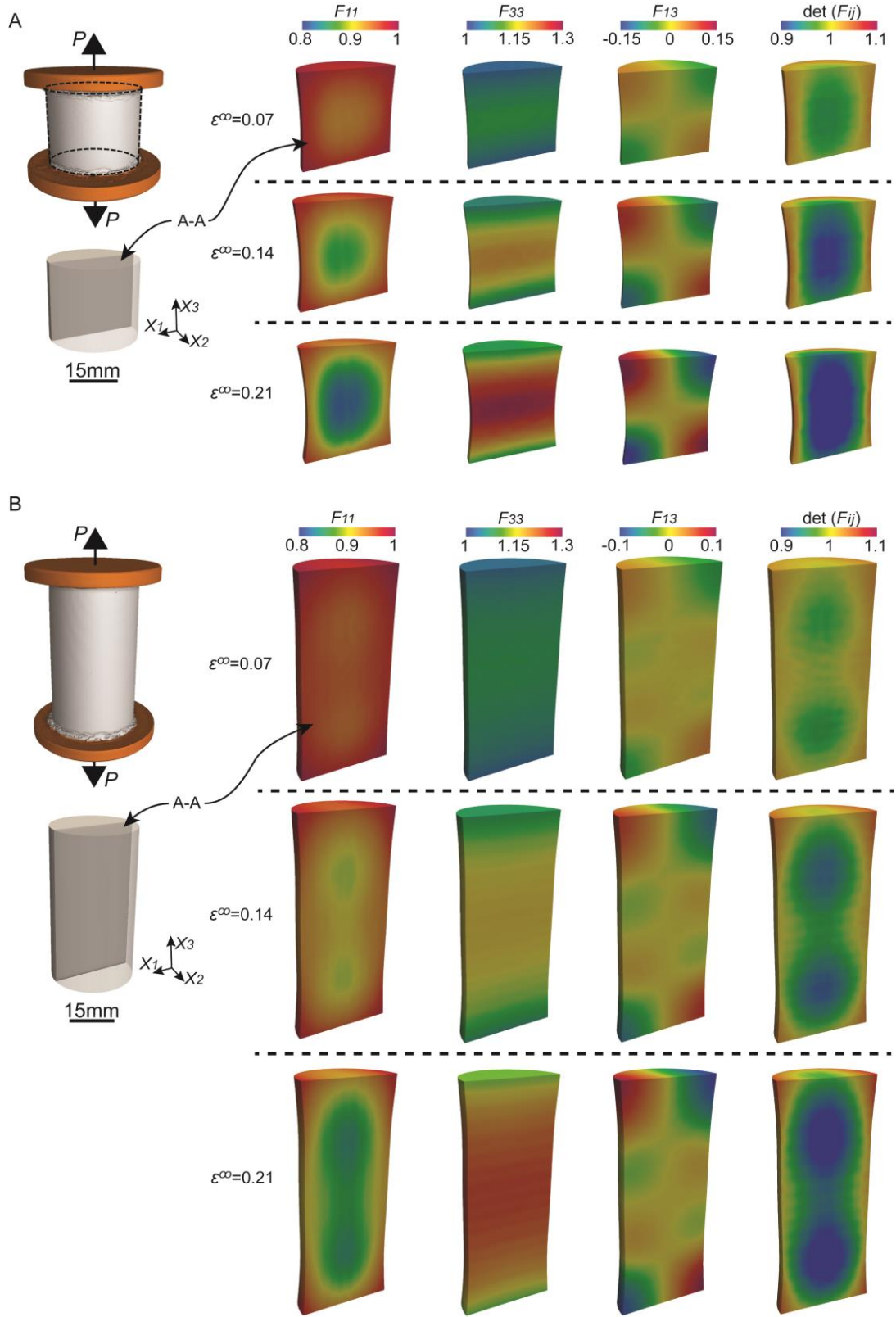

**Extended Data Figure 2: Spatial distributions of the measured deformation gradients.** Spatial distribution of selected components of $F_{ij}$ on a diametrical plane within the cylindrical Silicone rubber specimens subjected to tension. We include the components $F_{11}, F_{33}, F_{13}$ and $\det(F_{ij})$ at three levels of $\varepsilon^\infty$ for the (a) $H/D = 1$ and (b) $H/D = 2$ specimens. The deformation gradients are plotted on the deformed configuration.

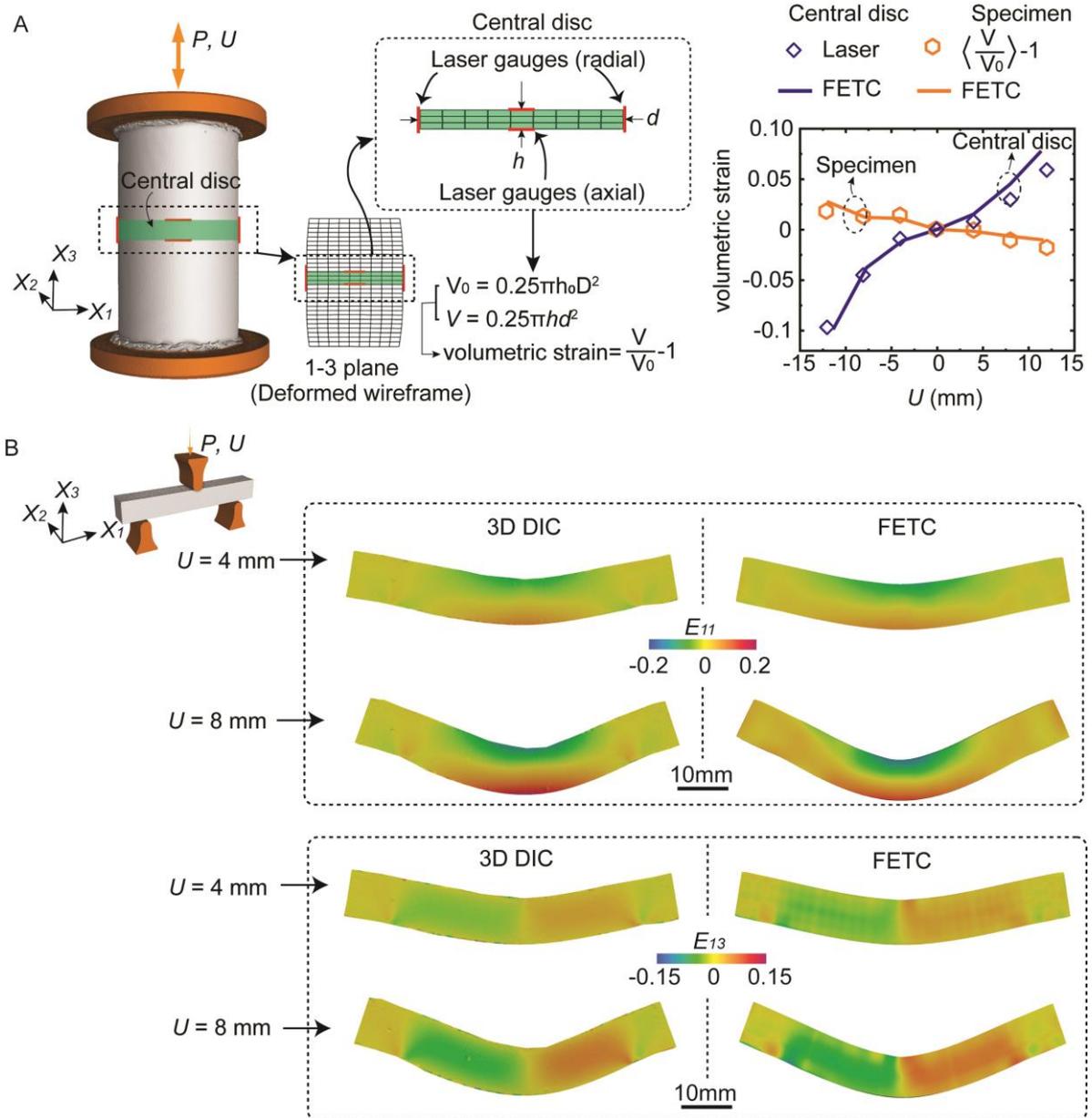

**Extended Data Figure 3: Measurements independent of X-ray setup to validate the FETC observations.** (a) Tension/compression tests on the $H/D = 2$ specimen measuring the axial and radial strains on a central disc of height $h_0 = 1.2$ mm using laser gauges. The insets show a wireframe of the deformed configuration (extracted from the FETC measurements to demonstrate that the central section remains planar) as well as the definition volume change as measured via the laser extensometers. The volumetric strains of the central disc as a function of the imposed axial displacement $U$ are plotted and compared with the corresponding FETC measurements averaged over the same volume. The measurements of the volume of the entire specimen using the FETC measurements of $\det(F_{ij})$ and the specimen outline from the X-ray tomographs (labelled $\langle V/V_0 \rangle - 1$) are also included. (b) Comparison between distributions of the Green-Lagrange strains $E_{11}$ and $E_{13}$ on the surface of the Silicone rubber beam in 3-point bending (geometry identical to Fig. 4) using 3D digital image correlation (3D DIC) and FETC measurements. The good agreement not only confirms the fidelity of FETC but also demonstrates that there is negligible radiation damage on the Silicone rubber specimens due to the X-rays. See Supplementary S3 for details of the methods.

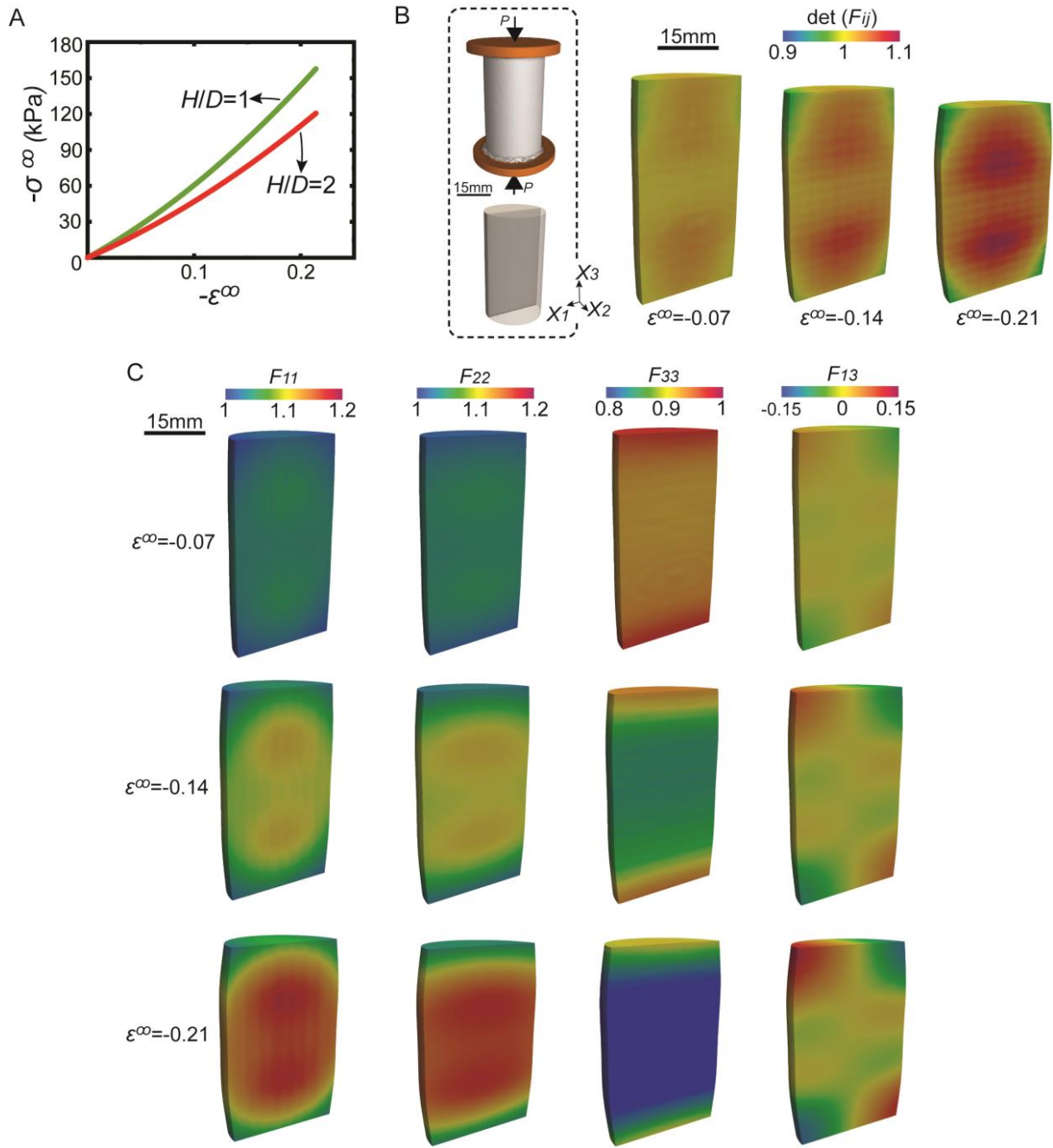

**Extended Data Figure 4: Compression of Silicone rubber.** (a) The compressive nominal stress $\sigma^{\infty}$ versus strain $\varepsilon^{\infty}$ responses of specimens glued to platens and with aspect ratios $H/D = 1$ and $H/D = 2$. (b) Spatial distribution of $\det(F_{ij})$ on a diametrical plane within the specimen for three values of $\varepsilon^{\infty}$. (c) Selected components of $F_{ij}$ on a diametrical plane within the cylindrical Silicone rubber specimen. We include the components $F_{11}, F_{22}, F_{33}$ and $F_{13}$ at three levels of compressive strain $\varepsilon^{\infty}$ within the $H/D = 2$ specimen. All spatial distributions are plotted on the deformed configuration.

# Supplementary Text

*S1: Flux-enhanced tomography for digital volume correlation*
*S2: Fitting of hyperelastic models to Silicone rubber data*
*S3: Additional experimental results & techniques*
*S4: A thermodynamically consistent rubber elasticity framework with a mobile species*

## Supplementary S1: Flux-enhanced tomography for digital volume correlation

*S1.1: Tomography to enhance contrast for digital volume correlation*

A two-dimensional (2D) phantom comprising 3 phases arranged in a checkboard (Extended Data Fig. 1) is used to illustrate the principle of flux enhanced tomography to magnify gray value contrast in nominally homogeneous materials. The 3 phases of the phantom are standard polymers viz. Silicone rubber, polytetrafluoroethylene (PTFE) and polyethylene (PE). Then given a X-ray photon intensity $i_0(E)$ per unit photon energy $E$ incident on the phantom, the transmitted intensity per unit energy at location $x$ where phase $j$ is present is given in terms of X-ray attenuation co-efficient $\mu_j$ of phase $j$ by Beer-Lamberts law as $i(x, E) = i_0(E) \exp(-\mu_j h)$ where we do not sum over phases as in the 2D phantom each photon only travels through one phase of thickness $h \approx 100$ μm (Extended Data Fig. 1a). The variation of the linear attenuation co-efficients $\mu_j$ with X-ray photon energy $E$ for the three phases is known [1] and illustrated in Extended Data Fig. 1b. The corresponding normalized transmitted intensities $\hat{\imath} \equiv i/i_0$ are included in Extended Data Fig. 1c with the upper and lower bound corresponding to phase $j = 1$ (Silicone rubber) and $j = 3$ (PE), respectively. Given a monochromatic X-ray source (e.g., with a photon energy of 25 keV) an ideal detector (with 100% detection efficiency) will measure a spatial distribution of intensities over $j = 1, \ldots 3$ to give the spatial distribution of phases in the 2D phantom.

However, all practical detectors have a finite quantum efficiency and the detection quantum efficiency (DQE) of a 100 μm thick Silicon detector [2] is included in Extended Data Fig. 1b (a DQE of x % means that only the fraction x/100 of the photons incident on the detector contribute to the electron current from the detector with a sufficiently low noise to signal ratio). Thus, for a monochromatic source, the detector is unable to produce a sufficient detection current (i.e., detection of incident photons does not have a sufficiently low noise to signal ratio) when the transmitted X-ray intensity $i < i_{cr} = i_{ref}(1 - \text{DQE})$ where $i_{ref}$ is a constant related to the exposure time and electrical system measuring the electron flux from the detector. We estimated $i_{ref}$ for our Nikon XTH-225 ST and the corresponding detection cut-off threshold is included in Extended Data Fig. 1c. We now observe that at $E = 25$ keV the measurable transmitted flux is severely limited by the detector sensitivity such that the high attenuation phases $j = 2,3$ are expected to be "blacked out". We employed a copper target with a 25 kVp source and a 50 μA electron flux to produce a near monochromatic X-ray beam with a photon energy $E \approx 20$ keV. The resulting measured radiograph (Extended Data Fig. 1d) indicates that the richness of the phase differences within the 2D phantom are not captured: the "speckle" on this radiograph is very coarse compared to the ground truth (Extended Data Fig. 1a) as the $j = 2,3$ phases with a high attenuation are blacked-out and thereby merged in the radiograph.

Detector sensitivity is thus the limiting factor to observe the speckle that is inherent in the low phase contrast 2D phantom that we have designed to illustrate the FETC principle. An obvious solution to this problem is to increase the transmitted flux by enhancing the photon flux $i_0$ incident on the phantom (this can be achieved by increasing the flux of electrons impacting the X-ray target). Then while the normalized transmitted flux $\hat{i}$ remains unchanged, the normalized critical detection flux $i_{cr}/i_0$ reduces by the factor that $i_0$ is increased by since $i_{ref}$ is a detector property that is unaffected by $i_0$. Increasing $i_0$ by a factor of 2 is expected to bring the all the phases in the phantom within the detection limits (Extended Data Fig. 1e). We then repeated radiograph of the phantom (Extended Data Fig. 1a) by keeping the copper target and source voltage unchanged but increasing the electron flux from 50 µA to 150 µA. The corresponding increase in the X-ray photon flux allows the detector to differentiate between all the phases in the phantom; see radiograph in Extended Data Fig. 1f.

While the copper target generates a near monochromatic X-ray beam, the energies are low (soft X-rays) and insufficient to penetrate most realistic specimens. Common laboratory-based X-ray sources therefore use Tungsten (W) which generates a polychromatic X-ray beam including high energy X-rays. The wide spectrum of X-ray photon energies for a standard source excitation voltage of 40 kVp and a W target is shown in Extended Data Fig. 1g [3]. The transmitted flux through a location $x$ of the 2D phantom where phase $j$ is present, as measured by the detector, is then

$$i_T = \int_0^\infty \{i_0 \exp(-\mu_j h_j) - i_{cr}\} dE. \quad (1.1)$$

Here $\{\cdot\}$ denotes the Macaulay bracket which ensures that a transmitted current below the detection threshold is not included (note we are neglecting the negligible attenuation due to air). A radiograph with a 40 kVp source and a current of 150 µA is shown in Extended Data Fig. 1h. However, regions of "whiteouts" are now observed. The whiteout effect is understood as follows. Observe from the Tungsten spectrum in Extended Data Fig. 1g that there is a high flux at low energies. Therefore, at these low energies $i_T$ is high for the low attenuation phases since both transmission is high and $i_{cr}$ is low. Then recalling that the detector (for a given exposure time) saturates when the flux exceeds a value $i_{sat}$, the low attenuation phases get merged as the detector saturates. This problem was avoided in the monochromatic beam as low photon energies that contribute to the whiteout did not exist. The whiteout drawback of the polychromatic beam cannot be solved by reducing the flux $i_0$ as that leads to blackout as seen in Extended Fig. 1d. Rather it is to filter out the low energy X-ray photons impacting the 2D phantom. We did this using a 0.25 mm thick copper filter placed between the X-ray source and the 2D phantom (Extended Data Fig. 1a). The corresponding radiograph (Extended Data Fig. 1i) using a 40 kVp source and current of 150 µA shows that the whiteout problem is now eliminated, and the detector differentiates between all the phases in the phantom. The combination of high X-ray fluxes along with the filtering of the low energy X-ray photons enables us to generate tomographs suitable for correlation in nominally homogenous materials. We call this method flux enhanced tomography for correlation (FETC).

We illustrate FETC using a cylindrical Silicone rubber specimen (Methods) of diameter 28 mm and height 56 mm using a W target, source voltage 150 kVp, 0.25 mm Cu filter and currents ranging from 100 µA to 200 µA. The computed grayscale tomograph of a 600 µm cube within the cylindrical specimen is shown in Fig. S1 for increasing currents along with the corresponding histograms of gray value pixel counts within the cube. Clearly increasing current significantly enhances contrast within the specimen by differentiating between the phases with mild X-ray contrast in the nominally homogenous Silicone rubber. This contrast obtained via FETC serves as a speckle for digital volume correlation (DVC).

*S1.2   Large deformation digital volume correlation (DVC)*

The problem in DVC to compute strain distributions in a large deformation process is that the correlation algorithm fails to converge much like in digital image correlation (DIC). Both DIC and DVC rely on finding the maximum correlation array between pixel intensity array subsets on two or more corresponding images. This gives the translational and rotational shift between the images. However, extreme deformations means that pixel intensity array subsets (known as the correlation boxes) themselves distort so that images of two consecutive deformation states can no longer be accurately correlated: in DIC the probability of correlation drops to less than 10% for deformations more than 30% [4]. We therefore developed a large deformation DVC procedure which involves only correlation between sequential states of deformation where the incremental deformation is small enough to allow for a high correlation co-efficient. These incremental deformations are then combined to give the overall large deformations. However, in each increment of deformation the displacements are computed in a Eulerian frame that is fixed in space (i.e., the laboratory frame within which the X-ray scans are conducted) and thus displacements are computed at a different set of material points for each increment of deformation. An algorithm akin to remeshing in finite element (FE) computations is therefore employed to interpolate the displacements and additively compute the displacements over the full deformation on a fixed set of material points.

As a first step we computed the displacement between consecutive loading increments using the global correlation algorithm [5-6] implemented in VGStudioMax 2022.4. These incremental displacement fields were then fed into the multi-step DVC algorithm. The principle of the multi-step DVC approach is demonstrated in Fig. S5. The displacement field of the $(n-1)$ step comes from the DVC analysis between the $(n-1)$ and $n$ steps. However, the DVC analysis outputs the displacement field in the form of the three orthogonal displacements $u_i$ at individual points in a 3D grid with the fixed Eulerian coordinates. This means that the discrete displacement field outputted by the DVC analysis for step $(n-1)$ (which is calculated as the increment in displacement between steps $(n-1)$ and $n$) need not be at the same material point as that for step $n$ that is calculated as the increment in displacement between steps and $n$ and $(n+1)$. Consequently, the displacement field of the material points cannot be tracked through multiple steps.

We now illustrate the multi-step DVC approach using a Lagrangian/Eulerian finite element setting as a basis to explain the basic idea. Prior to deformation the material is discretised into fixed material points akin to a Lagrangian finite element mesh with the nodes of the mesh following material points. Similarly, space is also discretised into a grid that is fixed in space (i.e., a Eulerian grid) on which the DVC analysis between two consecutive steps outputs the displacement field. The nodes of the Lagrangian mesh are then located within the Eulerian DVC mesh and the displacement of the nodes of the Lagrangian mesh incremented by the incremental DVC displacement. This then provides not only the total displacement of the material points associated with nodes of the Lagrangian mesh but also the spatial position of the Lagrangian nodes to locate within the Eulerian grid for the next increment of the DVC analysis. To visualise this procedure, we show in Fig. S5 a cube of material (nodes labelled A through E) in step $(n-1)$ and the deformation and motion of this cube through steps $n$ and $(n+1)$. The DVC analysis between steps $(n-1)$ and $n$ give the displacements on the Eulerian grid. The incremental displacement for example of Lagrangian node E of the cube between steps $(n-1)$ and $n$ is then determined by first locating node E in step $(n-1)$ within the Eulerian grid. It lies within the Eulerian grid surrounded by Eulerian grid nodes 1 through 8 as shown in the inset of Fig. S5. Then the incremental displacement $\{\Delta u^{(E)}\}^{(n-1)}$ in vector form of the material point associated with node E between steps $(n-1)$ and $n$ is given by

$$\{\Delta u^{(E)}\}^{(n-1)} = [N(\zeta^{(E)})]\{d\}^{(n-1)}, \quad (1.2)$$

where $[N]$ is the matrix of shape function of the Eulerian grid element with nodes 1-8, $\zeta^{(E)}$ the parametric co-ordinates of the material point E within this grid element and $\{d\}^{(n-1)}$ the vector of incremental DVC obtained displacements of the Eulerian nodes 1-8 between steps $(n-1)$ and $n$. The total displacement of node E is then

$$\{u^{(E)}\}^{(n-1)} = \{u^{(E)}\}^{(n-2)} + \{\Delta u^{(E)}\}^{(n-1)}, \quad (1.3)$$

and the location $\{X^{(E)}\}^{(n)}$ of node E for the analysis between steps $n$ and $(n+1)$ is given by

$$\{X^{(E)}\}^{(n)} = \{X^{(E)}\}^{(0)} + \{u^{(E)}\}^{(n-1)}, \quad (1.4)$$

where $\{X^{(E)}\}^{(0)}$ is the location of node E in the undeformed state. This allows for the process to be repeated for the incremental displacements between steps $n$ and $(n+1)$ with node E now again located within the Eulerian grid for step $n$ (it is now seen to lie in the Eulerian grid surrounded by grid nodes 11-18) as shown in Fig. S5. While here for the sake of clarity we have explained the multi-step DVC in terms of finite element shape functions, we implemented this procedure using 3D cubic B-splines [5] to ensure smooth deformation gradients.

*S1.3  Description of the in-situ X-ray setups*
We constructed a bespoke loading rig to conduct a range of in-situ mechanical tests. The driver behind this was to ensure minimal interference of the loading frame with the X-rays while acquiring the scans for the tomographic reconstruction: FETC requires especially accurate reconstructions to ensure accurate inference of the displacement field. A sketch of the loading fixture along with important dimensions and materials used is shown in Fig. S3. The rig was fixed to the loading stage of the XCT setup and rotated about the vertical axis to acquire the scans for the reconstructions. Another important feature of the setup is that we could modify it to change the type of loading we wanted to impose (we used it to perform tensile tests, compression tests, 4-point bending and 3-point bending). In addition, the setup permitted us to bring the specimens very close to the X-ray source (Fig. S3) to ensure maximum geometric magnification and thereby resolution. The setup was able to measure axial loads and imposed displacements as described in the Methods Section.

Sketches of the three main setups used in this study are shown in Fig. S3, viz. tension/compression of cylinders, 4-point bending and 3-point bending. Leading dimensions and materials used for these setups are also indicated. As discussed in the Methods section the bend tests were conducted with the longitudinal axis of the beam vertical: this ensured a near axisymmetric geometry about the vertical axis which significantly enhanced the quality of the reconstructions. Consequently, while the imposed bending displacements were measured in the in-situ setup we did not measure the corresponding loads. The measured moments/loads shown in Figs. 3 and 4 were performed independently with the loading fixtures rotated so that the longitudinal axis of the beam is horizontal. No X-ray reconstructions were conducted during these tests when the beams were horizontal. However, repeat tests confirmed the near complete reproducibility of the measured load versus displacement responses implying that it suffices to use the load measured from independent tests to relate the FETC measurements of the strain field to the measured macroscopic load-displacement responses of the beams.

**Supplementary S2: Fitting of hyperelastic models to Silicone rubber data**

Full field measurements of displacements (and thereby strains) can be used to determine material parameters and/or constitutive models. This has been done extensively using surface displacement data obtained via digital image correlation (DIC) [7]. The methods to solve the inverse problem of material parameter determination include virtual fields method (VFM) [8] and finite element model updating (FEMU) [9]. Here we use FEMU to fit 3 widely used hyperelastic models for rubber, viz. Neo-Hookean [10], Arruda-Boyce [11] and Mooney-Rivlin [12] to the FETC measurements of the 3D full-field displacement fields.

The experiments we consider in this section are the tension experiments on the $H/D = 1$ and $H/D = 2$ Silicone rubber cylinders glued to platens (Fig. 1b). This glued boundary condition results in a spatially varying strain field through which an appropriate constitutive model and parameters can be determined. The primary observables in the experiments are the point-wise displacement fields within the specimen as a function of the imposed platen displacement $U$ and the corresponding load $P$ versus displacement $U$ response. An illustration of this data for for the $H/D = 1$ specimen is shown in Fig. S6 where we include spatial distributions of the three components $u_i$ of the displacement for an imposed displacement $U = 6$ mm and measured $P$ versus $U$ response. The full-field measurements of $u_i$ were conducted at $\mathcal{M}$ loading stages and we define displacement and force residuals based on these measurements. The displacement residual[1] at loading stage $j$ is given by

$$\mathcal{R}_u^{(j)} \equiv \frac{1}{\mathcal{N}} \sum_{k=1}^{\mathcal{N}} \sqrt{\frac{\sum_{i=1}^{3}\left(u_i^{(k,j),\exp} - u_i^{(k,j),\sin}\right)^2}{\sum_{i=1}^{3}\left(u_i^{(k,j),\exp}\right)^2}}, \qquad (2.1)$$

where $u_i^{(k,j),\exp}$ and $u_i^{(k,j),\sin}$ are the measured and simulated displacement components in the $i$ direction at the $k = 1, \ldots \mathcal{N}$ material points and at loading stage $j$. The total displacement residue then follows as

$$\mathcal{C}_u \equiv \frac{1}{\mathcal{M}} \sum_{j=1}^{\mathcal{M}} \mathcal{R}_u^{(j)}. \qquad (2.2)$$

Similarly, the force residual is

$$\mathcal{C}_P \equiv \sqrt{\frac{\sum_{j=1}^{\mathcal{M}}(P^{(j),\exp} - P^{(j),\sin})^2}{\sum_{j=1}^{\mathcal{M}}(P^{(j),\exp})^2}}, \qquad (2.3)$$

where $P^{(j),\exp}$ and $P^{(j),\sin}$ are the measured and simulated loads, respectively at loading stage $j$. The total cost function is then [13]

$$\mathcal{C} \equiv \mathcal{C}_u + \mathcal{C}_P. \qquad (2.4)$$

*S2.1: The hyperelastic models*

We considered 3 hyperelastic models widely used in the literature to describe the deformation of rubber. These models are defined in terms of the invariants of the deformation given by

$$J = \lambda_1\lambda_2\lambda_3; \quad I_1 = \lambda_1^2 + \lambda_2^2 + \lambda_3^2; \quad I_2 = \lambda_1^2\lambda_2^2 + \lambda_2^2\lambda_3^2 + \lambda_1^2\lambda_3^2, \tag{2.5}$$

where $\lambda_1, \lambda_2, \lambda_3$, are the 3 principal stretches. It is then useful to define invariants of the isochoric deformation as $\bar{I}_1 \equiv J^{-2/3}I_1$ and $\bar{I}_2 \equiv J^{-4/3}I_2$. The strain energy density function $W$ is then written in terms of $(J, \bar{I}_1, \bar{I}_2)$. The first Piola Kirchhoff stress $P_{ij}$ (or nominal stress) and then determined from $W$ by

$$P_{ij} \equiv \frac{\partial W}{\partial F_{ij}}, \tag{2.6}$$

where $F_{ij}$ is the deformation gradient $F_{ij}$. We used the hyperelastic models and the associated parameters as implemented in commercial finite element (FE) package ABAQUS [14] and here we summarise these models:

(i) A compressible Neo-Hookean model which has no $\bar{I}_2$ dependence and two material parameters $C_{10}$ and $D_1$ such that

$$W \equiv C_{10}(\bar{I}_1 - 3) + \frac{1}{D_1}(J - 1)^2. \tag{2.7}$$

---

[1] If $\sqrt{\sum_{i=1}^{3}\left(u_i^{(k,j),\exp}\right)^2} \leq 10^{-5} U$ we neglect the contribution from material point $k$ at loading stage $j$.

(ii) The Mooney-Rivlin model [12] which is a generalisation of the Neo-Hookean model. It has an $\bar{I}_2$ dependency and three material parameters $C_{10}, C_{01}$ and $D_1$ so that

$$W \equiv C_{10}(\bar{I}_1 - 3) + C_{01}(\bar{I}_2 - 3) + \frac{1}{D_1}(J - 1)^2. \tag{2.8}$$

(iii) The Arruda-Boyce model [11] is a micro-mechanically motivated model again with no $\bar{I}_2$ dependency. Several of the parameters follow directly from micro-mechanical considerations, and this reduces the numbers of fitting parameters to three labelled here as $C_1, D_1$ and $\lambda_m$. The strain energy density function is given by

$$W \equiv C_1 \sum_{i=1}^{4} \alpha_i \left(\frac{1}{\lambda_m^2}\right)^{(i-1)} \left(\bar{I}_1^{\,i} - 3^i\right) + \frac{1}{D_1}\left(\frac{J^2 - 1}{2} - \ln J\right), \tag{2.9}$$

where $\alpha_1 = 1/2, \alpha_2 = 1/20, \alpha_3 = 11/1050$ and $\alpha_4 = 19/7000$.

*S2.2: Implementation of the FEMU method*

The FEMU method was implemented by coupling the package ABAQUS which performed the FE calculations to a Python script where a gradient-free Nelder-Mead [15] algorithm was implemented. The Nelder-Mead algorithm optimised the hyperelastic model parameters to minimize the cost function (2.4). The ABAQUS hyperelastic calculations were performed using 8-noded brick elements (C3D8 in the ABAQUS notation) with elements of size ~ 0.1mm. All displacements degrees of freedom were constrained to zero on the bottom surface of the cylinders and displacements $u_1 = u_2 = 0$ along with $u_3 = U$ applied on all the nodes on the top surface. The work conjugate to the displacement $U$ is the applied force $P$.

The calculations were performed as follows. An initial guess of the material parameters for the chosen hyperelastic model was made and the displacements and corresponding loads computed via FE calculations performed in ABAQUS. These displacements and loads were exported to a Python script running the Nelder-Mead algorithm to minimise the function (2.4). At every step the Nelder-Mead algorithm then provides new guesses of the hyperelastic model parameters, and the steps are repeated until a converged minimum is obtained. The values of the hyperelastic model parameters at this minimum are taken as the parameters that best fit the measured data. This procedure was used to infer the optimal model parameters from the measured data for tensile loading of the $H/D = 1$ and $H/D = 2$ specimens. These optimal parameters along with corresponding values of the cost function are listed in Tables S2a and S2b for the $H/D = 1$ and $H/D = 2$ specimens, respectively. To give a sense of the dominant residual we include in Table S2 both $C_u$ and $C_P$ in addition to the sum $C$: in all cases the displacement residuals are significantly greater than the force residue for reasons that we now proceed to discuss.

*S2.3: Summary of results*

The best fit predictions of the nominal stress $\sigma^\infty$ versus nominal strain $\varepsilon^\infty$ for the $H/D = 1$ and $H/D = 2$ specimens using the best fit parameters for the 3 hyperelastic models are shown in Figs. S7a and S7b and compared with the measurements. All three models agree to a very high level of accuracy with the measurements as already suggested by the force residuals $C_P$ in Table S2. The displacement residuals $C_u$ are $\gg C_P$ and we thus anticipate poor agreement of the deformation fields within the specimen.

Comparisons between the measured and predicted distributions of $F_{33}$ and $F_{23}$ on a diametrical plane through the $H/D = 1$ specimen are shown in Figs. S8 and S9, respectively for three stages of loading. Overall, the agreement between measurements and predictions seems reasonable although there are some discrepancies especially near the platens. However, a major discrepancy is observed when comparing predictions and measurements of $\det(F_{ij})$; see Fig. S10. While the measurements show a mixture of dilation and contraction as discussed in the main text the predictions are pointwise nearly incompressible with $\det(F_{ij}) \approx 1$ throughout the specimen for all three hyperelastic models. In fact, the predictions show that the hydrostatic stress is tensile throughout the specimen and even if we reduced the bulk modulus to allow volumetric straining, a mixture of contraction and dilation would not be predicted by these hyperelastic models. The corresponding predictions for the $H/D = 2$ specimen shown in Figs. S11-S13 are qualitatively similar with again the main discrepancy being in $\det(F_{ij})$. This discrepancy is not particularly surprising as these hyperelastic models will always only predict either negligible volume change (for a high bulk modulus) or volumetric dilation under tensile hydrostatic stress states.

The fact that the predictions of $\sigma^\infty$ versus $\varepsilon^\infty$ agree with measurements (Fig. S7) is somewhat misleading too. Recall that the hyperelastic model parameters predicted by FEMU differ somewhat for the $H/D = 1$ and $H/D = 2$ specimens; see Tables S2. To illustrate the discrepancy, we include in Fig. S14a predictions of the $\sigma^\infty$ versus $\varepsilon^\infty$ of the $H/D = 1$ specimen with parameters inferred from the $H/D = 2$ specimen for all the 3 hyperelastic models. Similarly, we also use the models (with parameters extracted from the $H/D = 2$ tensile test) to predict the 3-point bending response of the Silicone rubber beam (Fig. 4). This comparison shown in Fig. S14b again emphasises that the hyperplastic models cannot capture even the load versus displacement measurements with sufficient accuracy over a range of loading states (or specimen geometries).

**Supplementary S3: Additional experimental results & techniques**

*S3.1: Isotropy of the Silicone rubber and homogeneity of the specimens*
To confirm the isotropy of the as-cast Silicone rubber we cut 10 mm cubes from the $H/D = 2$ ($D = 28$ mm) cylindrical specimens from three locations within the cylinder (Fig. S15a). These specimens were then glued to platens similar to the cylindrical specimens of Fig. 1b and tested in compression. The compression tests were also conducted in the 3 orthogonal directions marked in Fig. S15a. The compressive responses in the $x_1'$ −directions for three cubes from the 3 locations are plotted in Fig. S15b while in Fig. S15c we include the response of the cube from location 2 in the three orthogonal directions. The data in Figs. S15b and S15c confirms that the as-cast Silicone rubber displays no orientational bias and that the specimen is spatially homogeneous.

*S3.2: Absence of size dependence of the response of Silicone rubber*
The rather strange behaviour we report in terms of large local volumetric strains, but overall incompressibility of the Silicone rubber specimen been shown to be associated with gradients in the strain field. Dependence of material response on not only the strains but also the gradient of strain is well-established especially in the context of metal plasticity [16-17]. Such gradient dependence introduces a length-scale into the material properties. This usually implies that tests on geometrically self-similar specimens of varying sizes result in deformation fields that do not self-similarly scale. As a consequence, the macroscopic stress versus strain responses are not independent of specimen size. While such size dependence of the properties of Silicone rubber at mm length scales has never to-date been reported, the self-similarity of the fields has never been confirmed due to the lack of suitable 3D measurement techniques. Given our FETC technique we decided to confirm this self-similarity for the tensile response of the glued $H/D = 2$ specimen.

Consider two geometrically self-similar $H/D = 2$ specimens: (i) $D = 28$ mm which is the standard size specimen in Fig. 1 and (ii) a smaller $D = 14$ mm specimen. The measured tensile $\sigma^\infty$ versus $\varepsilon^\infty$ responses for these two specimens are included in Fig. S16a. To within experimental variabilities the responses of the two specimens are identical. This is very much in line with expectation from usual rubber elasticity models. Next consider the deformation fields. Spatial distributions of selected components of $F_{ij}$ as well as $\det(F_{ij})$ on a diametrical plane within the cylindrical specimens are included in Fig. S16b for both specimen sizes and an imposed strain $\varepsilon^\infty = 0.21$. The deformation gradients are shown on the deformed configuration and the specimen images scaled so that images for both specimens are of the same size. This allows an easy check of the self-similarity of the fields. We clearly observe that the fields are indeed self-similar, and this confirms that the strange behaviour we observe cannot be attributed to usual gradient elasticity models [18].

*S3.3: Verification of the FETC measurements and confirmation of lack of radiation damage*
The FETC is a novel technique to measure strains in 3D volumes of nominally homogeneous materials without using tracer particles. Given that this is a new technique we performed independent measurements to provide confidence on the fidelity of the FETC measurements. These measurements conducted outside the X-ray setup also provide evidence that there is negligible X-ray damage to the Silicone rubber. Two types of independent measurements were performed:
  (i) Measurement, using laser extensometers, of the volume change of a central disc of height $h_0$ during tension/compression loading of the $H/D = 2$ ($D = 28$ mm)

(ii) Comparison between 3D digital image correlation (3D DIC) and FETC measurements of the surface strains for 3-point bending of the Silicone rubber specimen (Extended Data Fig. 3b).

First consider the cylindrical specimen that we tested in both compression and tension but now outside the X-ray tomography setup. Symmetry of the loading implies that on the central horizontal plane through the cylinder (parallel to the $X_1 - X_2$ plane), the shear strains vanish. This implies that a central cylindrical disc of height $h_0 \ll H$ (Extended Data Fig. 3a) remains cylindrical under tension/compression loading (we shall a posteriori confirm this). We measured the change in diameter and height of a central cylindrical section of initial height $h_0 = 1.2$ mm during compression/tension loading. The change in height was measured using a reflective axial laser extensometer with an accuracy of $\pm 10$ μm while the change in diameter $D$ of the specimen was measured with an emissive-receptive laser also with an accuracy of $\pm 10$ μm. Denoting the diameter of the disc in the deformed state as $d$ (with $d = D$ in the undeformed state) the change in volume of this central cylindrical disc for a given imposed axial displacement $U$ on the cylindrical specimen is given by

$$\Delta V = \frac{\pi}{4}(d^2 h - D^2 h_0), \qquad (3.1)$$

where $h$ is the current height of the central cylindrical disc at the imposed displacement $U$. The volumetric strain then follows as $\Delta V/V_0$ where $V_0 = \pi D^2 h_0/4$. These measurements of volumetric strain of this disc are plotted in Extended Fig. 3a as a function of $U$ (for both compression and tensile loading) and marked as "Laser". The disc loses volume under compressive loading and gains volume under tensile loading. Also included in Extended Fig. 3a is a measurement of the volumetric strain of the entire specimen based on the outline of the specimen as determined from the X-ray tomographic reconstruction (marked $\langle V/V_0\rangle - 1$). While the volume change of the central disc over the loading history varies from approximately $-10\%$ to $\sim 10\%$, the specimen volume changes by less than $\pm 1\%$. Clearly, this implies that when the central disc loses volume other parts of the specimen gain volume and vice versa in line with the full-field FETC measurements.

To confirm the accuracy of the FETC results we include in Extended Data Fig. 3a measurements of the volumetric strain of the central cylindrical disc and the specimen using the FETC measurements of $\det(F_{ij})$. Specifically, we calculate the volumetric strain of the central disc as $\langle \det(F_{ij})\rangle - 1$ where $\langle \cdot \rangle$ is now the volume average over the material points within the disc (this is marked FETC in Extended data Fig. 3a). Similarly, the volumetric strain of the entire specimen is also calculated from the FETC as $\langle \det(F_{ij})\rangle - 1$ with $\langle \cdot \rangle$ now denoting the volume average over the material points within the entire specimen (again marked EFTC in Extended Data Fig. 3a). Excellent agreement between all the sets of measurements provides confidence in the FETC measurements of strains within the specimens and also shows that the strange behaviour, which is also observed outside the X-ray setup, is not associated with X-ray radiation damage.

An important assumption made in the laser measurements of the volume change of the central disc is that it remains cylindrical under the imposed deformations. To confirm this assumption, we revert to the FETC measurements and include in Extended Data Fig. 3a a wiremesh view of a diametrical plane through the deformed specimen ($U = -12$ mm): in the wiremesh each node of the wiremesh denotes a material point. The wiremesh in the undeformed state is a rectangular grid and so the deformed grid provides an immediate visualisation of the material strain. The zoom-in of the deformed mesh around the central disc (shown to better illustrate any shear strains) confirms that the disc remains cylindrical under the imposed deformation. The vanishing shear deformation gradients in Extended Data Fig. 3b also confirms the symmetry of the deformation.

We now consider the 3-point bending of the Silicone rubber beam, again outside the X-ray setup but using the loading fixture for the data reported in Fig. 4 (see Supplementary Section S1 for details of the loading fixture). The 3-point bending test was conducted in conjunction with 3D digital image correlation (3D DIC) performed using the GOM ARAMIS dual camera system. The measurements provide the displacements $u_i$ ($i = 1,2,3$) of material points on the surface of the beam parallel to the $X_1 - X_3$ plane (Extended Data Fig. 3b). From these measurements we calculate the surface Green-Lagrange strains as

$$E_{ij} \equiv \frac{1}{2}\left[\frac{\partial u_i}{\partial X_j} + \frac{\partial u_j}{\partial X_i} + \frac{\partial u_k}{\partial X_i}\frac{\partial u_k}{\partial X_j}\right]. \tag{3.2}$$

The 3D DIC measurements of $E_{11}$ and $E_{13}$ are included in Extended Data Fig. 3b at two imposed displacements $U$. These measurements show that the deformation of the beam is via a combination of bending and shearing as $E_{11}$ and $E_{13}$ are of the same order. The 3D DIC measurements (which include contributions from all three displacement components on the surface) can then then directly be compared with the volumetric FETC measurements. The FETC measurements give all components of $E_{ij}$ throughout the beam volume and we extracted the distributions of $E_{11}$ and $E_{13}$ on the specimen surface corresponding to the 3D DIC measurements. The excellent agreement between the 3D DIC and FETC measurements shown in Extended Data Fig. 3b provide further evidence of the fidelity of the FETC measurements and the absence of X-ray radiation damage on the surfaces of the Silicone rubber.

*S3.4: Tracking the front of the tracers in the fast radiography measurements*
The fast radiography measurements are discussed in the main text with the technique described in the Methods section. Here we describe (i) the protocol for generating the colour-toned radiographs as presented in Fig. 3d ($\dot{\kappa}h = 0.12 \text{ s}^{-1}$ loaded to $\kappa^\infty h = 0.48$) and (ii) the procedure for extracting the tracer front from the radiographic images (Fig. 3e).

The radiographs generate grayscale images (Fig. S17a), showing distinctive grayscale distributions corresponding to the two constituent materials of the composite beam: $ZnI_2$ (with gray values in the range ~5000 to ~25000) and Silicone rubber (gray values of ~65000). The air around the beam, being oversaturated, does not manifest in the histogram. This oversaturation of air is intentional, allowing for a large separation in the gray values between $ZnI_2$ and Silicone rubber within the detector's range (0 to 65535). To enhance visual representation in the main text (Fig. 3d), we employed a tonal colour scheme for the radiographs, utilizing a feature provided by the Xtek acquisition software from Nikon. The colour scheme is as follows: blue represents the oversaturated air, red denotes Silicone rubber, and grey signifies the presence of $ZnI_2$. Notably, no opacity is applied to the gray value of $ZnI_2$, preventing the red colour of Silicone rubber from overlaying in areas where $ZnI_2$ is predominant in the composite beam. This deliberate choice enables a clear distinction of the moving $ZnI_2$ tracer front within the Silicone rubber beam.

To determine the tracer front location, we analyse the grayscale radiographs without the tonal colour scheme (Fig. S17a). Consider the radiograph depicted in Fig. S17b, where we introduce a local coordinate system, denoted as $x_1' - x_3'$, positioned at the midpoint of the beam (i.e., the neutral axis) and fixed to that specific material point. This coordinate system translates with the beam motion. The variation of gray values with $x_3'$ along $x_1' = 0$ is plotted in Fig. S17b for the state of the beam at $t = 0$ s. At $t = 0$ s, the gray values are ~ 20000 for $x_3' < 0$ (below the neutral axis). A distinct jump in gray value occurs for the portion of the beam without $ZnI_2$ and reaches ~65000. The state of the beam and the corresponding gray value distribution is shown in Fig. S17c at $t = 2$ s. At this time the $ZnI_2$ front is progressing upwards (towards the tensile side of the beam) and is clear from the observed drop in gray value from ~65000 to ~20000; compare the $t = 0$ s and $t = 2$ s gray value distributions shown in Fig. S17c. Around the tracer front, the gray values fluctuate between ~65000 and ~20000 corresponding to the non-uniform progression of $ZnI_2$ through the beam thickness (i.e., in the $x_2'$ −direction). We define the front position as lowest value of $x_3'$ where the fluctuations commence, and this is labelled as "q" in Fig. S17c. This definition is used consistently for all other times (Fig. S17d). At each time, we repeated this procedure at 10 locations $x_1'$ along the imaged portion of the beam (Fig. S17a) and define the front position $X_f$ in Fig. 3e as the average location of "q" over these 10 locations. The error bars indicate the variation of the front position over these 10 locations. It is worth emphasizing that the tracer front remains stationary for times $t > t_R$ in line with the idea that the mobile phase keeps pace with the loading rate.

**Supplementary S4: A thermodynamically consistent rubber elasticity framework with a mobile species**

Most synthetic rubbers are synthesised by a polymerisation and/or crosslinking reaction which transforms the highly viscous polymer into an insoluble rubber-like solid. Typically, the crosslinking or curing reaction is initiated by addition of a curing agent or a catalyst. Peroxide, condensation and addition curing are the three mechanisms by which Silicone can be crosslinked [19-20]: the commercial Silicone rubber we employed is a two-part condensation curing Silicone. The Silane crosslinker and catalyst are packaged together as one reactive component, with the mixture of polymer and filler as the second unreactive component. Once mixed they produce a crosslinking reaction. However, this crosslinking is never complete with some fraction of the molecules remaining un-crosslinked. Based on observations reported in the main text these small and un-crosslinked molecules are mobile within the crosslinked polymer network. This physics is absent in the classical models for rubbers and is the source of their poor predictive capability (Supplementary S2). Further, the measurements (Fig. 2) strongly suggest that the hydrostatic component of the mechanical work can be negative in Silicone rubber: such behaviour is absent in classical hydrogel theories which also include mobile species. This motivates us to develop a *new* thermodynamically consistent model for the deformation behaviour of rubbers.

*S4.1:  Summary of key experimental observations with a discussion on mechanisms*

The first key surprising finding made in this study is that there exists a mobile species within the Silicone rubber. It seems reasonable to assume that this mobile species comprises the un-crosslinked molecules within the rubber. However, there are multiple observations associated with this mobility which require interpretation. These include:
  (i)    The volumetric strain is negative at material points where the mobile species fluxing into (compare Figs. 2d and 3b). Conversely, the hydrostatic strain is positive at locations where mobile species fluxes out.
  (ii)   From the 4-point bend experiment we deduce that the hydrostatic stress is positive at locations where the volumetric strain is negative and vice versa. Thus, the hydrostatic mechanical work is negative in this case.

(iii) The transport rate of the mobile species is surprisingly high with direct observations showing it moves a distance $\ell \approx 2.7$ mm in time $t = 4$ s (the $\dot{\kappa}h = 0.12$ with $\kappa^\infty h = 0.48$ case in Fig. 3e). This implies an effective linear diffusion co-efficient $D = \ell^2/t = 1.8$ mm$^2$s$^{-1}$.

Before proceeding to develop a continuum model to show that these observations fit within a thermodynamically consistent framework, we hypothesise on possible molecular mechanisms associated with these surprising findings. We emphasize here that the molecular mechanisms are hypotheses: a detailed understanding of the molecular origins of these mechanisms is beyond the scope of the current study. The microstructure of the rubber is envisaged to be an open-celled porous deformable elastic network comprising the crosslinked polymer with the un-crosslinked molecules partially filling the pores as sketched in Fig. S18a. These pores are expected to be a few nanometers in size and the un-crosslinked molecules move through these pores in a manner akin to fluid flowing through a porous media [21]. However, the nano size of the pores brings in additional physics that is usually neglected or plays little role for flow in a porous medium with mm scale pores. We shall now discuss the above three observations in the context of this microstructural model.

The first surprising finding is that the rubber volumetrically contracts when the concentration of the mobile species increases. This is opposite to expectations where we typically anticipate an increasing concentration of a mobile species results in volumetric expansion, e.g. diffusion of hydrogen in Fe [22]. However, there also exist examples of the opposite effect including for Li ions in layered materials such as NMC (Nickel Manganese Cobalt oxides) [23] or NWO (Niobium Tungsten Oxides) [24]. These materials exhibit a non-monotonic variation of the volumetric strain with increasing Li concentration. However, these battery materials have a crystallographic structure and therefore the mechanisms associated with this behaviour differ from the rubber being investigated. Nevertheless, they serve as examples to show that volumetric strain need not increase with the concentration of a mobile species. To explain the volumetric contraction with increasing concentration of the mobile species in Silicone rubber recall our hypothesised microstructure (Fig. S18a) where the un-crosslinked molecules are within the pores in the crosslinked network. Given the nano size of the pores, capillary forces between the fluid-like un-crosslinked phase and the crosslinked network are expected to play an important role. Capillary forces are known to deform micro/nanostructures and even result in their collapse [25]. Here we envisage that as this mobile phase flows into the empty pores, capillary forces collapse these pores resulting in the observed macroscopic volumetric contraction (Fig. S18b).

Next consider the observation that the volumetric contraction occurs in the presence of hydrostatic tension (and vice versa) resulting in negative hydrostatic work. Such negative hydrostatic work is not forbidden by thermodynamic constraints so long as the total work remains positive. In fact, volumetric expansion under hydrostatic compression is often observed in granular media [26]: in this case the negative hydrostatic work is compensated by the positive deviatoric work. We thus envisage that the strain associated with a change in concentration of the un-crosslinked phase is not purely hydrostatic but also has a deviatoric component. In fact, similar observations have also been reported for Li in the NMC [23] and NWO [24] crystals referred to above. Again, the mechanism causing deviatoric (shear) strains in rubber cannot be those associated with Li intercalation in layered crystals. Rather here we envisage the mechanism of shearing to be that sketched in Fig. S18. The mobile fluid-like un-crosslinked phase is assumed to be comprised of non-spherical molecules. The fluid-like nature of the phase combined with the non-spherical molecular shapes implies that the phase has a strong tendency to nematically order [27]. This implies that strains associated with this phase are not purely hydrostatic but also have a large deviatoric component as illustrated in Fig. S18; compare the states in Figs. S18a and S18b where we show that the pores in the deformed state are elongated along the long axis of the un-crosslinked molecules.

Finally consider the high transport rate of the un-crosslinked phase. The inferred linear diffusion co-efficient is $D = 1.8$ mm$^2$s$^{-1}$. This is at-least 2 orders of magnitude higher than the diffusion rate of water in hydrogels [28-29]. The flow of a fluid through a porous medium is usually described by Darcy's law where the flux $j_i$ is given in terms of the permeability $\hbar$ of the porous medium and the Newtonian viscosity $\omega$ of the fluid as

$$j_i = -\frac{\hbar}{\omega}\frac{\partial p}{\partial X_i}, \qquad (4.1)$$

where $\partial p/\partial X_i$ is the pressure gradient. The permeability $\hbar$ is usually assumed to depend on the square of the pore size [30] and thus the flux for a given pressure gradient decreases with decreasing pore size and increasing viscosity. The basic physics that motivates the form of (4.1) is that with decreasing pore size, the strain rate within the fluid increases and a larger pressure drop is required to impose this strain rate. Equivalently, if the pore size increased or the fluid viscosity is decreased the pressure drop is reduced for the same flux. The pore size in Silicone rubber is the spacing of the crosslinked network and is thus expected on be on the order of a few nanometers. At this length scale, if the un-crosslinked polymer behaved akin to a Newtonian fluid, its viscosity would need to be about 2 orders of magnitude less than that of water to explain the high transport rates observed in Figs. 3d and 3e. This is unrealistic. Rather a more realistic explanation is that the un-crosslinked polymer behaves as a non-Newtonian shear thinning fluid [31] whose viscosity decreases with increasing strain rate. In the case of the crosslinked network with nanopores, the strain rates are very high for any realistic flux and thus the effective viscosity will be very low. This implies a significantly larger *effective* diffusion co-efficient. The question is how realistic is it that the un-crosslinked polymer behaves like a shear thinning fluid? Shear thinning behaviour is often observed in polymeric systems including for example paint. The precise physics of shear thinning in polymers is not fully understood but one explanation is as follows. At rest, the polymers molecules are entangled but when undergoing deformation, the highly anisotropic polymer molecules start to disentangle. This leads to reduced molecular interactions and a reduction in the viscosity. This overall picture of anisotropic molecules of the un-crosslinked phase is consistent with the idea of a deviatoric strain associated with a change in the concentration of the un-crosslinked phase as discussed above.

*S4.2: Thermodynamic framework for the deformation of Silicone rubber*
Here we present a thermodynamically consistent model for the deformation response of Silicone rubber. The aim is not to capture the measurements to a very high degree of accuracy but rather to demonstrate that the rather counter-intuitive observations fit within a framework that does not violate basic thermodynamic considerations. For simplicity and clarity, we present the model in a small strain framework, but it can be readily extended to a finite strain setting.

In addition to temperature $\mathcal{T}$, we write the Helmholtz free-energy in terms of two state variables: (i) the total strain $\varepsilon_{ij}$ and (ii) the molar concentration $c$ (per unit undeformed volume of the rubber) of the un-crosslinked phase. Let $c_R$ denote the molar concentration of the sites for the un-crosslinked molecules in the undeformed (reference) state and we define an occupancy $\theta \equiv c/c_R$ (with $\theta = \theta_0$ in the reference state) of the sites for the un-crosslinked molecules. This mobile un-crosslinked phase is assumed to induce a "swelling" strain. Our measurements show that that the hydrostatic strain can be tensile in locations where the hydrostatic stress is compressive (Fig. 2). To create a thermodynamically consistent theory we hypothesize that the mobile phase induces a combination of volumetric and deviatoric strains. Denote the three principal swelling strains by $e^{[1]}, e^{[2]}$ and $e^{[3]}$ along their corresponding principal directions $q_k^{[p]}$ ($p = 1,2,3$). These are assumed to co-incide with the principal directions $n_k^{[p]}$ ($p = 1,2,3$) of $\varepsilon_{ij}$ such that in modular arithmetic[2] notation

$$q_k^{[p+\ell \ (\text{mod } 3)]} = n_k^{[p]} \quad , \tag{4.2}$$

and the shift $\ell$ takes values $\ell = (0,1,2)$. The swelling strain $\varepsilon_{ij}^s$ then follows as $\varepsilon_{kj}^s n_j^{[p]} = e^{[p+\ell \ (\text{mod } 3)]} n_k^{[p]}$. The total strain is additively decomposed into an elastic component $\varepsilon_{ij}^{el}$ and the swelling component such that $\varepsilon_{ij} = \varepsilon_{ij}^{el} + \varepsilon_{ij}^s$ and the shift $\ell$ chosen to minimise $\varepsilon_{ij}^{el}\varepsilon_{ij}^{el}$. This choice sets the alignment of the principal directions of the swelling strains with the total strains. The Helmholtz energy of the rubber is written as a sum of two components: (i) elastic energy due to the deformation of the crosslinked network and (ii) an entropy of mixing of the un-crosslinked molecules with the sites for these molecules such that

$$F(\varepsilon_{ij}, c, \mathcal{T}) \equiv \left[\frac{\lambda}{2}\left(\varepsilon_{kk}^{el}\right)^2 + G \, \text{erf}\left(\frac{\eta}{\eta_0}\right) \varepsilon_{ij}^{el}\varepsilon_{ij}^{el}\right] \\ + R\mathcal{T}c_R[\theta \ln \theta + (1-\theta)\ln(1-\theta)], \tag{4.3}$$

where $R$ is the universal gas constant and $\mathcal{T}$ the absolute temperature. In (4.3) $\lambda$ and $G$ are the first and second Lame parameters, respectively such that the first term represents the elastic energy while the second term is the contribution from the entropy of mixing. In the elastic energy there is a term associated with the strain triaxiality

$$\eta \equiv \frac{\sqrt{(2/3)\varepsilon_{ij}^d \varepsilon_{ij}^d}}{|\varepsilon_{kk}|}, \tag{4.4}$$

where the deviatoric strain $\varepsilon_{ij}^d \equiv \varepsilon_{ij} - \delta_{ij}\varepsilon_{kk}/3$ with $\delta_{ij}$ the Kronecker delta. This term was introduced for the following reason. Recall from Supplementary S3 that the rubber is an isotropic material and thus in the absence of imposed deviatoric strains we require the rubber to have a purely hydrostatic response. To ensure that this remains the case independent of the choice of $e^{[p]}$ we introduce the strain triaxiality term in (4.3) such that the contribution to the elastic energy due to the shear modulus vanishes for vanishing total deviatoric strains. In (4.3), $\eta_0$ sets the triaxiality range over which we apply these additional considerations to ensure isotropy come into play and a choice $\eta_0 \ll 1$ implies that this additional term only serves as a regularising parameter. Given the Helmholtz free-energy $F$, the material stress and the chemical potential of the un-crosslinked species follow as

$$\sigma_{ij} \equiv \left.\frac{\partial F}{\partial \varepsilon_{ij}}\right|_{c,\mathcal{T}}, \tag{4.5}$$

and

$$\mu \equiv \left.\frac{\partial F}{\partial c}\right|_{\varepsilon_{ij},\mathcal{T}}, \tag{4.6}$$

respectively.

---

[2] Modular arithmetic is the system of arithmetic for integers where the numbers "wrap around" when reaching a certain value, called modulus. For example, in a 12-hour clock if it is 7:00 now in 8 hours it will be 3:00 since the clocks "wraps around" every 12 hours. Thus, in (4.2) if $p = 3 \ \& \ \ell = 2$; $p + \ell \ (\text{mod } 3) = 2$.

Now consider a rubber specimen occupying a domain $\Omega$ subjected to surface tractions $T_i = T_i^0$ on a portion $S_T$ and displacements $u_i = u_i^0$ on a portion $S_U$ of the boundary $S$ of the domain $\Omega$. The potential energy of this system then

$$\Pi(u_i, c) = \int_\Omega F d\Omega - \int_{S_T} T_i^0 u_i dS, \qquad (4.7)$$

and at equilibrium we then require that the total variation $d\Pi = 0$. Then using (4.5) and (4.6) it follows that under *isothermal* conditions

$$\int_\Omega \sigma_{ij} \delta\varepsilon_{ij} d\Omega + \int_\Omega \mu \delta c d\Omega = \int_{S_T} T_i^0 \delta u_i dS, \qquad (4.8)$$

where the strains are related to displacement via $\varepsilon_{ij} \equiv 0.5(u_{i,j} + u_{j,i})$. Since the variation in displacement and concentration are arbitrary it follows that

$$\int_\Omega \sigma_{ij} \delta\varepsilon_{ij} d\Omega = \int_{S_T} T_i^0 \delta u_i dS, \qquad (4.9)$$

and

$$\int_\Omega \mu \delta c d\Omega = 0. \qquad (4.10)$$

The total amount of the un-crosslinked phase in the specimen is a constant (i.e., the un-crosslinked phase does not leave or enter the specimen). Therefore, $\int_\Omega \delta c d\Omega = 0$. It then follows from (4.10) that at equilibrium the chemical potential of the un-crosslinked phase throughout the specimen is a constant such that $\mu = \mu_0$ with $\mu_0$ set by the conservation law $\int_\Omega c d\Omega = c_R \theta_0 \Omega$. Similarly, applying the divergence theorem to (4.9) we recover the usual stress equilibrium relation $\sigma_{ij,j} = 0$ with boundary conditions $T_i^0 = \sigma_{ij} m_j$ on $S_T$ where $m_j$ is the outward normal to $S_T$.

To complete the formulation, we now need to specify a constitutive relation between $e^{[p]}$ and the concentration $c$ of the un-crosslinked phase. Physical considerations impose important restrictions on the relation between $e^{[p]}$ and $c$ as follows. Recall that the measurements suggest that the overall volume of the specimen is conserved such that $\int_\Omega \varepsilon_{kk} d\Omega = 0$ irrespective of the nature of the loading. Given that the elastic strains are strongly dependent on the nature of the loading it is therefore reasonable to assume that $\varepsilon_{kk}^{el}$ contributes negligibly to $\varepsilon_{kk}$ (i.e., the rubber is elastically nearly incompressible implying that $G/\lambda \to 0$) with the volumetric strains predominantly set by the swelling strains $e^{[p]}$. Then the constraint $\int_\Omega \varepsilon_{kk} d\Omega = 0$ reduces to $\int_\Omega \varepsilon_{kk}^s d\Omega = 0$. Let $\varepsilon_{kk}^s = f(c)$ so that

$$\int_\Omega f(c) d\Omega = 0, \qquad (4.11)$$

and $\int_\Omega c\, d\Omega = c_R \theta_0 \Omega$ as the total number of molecules of the un-crosslinked phase is conserved. Then

$$\int_\Omega \frac{\partial f}{\partial c} \delta c\, d\Omega = 0, \qquad (4.12)$$

where $\delta c$ is an arbitrary variation. But the conservation of un-crosslinked phase implies that $\int_\Omega \delta c\, d\Omega = 0$ and thus (4.12) can only hold for arbitrary $\delta c$ if $\partial f / \partial c$ is a constant, i.e. $\varepsilon_{jj}^s$ is a linear function of $c$. We thus choose the form $f(c) = \varepsilon_{jj}^s = k(\theta_0 - \theta)$, where $k$ is a material constant so that (4.11) is always satisfied.

Given the above constraint on the form of the volumetric swelling strain we propose that the principal swelling strains written in terms of $\bar{\theta} \equiv \theta/\theta_0$ take the form

$$e^{[1]} = k\theta_0 \alpha_1 (1 - \bar{\theta}); \quad e^{[2]} = k\theta_0 \alpha_2 (1 - \bar{\theta}); \quad (4.13)$$
$$e^{[3]} = k\theta_0 (1 - \alpha_1 - \alpha_2)(1 - \bar{\theta}),$$

where the material parameters $(\alpha_1, \alpha_2)$ govern the extent of the deviatoric swelling strains. For example, with the choice $\alpha_1 = \alpha_2 = 1/3$ the swelling is purely volumetric (as typically assumed for hydrogels). This completes the specification of the constitutive model for rubber in the equilibrium setting.

*S4.3: Extension to the non-equilibrium setting*

We now wish to determine the kinetic laws for the case where loading is sufficiently fast that the specimen may not be under equilibrium conditions at every stage of the loading. Recall that the rate of change of potential energy $\dot{\Pi}$ of the system satisfies the inequality $\dot{\Pi} \leq 0$ so that

$$\dot{\Pi} = \int_\Omega (\sigma_{ij} \dot{\varepsilon}_{ij} + \mu \dot{c}) d\Omega - \int_{S_T} T_i^0 \dot{u}_i dS \leq 0. \tag{4.14}$$

The decrease in potential energy drives a dissipation rate $\dot{d}$ due to the transport of the mobile phase with a flux $j_i$, i.e. $\dot{\Pi} = -\int_\Omega \dot{d}\, d\Omega$. We define a dissipation potential $\Phi_D$ defined such that

$$\dot{d} = j_i \frac{\partial \Phi_D}{\partial j_i}. \tag{4.15}$$

To motivate the functional form of $\Phi_D$ we extend the usual Darcy flow mechanism of flow of a Newtonian fluid through a porous medium to the case where the fluid is shear thinning. Analogous to Darcy flow, the gradient of the chemical potential drives the flux of the mobile phase. The flux in case of a shear thinning fluid is then assumed to be given in terms of a reference flux $j_0$ and a mobility $M$ by

$$j_i = -j_0 \left( \frac{M c_R \theta}{j_0} \left| \frac{\partial \mu}{\partial X_i} \right| \right)^{\frac{1}{N}} \operatorname{sign}\left( \frac{\partial \mu}{\partial X_i} \right), \tag{4.16}$$

where $0 < N \leq 1$ is the shear thinning exponent of the mobile phase. The relation (4.16) reduces to a Newtonian fluid for $N = 1$ while in the limit $N \& j_0/M \to 0$ the flux $j_i \to \infty$ for a finite spatial gradient of the chemical potential, i.e., we expect equilibrium corresponding vanishing gradients of the chemical potential to be established nearly instantaneously.

To establish the form of $\Phi_D$ we first relate $\mu$ and $j_i$ to the dissipation rate. Taking the rate form of the potential energy (4.7) we obtain

$$\dot{\Pi}(\dot{u}_i, \dot{c}) = \int_\Omega \sigma_{ij} \dot{\varepsilon}_{ij} d\Omega + \int_\Omega \mu \dot{c}\, d\Omega - \int_{S_T} T_i^0 \dot{u}_i dS. \tag{4.17}$$

Then using stress equilibrium condition (4.9) this reduces to

$$\dot{\Pi} = \int_\Omega \mu \dot{c}\, d\Omega = \int_\Omega \frac{\partial \mu}{\partial X_i} j_i d\Omega, \tag{4.18}$$

where we have used the divergence theorem along with the conservation law $\dot{c} = -j_{i,i}$ and the boundary condition that there is no flux of the material in/out of the specimen such that $j_i m_i = 0$ on the boundary $S$ (with outward normal $m_i$) of the domain $\Omega$. Then recalling that

$$\dot{\Pi} = -\int_\Omega \dot{d}\, d\Omega, \tag{4.19}$$

it follows that

$$\dot{d} = -j_i \frac{\partial \mu}{\partial X_i} = \frac{j_0^2}{Mc_R\theta}\left(\frac{j_i j_i}{j_0^2}\right)^{(N+1)/2}. \tag{4.20}$$

Then from the definition (4.15) of $\Phi_D$ the dissipation potential follows as

$$\Phi_D = \frac{1}{(N+1)}\frac{j_0^2}{Mc_R\theta}\left(\frac{j_i j_i}{j_0^2}\right)^{(N+1)/2}. \tag{4.21}$$

This relation reduces to the well-established dissipation potential for Fickian diffusion, viz.

$$\Phi_D = \frac{j_i j_i}{Mc_R\theta}, \tag{4.22}$$

For the choice $N = 1$ with $M$ the linear mobility of the mobile species. Given $\Phi_D$ we can then re-obtain the flux relation (4.16) by following Onsager [32-33] (also see [34]), viz, define an augmented potential

$$\Psi(\dot{u}_i, j_i) \equiv \dot{\Pi} + \int_\Omega \Phi_D d\Omega, \tag{4.23}$$

such that the kinetic path that the system follows satisfies $\delta\Psi = 0$. Then taking variations with respect to the kinetic variables $(\dot{u}_i, j_i)$ and noting that the variations $\delta\dot{u}_i$ and $\delta j_i$ are independent it follows that

$$\int_\Omega \sigma_{ij}\delta\dot{\varepsilon}_{ij}d\Omega = \int_{S_T} T_i^0 \delta\dot{u}_i dS, \tag{4.24}$$

and the flux is given by (4.16) with boundary condition $j_i m_i = 0$ on the boundary $S$ and the flux related to the concentration via the conservation law $\dot{c} = -j_{i,i}$.

*S4.4: Analysis of the 4-point bend test*
In a 4-point bend test (Fig. 2a), the central section of length $s_1$ between the two inner rollers is under pure bending. We define a co-ordinate system $x_i'$ with the origin at the centre of the beam and use the above constitutive model to analyse a beam of thickness $h$, under plane stress conditions, in pure bending along the $x_1'$ direction (Fig. S19a). The total strain-rate fields are independent of $x_1'$ and given by

$$\dot{\varepsilon}_{11} = \dot{\kappa} x_3'; \quad \dot{\varepsilon}_{12} = \dot{\varepsilon}_{13} = 0, \tag{4.25}$$

in terms of the imposed curvature rate $\dot{\kappa}$. The problem specification is completed with the plane stress boundary condition $\sigma_{22} = 0$ and the traction conditions $T_i = 0$ on $x_3' = \pm h/2$. Under these conditions the principal directions coincide with the global co-ordinate system $x_i'$ such that the principal stresses are $\Sigma_1 = \sigma_{11}$, $\Sigma_2 = \sigma_{22}$ and $\Sigma_3 = \sigma_{33}$. Then the relation between principal stresses $\Sigma_i$ and corresponding principal strains $\mathcal{E}^{[i]}$ (defined by $\varepsilon_{kl}n_l^{[i]} = \mathcal{E}^{[i]}n_k^{[i]}$) is given by

$$\Sigma_i = \lambda\big[\mathcal{E}^{[1]} + \mathcal{E}^{[2]} + \mathcal{E}^{[3]} - k\theta_0(1-\bar{\theta})\big] + 2G\mathrm{erf}\left(\frac{\eta}{\eta_0}\right)\left(\mathcal{E}^{[i]} - e^{[i]}\right) \tag{4.26}$$

$$+ G\frac{\partial\left[\mathrm{erf}\left(\frac{\eta}{\eta_0}\right)\right]}{\partial \mathcal{E}^{[i]}}\left[\left(\mathcal{E}^{[1]} - e^{[1]}\right)^2 + \left(\mathcal{E}^{[2]} - e^{[2]}\right)^2 + \left(\mathcal{E}^{[3]} - e^{[3]}\right)^2\right],$$

and chemical potential is

$$\mu = \frac{\lambda k}{c_R}\left[\mathcal{E}^{[1]} + \mathcal{E}^{[2]} + \mathcal{E}^{[3]} - k\theta_0(1-\bar{\theta})\right]$$
$$-2G\mathrm{erf}\left(\frac{\eta}{\eta_0}\right)\left[(\mathcal{E}^{[1]} - e^{[1]})\frac{\partial e^{[1]}}{\partial c} + (\mathcal{E}^{[2]} - e^{[2]})\frac{\partial e^{[2]}}{\partial c}\right.$$
$$\left.+ (\mathcal{E}^{[3]} - e^{[3]})\frac{\partial e^{[3]}}{\partial c}\right] + R\mathcal{T}\ln\frac{\theta}{1-\theta}. \quad (4.27)$$

Recall that the principal directions of the swelling strains are chosen to coincide with the principal directions of total strains so as to minimise $\varepsilon^{el}_{ij}\varepsilon^{el}_{ij}$. Here we choose $\alpha_1 = -\alpha_2 = -1$ and therefore the choice $q^{[i]}_k = n^{[i]}_k$ minimises $\varepsilon^{el}_{ij}\varepsilon^{el}_{ij}$. This implies

$$e^{[1]} = -k\theta_0(1-\bar{\theta}); \; e^{[2]} = e^{[3]} = k\theta_0(1-\bar{\theta}). \quad (4.28)$$

With the swelling strains now known in terms of $\theta$ the expressions for the stresses and chemical potential can be further simplified. The equilibrium condition $\sigma_{ij,j} = 0$ combined with the traction boundary condition $T_3 = 0$ on $x'_3 = \pm h/2$ implies that $\Sigma_3 = 0$. Then from (4.26) it follows that $\mathcal{E}^{[2]} = \mathcal{E}^{[3]}$ such that (4.26) and (4.27) reduce to

$$\Sigma_1 = \lambda\left[\kappa x'_3 + 2\mathcal{E}^{[3]} - k\theta_0(1-\bar{\theta})\right] + 2G\mathrm{erf}\left(\frac{\eta}{\eta_0}\right)\{\kappa x'_3 + k\theta_0(1-\bar{\theta})\} \quad (4.29)$$
$$+ G\frac{\partial\left[\mathrm{erf}\left(\frac{\eta}{\eta_0}\right)\right]}{\partial \mathcal{E}^{[1]}}\left[\{\kappa x'_3 + k\theta_0(1-\bar{\theta})\}^2 + 2\{\mathcal{E}^{[2]} - k\theta_0(1-\bar{\theta})\}^2\right],$$

and

$$\mu c_R = \lambda k\left[\kappa x'_3 + 2\mathcal{E}^{[3]} - k\theta_0(1-\bar{\theta})\right]$$
$$- 2Gk\mathrm{erf}\left(\frac{\eta}{\eta_0}\right)\left[\{\kappa x'_3 + k\theta_0(1-\bar{\theta})\}\right.$$
$$\left.- 2\{\mathcal{E}^{[3]} - k\theta_0(1-\bar{\theta})\}\right] + R\mathcal{T}c_R\ln\frac{\theta}{1-\theta}, \quad (4.30)$$

respectively. The condition that $\Sigma_2 = \Sigma_3 = 0$ can then be used to determine $\mathcal{E}^{[3]}$ from the relation

$$\lambda\left[\kappa x'_3 + 2\mathcal{E}^{[3]} - k\theta_0(1-\bar{\theta})\right] + 2G\mathrm{erf}\left(\frac{\eta}{\eta_0}\right)\left[\{\mathcal{E}^{[3]} - k\theta_0(1-\bar{\theta})\}\right] \quad (4.31)$$
$$+ G\frac{\partial\left[\mathrm{erf}\left(\frac{\eta}{\eta_0}\right)\right]}{\partial \mathcal{E}^{[3]}}\left[\{\kappa x'_3 + k\theta_0(1-\bar{\theta})\}^2 + 2\{\mathcal{E}^{[3]} - k\theta_0(1-\bar{\theta})\}^2\right] = 0.$$

Finally, we need a condition to determine $\theta$. The evolution of $\theta$ depends on the kinetics of transport of the un-crosslinked species and we first consider the limiting case of loading with $\dot{\kappa} \to 0$. In this limit the chemical potential is given by the equilibrium condition $\mu = \mu_0$: this constant chemical potential is determined by writing $\theta$ in terms of $\mu_0$ from (4.30) and then calculating $\mu_0$ from the conservation constraint

$$\frac{1}{h}\int_{-h/2}^{h/2}\bar{\theta}dx'_3 = 1. \quad (4.32)$$

Now consider the case of a finite $\dot{\kappa}$ where the concentration of the un-crosslinked species may not be in equilibrium. In this case, the flux of the un-crosslinked species is given by

$$j_3 = -j_0 \left(\frac{Mc_R\theta}{j_0}\left|\frac{\partial\mu}{\partial x'_3}\right|\right)^{\frac{1}{N}} \text{sign}\left(\frac{\partial\mu}{\partial x'_3}\right), \tag{4.33}$$

which combined with the conservation law gives the overall differential equation for concentration of the un-crosslinked species as

$$c_R\frac{\partial\theta}{\partial t} = j_0\left(\frac{Mc_R}{j_0}\right)^{\frac{1}{N}}\frac{\partial}{\partial x'_3}\left[\theta\left|\frac{\partial\mu}{\partial x'_3}\right|^{1/N}\text{sign}\left(\frac{\partial\mu}{\partial x'_3}\right)\right], \tag{4.34}$$

with initial conditions $\theta = \theta_0 \ \forall \ x'_3$ at time $t = 0$ and boundary conditions $j_3 = 0$ at $x'_3 = \pm h/2$. Thus, we need to solve equations (4.31) and (4.34) together with the boundary conditions to obtain solutions for $\theta(x'_3)$ and $\mathcal{E}^{[3]}(x'_3)$.

*S4.5: Material parameters and comparison with measurements*
The model requires the specification of the following 5 parameters (in addition to the choice we have made of $\alpha_1 = -\alpha_2 = -1$) in the equilibrium limit of an infinitesimal loading rate. These are the shear modulus $G$, the Poisson ratio[3] $\nu$, the constant $R\mathcal{T}c_R$ governing the entropy of mixing, the swelling strain constant $k$ and the occupancy $\theta_0$ of the mobile phase in the undeformed configuration. These parameters were determined as follows. The Poisson ratio was selected to be $\nu = 0.45$ so that the rubber is nearly elastically incompressible and then $G$ chosen to match the moment versus curvature relations. The parameters $R\mathcal{T}c_R$, $k$ and $\theta_0$ were chosen together to obtain the measured strain ($\mathcal{E}^{[3]}$ and volumetric strain $\varepsilon_{kk}$) distributions within the beam.

Now consider the parameters governing the transport of the un-crosslinked polymer. In general, two additional parameters are required, viz. $N$ and $Mj_0^{(N-1)}$ to fully characterise the non-linear transport. However, the measurements show that the response of the Silicone rubber is nearly rate independent over the 3 decades of loading rates investigated here. This implies that equilibrium is attained much faster than the loading rates we have considered in this study. Within the framework presented here this implies that the un-crosslinked phase is an approximately rate independent shear thinning fluid with $N, j_0/M \to 0$. In this limit equilibrium with vanishing spatial gradients of $\mu$ is attained instantaneously, i.e., the problem reduces to the equilibrium problem discussed in Section S4.2. All the parameters inferred for Silicone rubber are listed Table S3 and comparisons between the simulations and measurements are included in Figs. 2 and 3 of the main text.

*S4.6: Simulation of the tracer mobility measurements*
The tracer radiographs (Fig. 3d) were the key measurements that provided evidence of the rapid transport of the un-crosslinked species. Those measurements (Fig. 3e) shows that the spatial distribution of mobile phase is independent of the loading rate. It is thus critical to simulate the motion of these tracers in the limit $N \to 0$. In this limit time is no longer an independent parameter and thus we need to re-cast the transport equations of Section S4.3.

Mass conservation of the un-crosslinked species is written as

$$\frac{\partial c}{\partial \kappa} = -\frac{\partial \hat{j}_3}{\partial x'_3}, \tag{4.35}$$

where time $t = \kappa/\dot{\kappa}$ and $\hat{j}_3 \equiv j_3/\dot{\kappa}$. Here $\partial c/\partial \kappa$ is known from the equilibrium solution procedure of Section S4.4, i.e. given by (4.25) - (4.32). The normalised flux is therefore

---
[3] $\nu$ is related to the first Lame parameter via $\lambda \equiv 2G\nu/(1-2\nu)$.

$$\hat{j}_3(x_3') = -\int_{-h/2}^{x_3'} \frac{\partial c(\bar{x}_3)}{\partial \kappa} \, d\bar{x}_3 , \qquad (4.36)$$

since $\hat{j}_3(x_3' = -h/2) = 0$. We shall now use this flux to determine the motion of the tracers. The tracer particles were assumed to be attached to a fraction $\chi_0$ of the un-crosslinked molecules and this concentration is approximately constant over the region $-h/2 \leq x_3' \leq h_T^0$ where $x_3' = h_T^0$ is initial position of the tracer front as seen in Fig. S17a. To avoid a step function, we assumed a smooth distribution of the initial tracer concentration given by

$$c_T^0 = \frac{\chi_0 c_R}{2}\left[\mathrm{erf}\left\{-\frac{x_3' - h_T^0}{h_\varepsilon}\right\} + 1\right], \qquad (4.37)$$

where the constant $h_\varepsilon \ll h$ is a regularising parameter that smoothens the step function of the initial concentration of the tracers; see Fig. S19b for a plot of (4.37) for the choice $h_T^0 = -0.6$ mm and selected values of $h_\varepsilon/h$ (in Fig. S19b we use the experimental parameters of the beam thickness and $h_T^0$ from Fig. 3d). Recall that the tracers are attached to the un-crosslinked molecules and thus their motion follows directly from the flux of the un-crosslinked species. In particular, given the current tracer concentration $c_T(x_3')$, the flux of the tracers is $j_T = j_3 c_T/c$ and the evolution of the tracer concentration follows from the conservation equation $\dot{c}_T = -\partial j_T/\partial x_3'$, i.e.

$$\frac{\partial c_T}{\partial t} = -\frac{\partial}{\partial x_3'}\left(j_3 \frac{c_T}{c}\right). \qquad (4.38)$$

We then re-cast this equation using the transformations $t = \kappa/\dot{\kappa}$ and $\hat{j}_3 \equiv j_3/\dot{\kappa}$ as

$$\frac{\partial c_T}{\partial \kappa} = -\frac{\partial}{\partial x_3'}\left(\hat{j}_3 \frac{c_T}{c}\right). \qquad (4.39)$$

This equation, with $\hat{j}_3$ given by (4.36) is solved with the initial condition (4.37) along with boundary conditions $j_T = 0$ on $x_3' = \pm h/2$ to determine the motion of the front of the tracer particles.

Solution of this transport equation requires the specification of three additional parameters, viz. $h_\varepsilon$, $h_T^0$ and $\chi_0$. The initial position of the tracer front $h_T^0$ follows directly from the radiographs (Fig. 3d) while $h_\varepsilon$ is a numerical parameter that we choose to be as small as possible but still allowing the numerical calculations to converge. Our tests demonstrated that $h_\varepsilon = 0.01h$ suffices. Finally, we need to choose $\chi_0$. This parameter is not precisely known but given that approximately 5% by wt. $ZnI_2$ was dissolved into the catalyst we take $\chi_0 = 0.05$: we shall subsequently show that the choice of $\chi_0$ over the range $0.005 \leq \chi_0 \leq 0.2$ has a minimal effect on the predictions.

Predictions of the spatio-temporal evolution of $c_T$ with the measured value of $h_T^0 = -0.6$ mm (and choice $\chi_0 = 0.05$) are shown in Fig. S20a for the experimentally applied loading protocol, viz. $\dot{\kappa}h = 0.12$ s$^{-1}$ over $0 \leq t \leq 4$ s and $\dot{\kappa}h = 0$ for $t > 4$ s. The tracer concentration evolves such that the tracer concentration reduces near the compressive side of the beam and increases near the front of the tracers. We define the position $x_f$ of the "front" as the value of $x_3'$ where $c_T$ is a maximum (with $x_f = h_T^0$ at $t = 0$). The predicted temporal variation of $x_f$ for this case is included in Fig. S20b. Comparisons between the measurements and these predictions of the tracer front are included in Fig. 3e and show excellent agreement. Calculations for predictions of the spatio-temporal evolution of $c_T$ for two additional values of $\chi_0$ are included in Figs. S21a and S21b. These results are nearly indistinguishable (and also nearly identical to the predictions in Fig. S20a). This demonstrates that the value of $\chi_0$ has a negligible influence on the predictions.

**Supplementary References**
[1] Hubbell, J.H., Seltzer, S.M., (2004). X-Ray Mass Attenuation Coefficients, *NIST Standard Reference Database 126*, dx.doi.org/10.18434/T4D01F.
[2] Pauwels, K., Douissard, P.A. (2022). Indirect X-ray detectors with single-photon sensitivity. *Journal of Synchrotron Radiation*, 29, 1394-1406.
[3] Kelsey, C.A. (1984). The Physics of Radiology. In Johns, H.E., Cunningham, R., *Medical Physics*, 11, 731-732.
[4] Gonzalez, J., Knauss, W.G. (1998). Strain inhomogeneity and discontinuous crack growth in a particulate composite. *Journal of the Mechanics and Physics of Solids*, 46, 1981-1995.
[5] Rueckert, D., Sonoda, L.I., Hayes, C., Hill, D.L., Leach, M.O., Hawkes, D.J. (1999). Nonrigid registration using free-form deformations: application to breast MR images. *IEEE Transactions on Medical Imaging*, 18, 712-721.
[6] Klein, S., Pluim, J.P.W., Staring, M., Viergever, M.A. (2008). Adaptive Stochastic Gradient Descent Optimisation for Image Registration. *International Journal of Computer Vision*, 81, 227–239.
[7] Hild, F., Roux, S. (2006). Digital Image Correlation: from Displacement Measurement to Identification of Elastic Properties – a Review. *Strain*, 42: 69 80.
[8] Avril, S., Pierron, F. (2007). General framework for the identification of constitutive parameters from full-field measurements in linear elasticity. *International Journal of Solids and Structures,* 44, 4978-5002.
[9] Ereiz, S., Duvnjak, I., Jiménez-Alonso, J.F. (2022). Review of finite element model updating methods for structural applications, *Structures,*41,684-723.
[10] Treloar, L.R.G. (1975). The Physics of Rubber Elasticity, Clarendon Press, Oxford, Third Edition.
[11] Arruda, E.M., Boyce, M.C. (1993). A three-dimensional model for the large stretch behavior of rubber elastic materials. *Journal of the Mechanics and Physics of Solids*, 41, 389-412.
[12] Rivlin, R.S. (1948). Large elastic deformations of isotropic materials. I. Fundamental concepts. *Philosophical Transactions of the Royal Society of London. Series A*, 240, 459-490.
[13] Mathieu, F., Leclerc, H., Hild, F., Roux, S. (2015). Estimation of elastoplastic parameters via weighted FEMU and integrated-DIC. *Experimental Mechanics*, 55, 105-119.
[14] ABAQUS (2012). Analysis User's Manual. Version 6.12, Dassault Systemes Simulia, Inc.
[15] Nelder, J.A., Mead, R. (1965). A simplex method for function minimization. *Computer Journal*, 7, 308-313.
[16] Fleck, N.A., Hutchinson, J.W. (1997). Strain gradient plasticity. *Advances in Applied Mechanics*, 33, 295-361.
[17] Chakravarthy, S.S., Curtin, W. (2011). Stress-gradient plasticity. *Proceedings of the National Academy of Sciences*, 108, 15716-15720.
[18] Mindlin, R.D. (1965). Second gradient of strain and surface tension in linear elasticity. *International Journal of Solids and Structures*, 1, 417-438.
[19] Harkous, A., Colomines, C., Leroy, E., Mousseau, P., Deterre, R. (2016). The kinetic behavior of Liquid Silicone Rubber: A comparison between thermal and rheological approaches based on gel point determination. *Reactive and Functional Polymers*, 101, 20-27.
[20] de Buyl, F., Hayez, V., Harkness, B., Kimberlain, J., Shephard, N. (2023). Advances in structural silicone adhesives, In Dillard, D.A. *Advances in Structural Adhesive Bonding (Second Edition)*, Woodhead Publishing, 179-219.
[21] Biot, M.A. (1941). General Theory of Three-Dimensional Consolidation. *Journal of Applied Physics,* 12, 155-164.
[22] Sofronis, P., McMeeking, R.M. (1989). Numerical analysis of hydrogen transport near a blunting crack tip. *Journal of the Mechanics and Physics of Solids*, 37, 317-350.

[23] de Biasi, L., Kondrakov, A.O., Geßwein, H., Brezesinski, T., Hartmann, P., Janek, J. (2017). Between Scylla and Charybdis: Balancing Among Structural Stability and Energy Density of Layered NCM Cathode Materials for Advanced Lithium-Ion Batteries, *The Journal of Physical Chemistry*, 121, 26163-26171.
[24] Merryweather, A.J., Jacquet, Q., Emge, S.P., Schnedermann, C., Rao, A., Grey, C.P. (2022). Operando monitoring of single-particle kinetic state-of-charge heterogeneities and cracking in high-rate Li-ion anodes. *Nature Materials*, 21, 1306-1313.
[25] Kim, I., Mun, J., Hwang, W., Yang, Y., Rho, J. (2020). Capillary-force-induced collapse lithography for controlled plasmonic nanogap structures. *Microsystems & Nanoengineering*, 6, 65.
[26] Rowe, P.W. (1962). The Stress-Dilatancy Relation for Static Equilibrium of an Assembly of Particles in Contact. *Proceedings of the Royal Society of London. Series A*, 269, 500-527.
[27] Onsager, L. (1949). The effects of shape on the interaction of colloidal particles. *Annals of the New York Academy of Sciences*, 51, 627-659.
[28] Hu, Y., Chen, X., Whitesides, G.M., Vlassak, J.J., Suo, Z. (2011). Indentation of polydimethylsiloxane submerged in organic solvents. *Journal of Materials Research,* 26, 785-795.
[29] Hu, Y., Zhao, X., Vlassak, J.J., Suo, Z. (2010). Using indentation to characterize the poroelasticity of gels. *Applied Physics Letters,* 96, 121904.
[30] Carman, P.C. (1937). Fluid flow through granular beds. *Transactions, Institution of Chemical Engineers, London*, 15, 150-166
[31] Vergne, P. (2007) Super Low Traction under EHD & Mixed Lubrication Regimes. In Erdemir, A., Martin, J.M. *Superlubricity*, Elsevier Science, 427-443.
[32] Onsager, L., (1931). Reciprocal relations in irreversible processes. I. *Physical Review*, 37, 405-426.
[33] Onsager, L., (1931). Reciprocal relations in irreversible processes. II. *Physical Review*, 38, 2265-2279.
[34] Suo, Z., (1997). Motions of microscopic surfaces in materials. *Advances in Applied Mechanics*, 33, 193-294.

**Supplementary Figures**

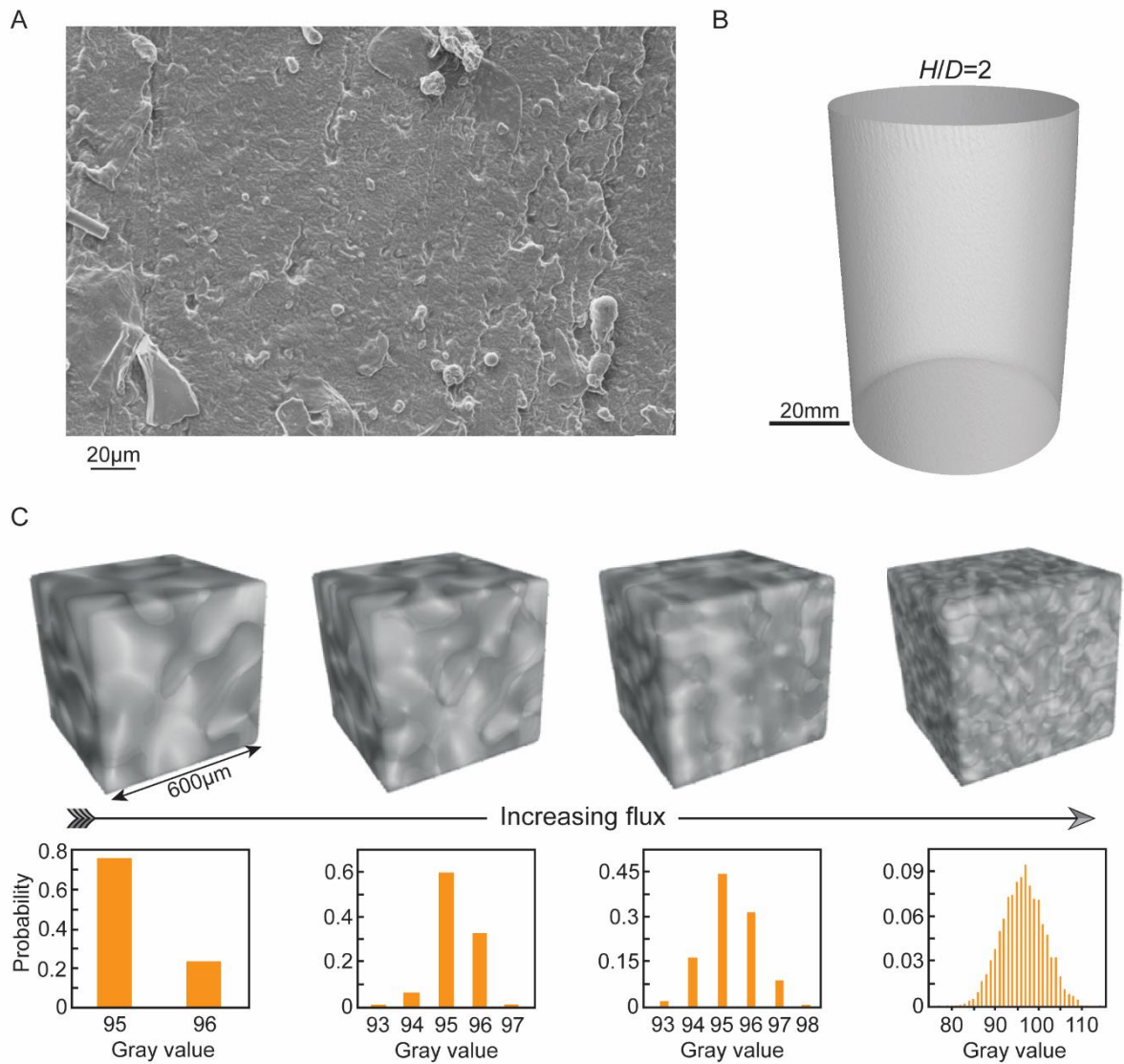

**Figure S1:** Images to show the absence of porosity in the as-cast Silicone rubber. (a) Scanning electron microscope (SEM) and (b) a 3D XCT reconstructed image. (c) The computed grayscale tomograph of a 600 μm cube within a cylindrical Silicone rubber specimen using a Tungsten target with a 150 kVp source and increasing currents. The corresponding histogram of grayscale pixel counts within the cube are also indicated.

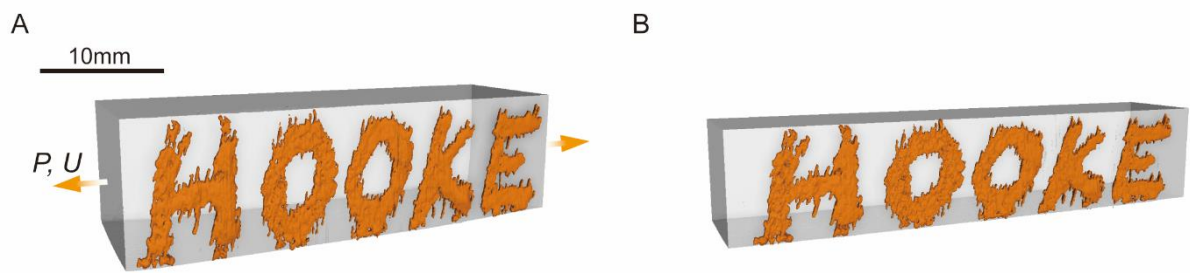

**Figure S2:** XCT reconstructed images of the Silicone rubber beam with a $ZnI_2$ pattern painted on the surface. (a) The undeformed with the "Halloween style" $ZnI_2$ lettering painted on the surface. (b) The beam stretched in uniaxial tension to 30% strain along the longitudinal axis. The reconstructed image shows that the $ZnI_2$ pattern deforms conformally with the beam.

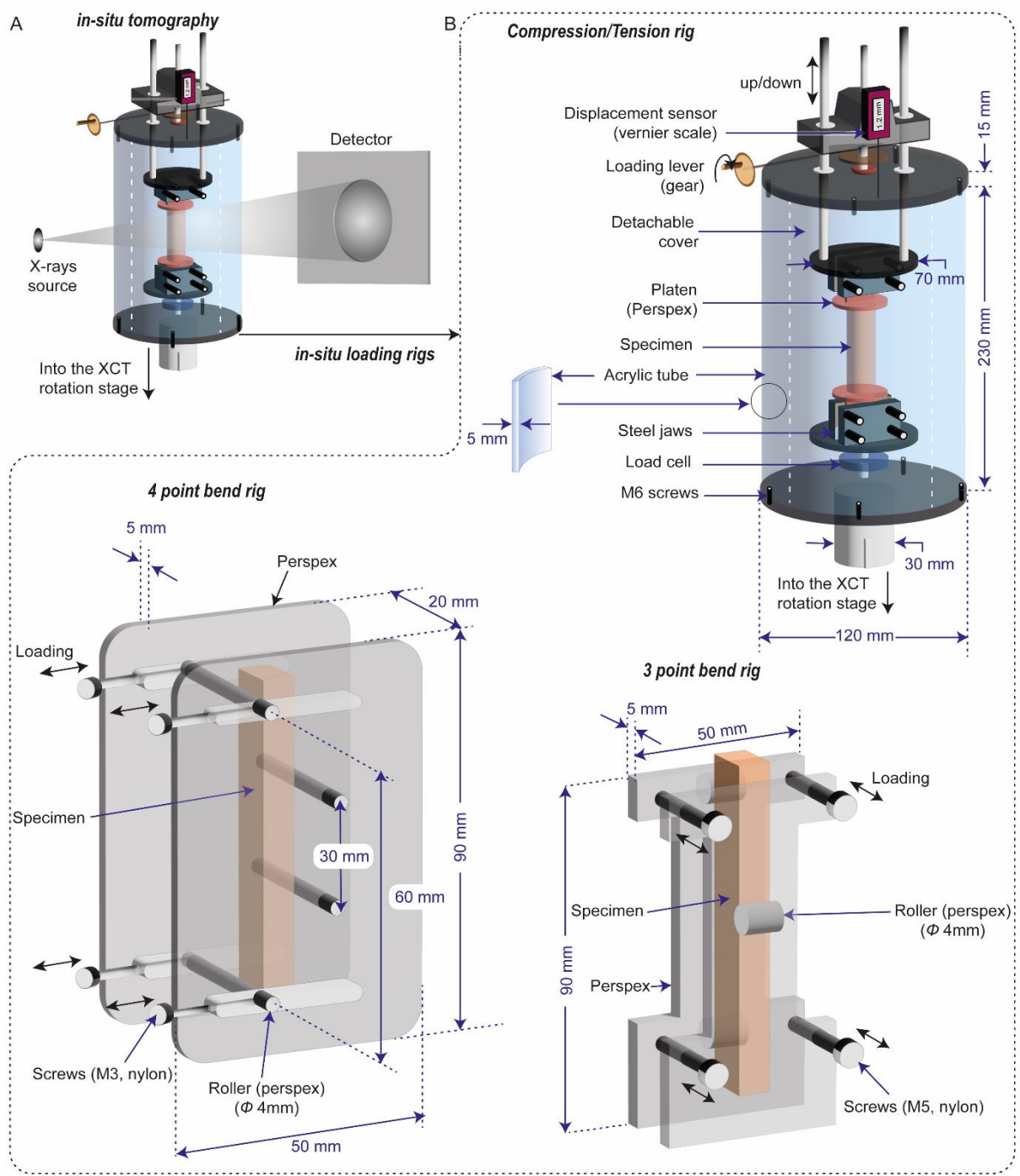

**Figure S3:** Sketches of the in-situ loading setups. (a) The loading rig within the XCT setup (b) the compression/tension, 4-point bending and 3-point bending test fixtures. In each case we have marked leading dimensions and the materials used.

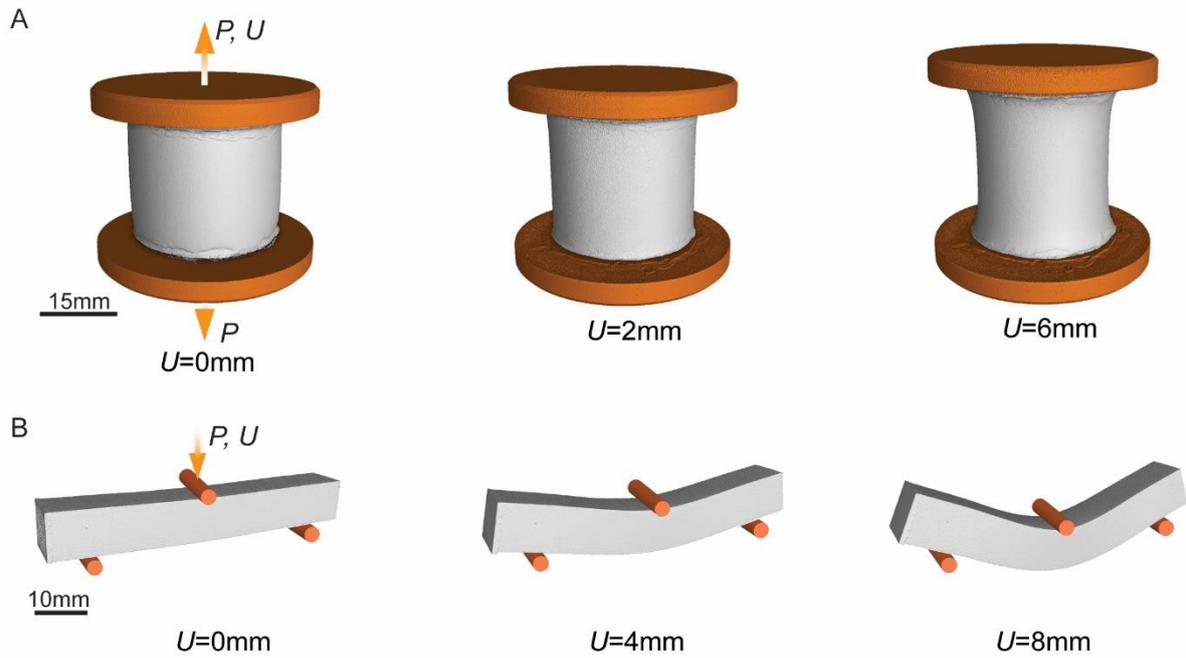

**Figure S4:** Examples of reconstructed X-ray tomographs at selected loading stages for the Silicone rubber (a) $H/D = 1$ cylindrical specimen loading in tension and (b) beam in 3-point bending.

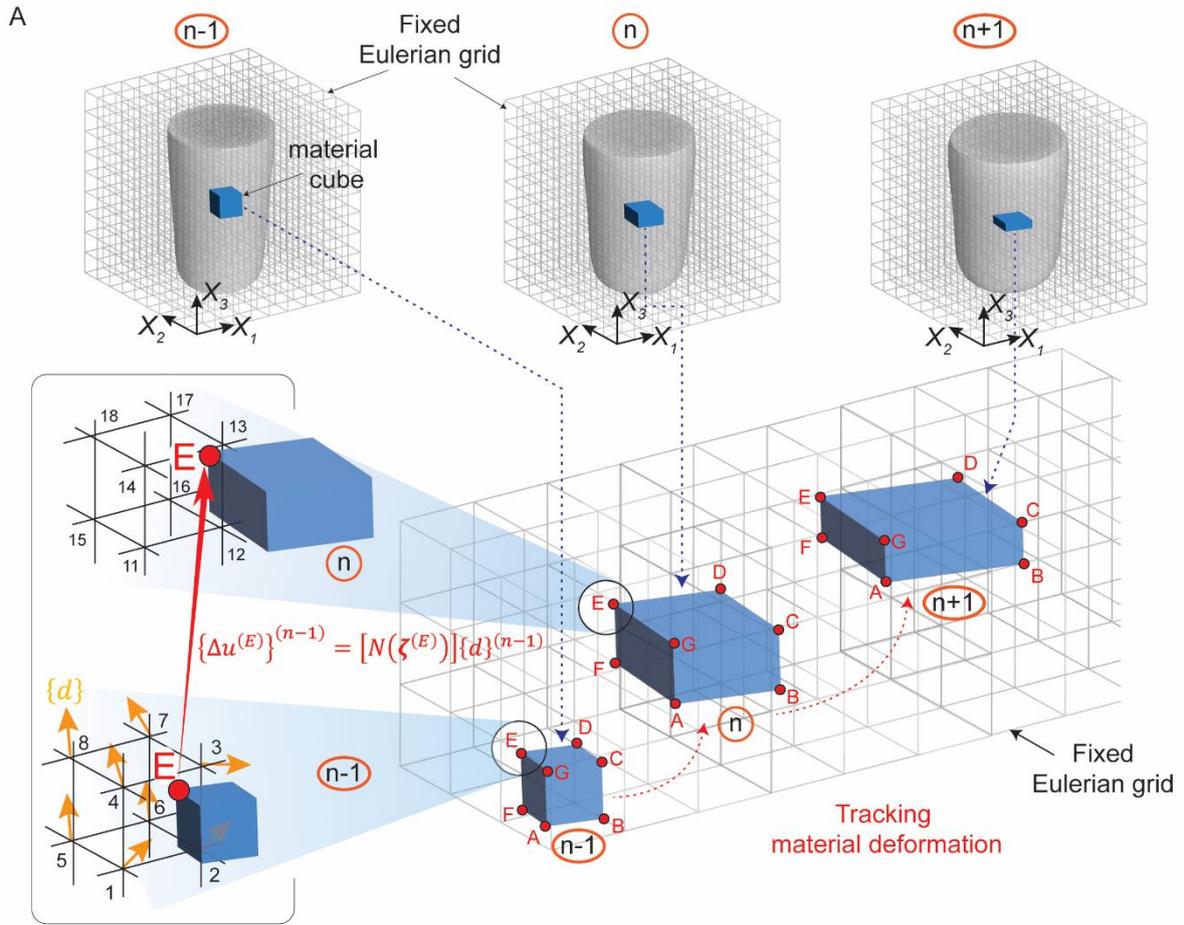

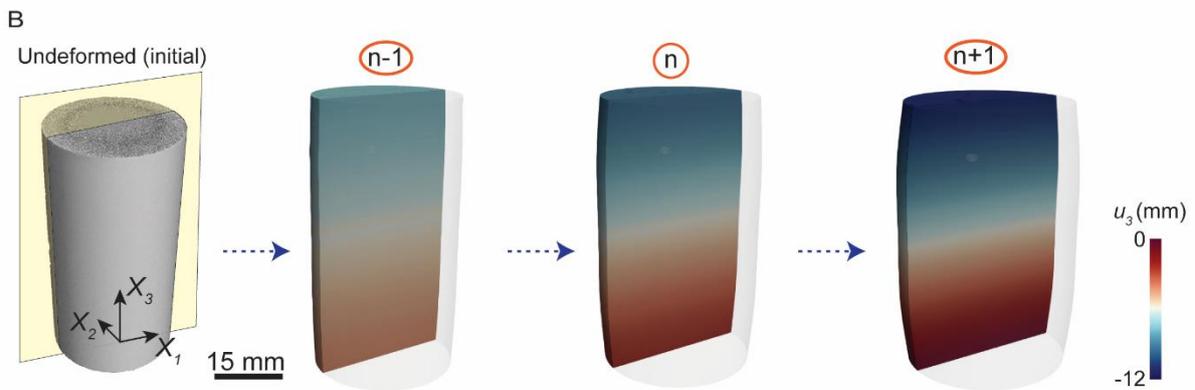

**Figure S5:** (a) Schematic of the multi-stage DVC analysis. A material cube (coloured blue) with nodes labelled A through E is marked and its deformation tracked through steps $(n-1)$, $n$ and $(n+1)$. The fixed Eulerian grid (coloured black) on which the incremental DVC displacements $\{d\}$ are evaluated is also shown and then converted to incremental material point displacements $\{\Delta u\}$. (b) An example of the material displacement field computed using the multi-stage DVC analysis for compression of the $H/D = 2$ cylindrical Silicone rubber specimen. We only show the displacement component $u_3$.

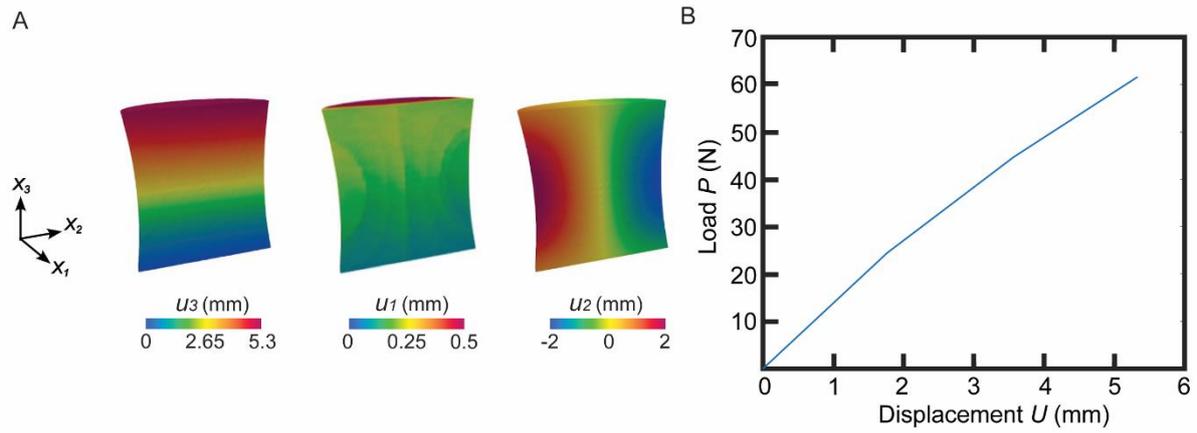

**Figure S6:** An example of the data used in the FEMU analysis is shown for the $H/D = 1$ cylindrical specimen ($D = 28$ mm) loaded in tension. (a) All three components of the displacement field shown in a diametrical plane through the specimen for an imposed displacement $U = 6$ mm and (b) the measured load $P$ versus displacement $U$ response.

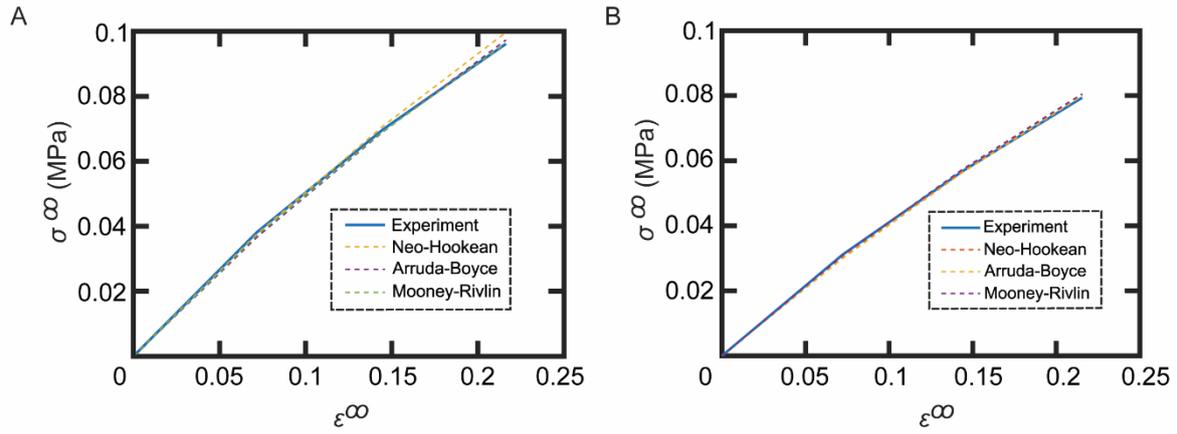

**Figure S7:** Comparisons between predictions of hyperelastic models and measurements of the nominal stress $\sigma^\infty$ versus nominal strain $\varepsilon^\infty$ for the tensile tests on the (a) $H/D = 1$ and $H/D = 2$ specimens ($D = 28$ mm). In (a) and (b) the models are calibrated by the $H/D = 1$ and $H/D = 2$ data, respectively.

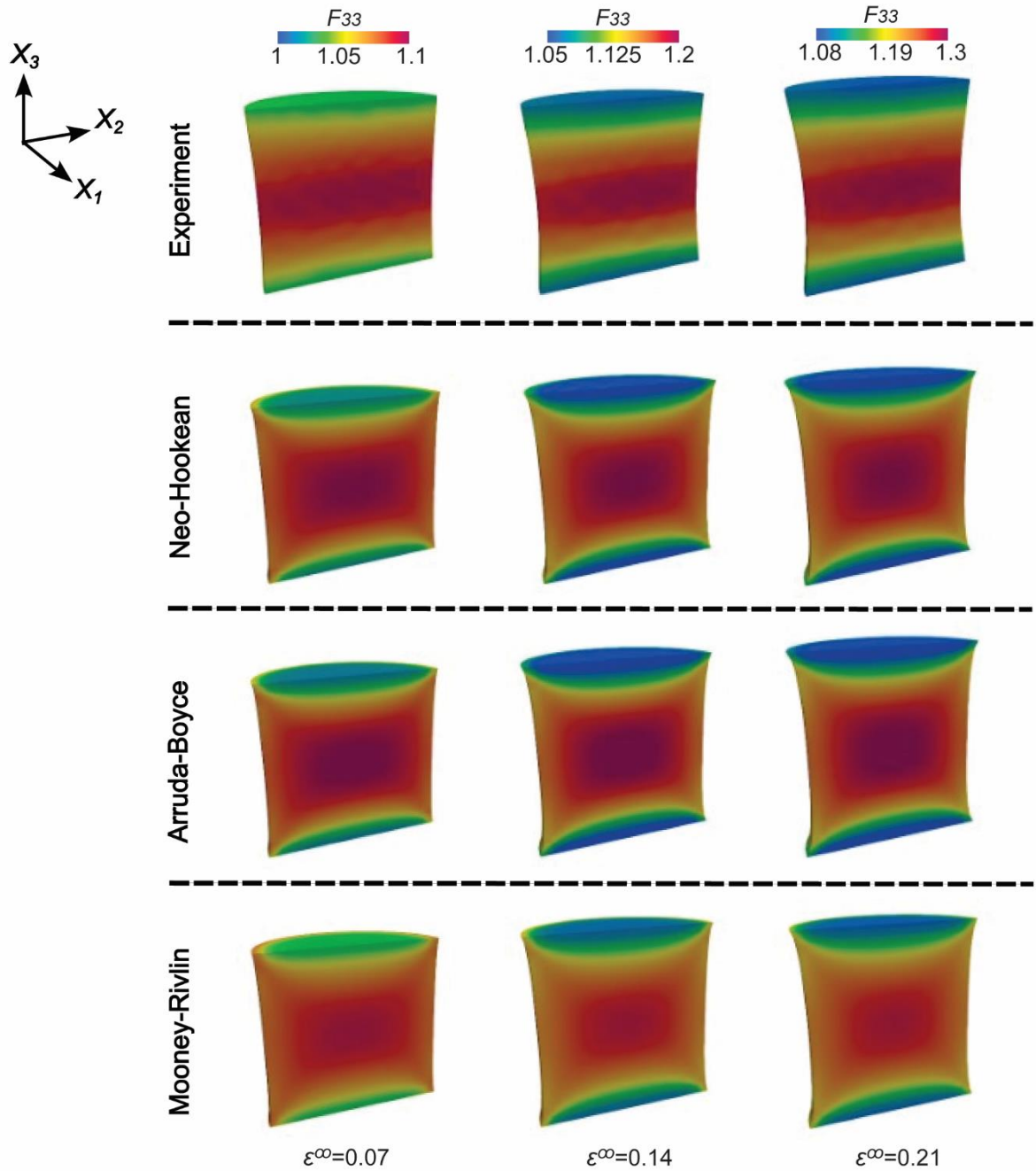

**Figure S8:** Comparisons between the measured and predicted distributions of $F_{33}$ on the central plane through the $H/D = 1$ specimen ($D = 28$ mm) for three stages of loading. The models are calibrated using the $H/D = 1$ measurements.

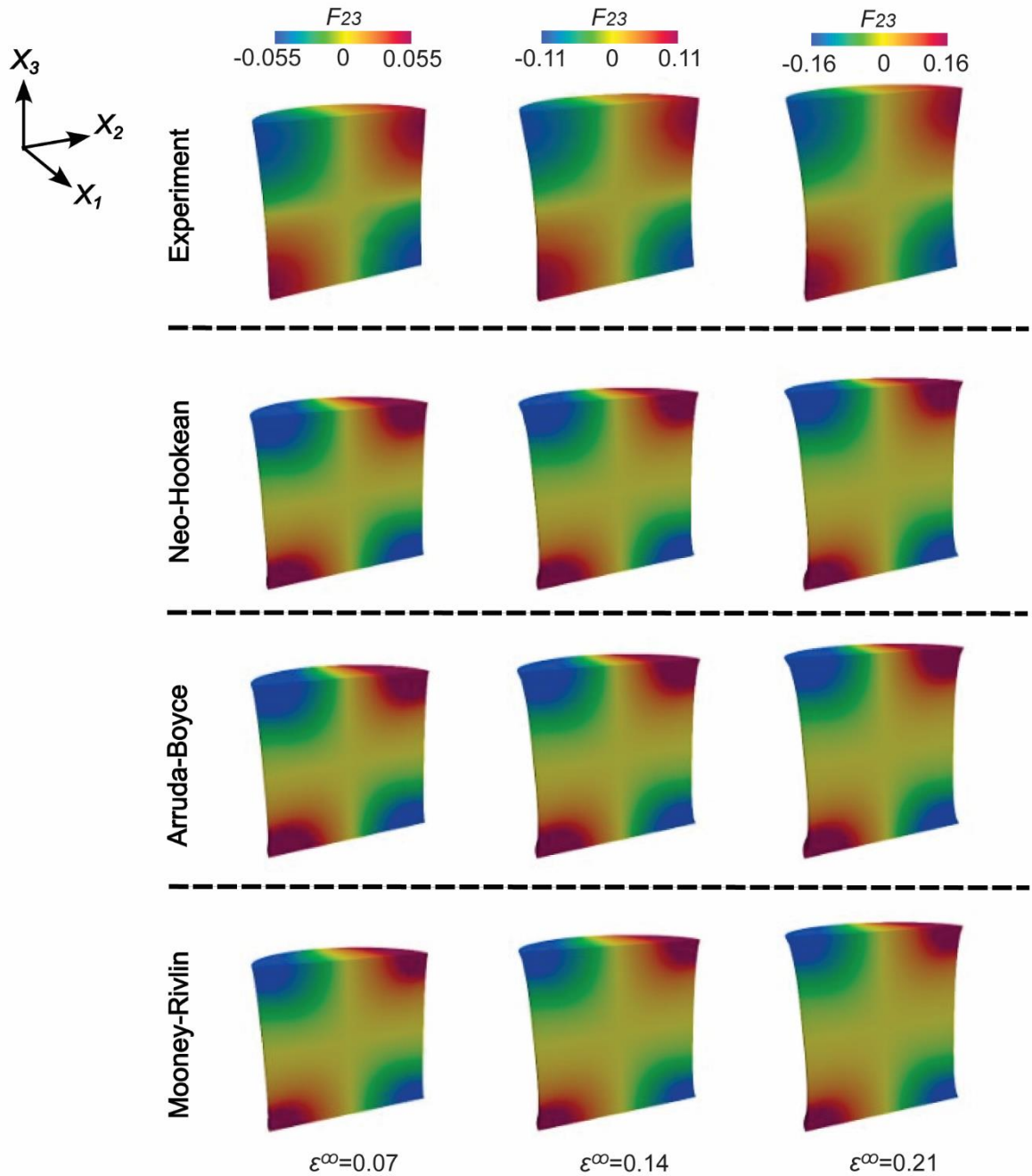

**Figure S9:** Comparisons between the measured and predicted distributions of $F_{23}$ on the central plane through the $H/D = 1$ specimen ($D = 28$ mm) for three stages of loading. The models are calibrated using the $H/D = 1$ measurements.

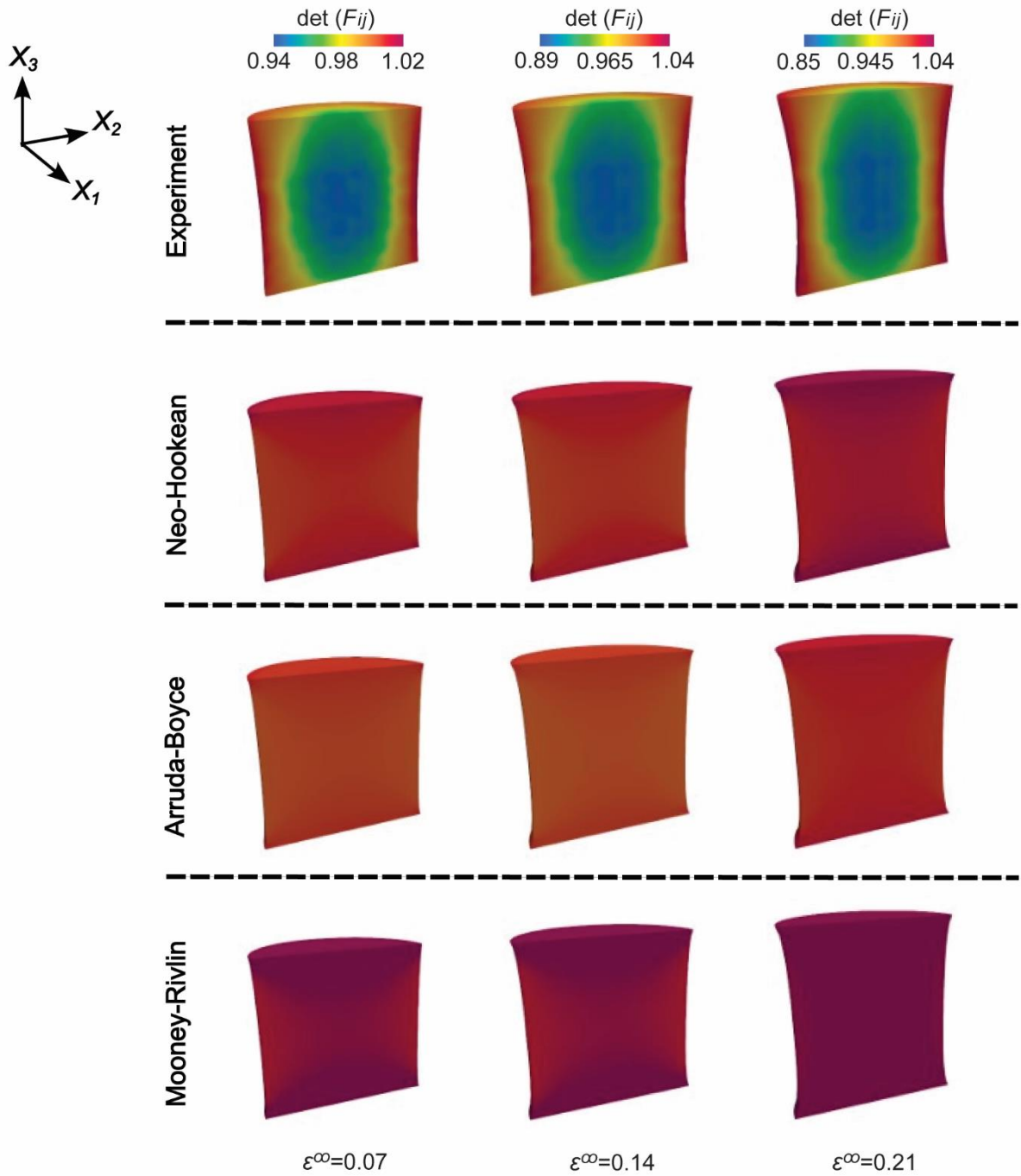

**Figure S10:** Comparisons between the measured and predicted distributions of $\det(F_{ij})$ on the central plane through the $H/D = 1$ specimen ($D = 28$ mm) for three stages of loading. The models are calibrated using the $H/D = 1$ measurements.

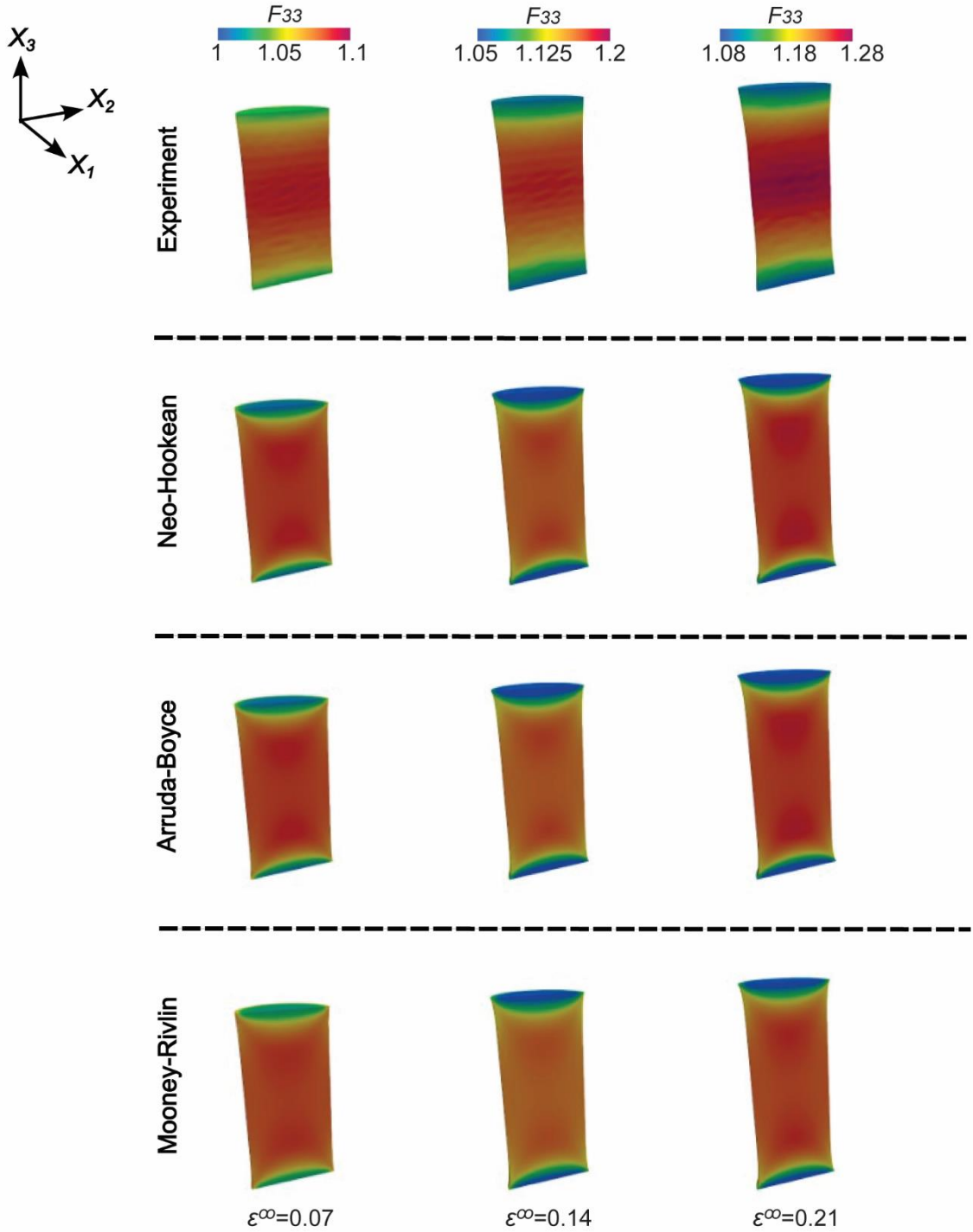

**Figure S11:** Comparisons between the measured and predicted distributions of $F_{33}$ on the central plane through the $H/D = 2$ specimen ($D = 28$ mm) for three stages of loading. The models are calibrated using the $H/D = 2$ measurements.

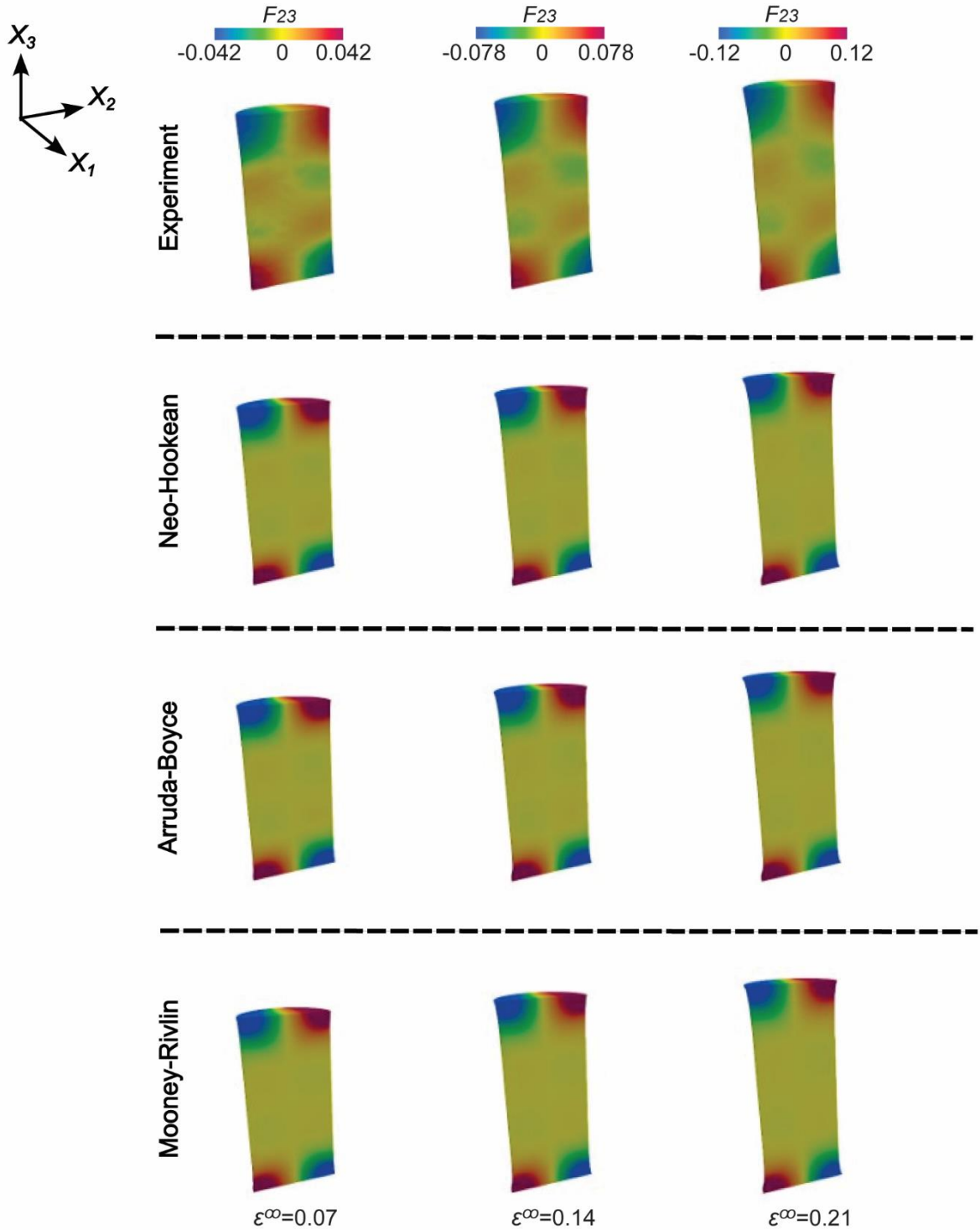

**Figure S12:** Comparisons between the measured and predicted distributions of $F_{23}$ on the central plane through the $H/D = 2$ specimen ($D = 28$ mm) for three stages of loading. The models are calibrated using the $H/D = 2$ measurements.

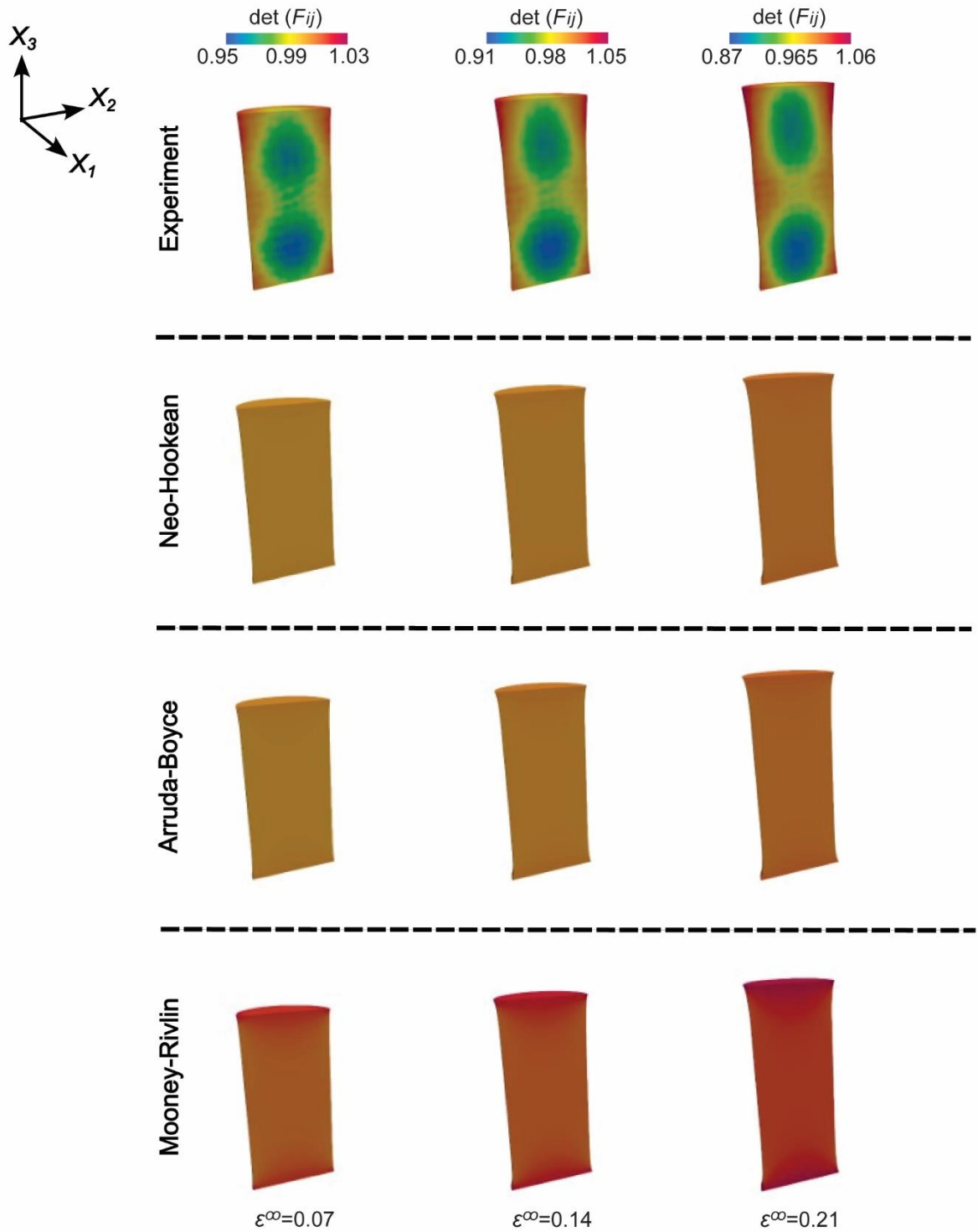

**Figure S13:** Comparisons between the measured and predicted distributions of $\det(F_{ij})$ on the central plane through the $H/D = 2$ specimen ($D = 28$ mm) for three stages of loading. The models are calibrated using the $H/D = 2$ measurements.

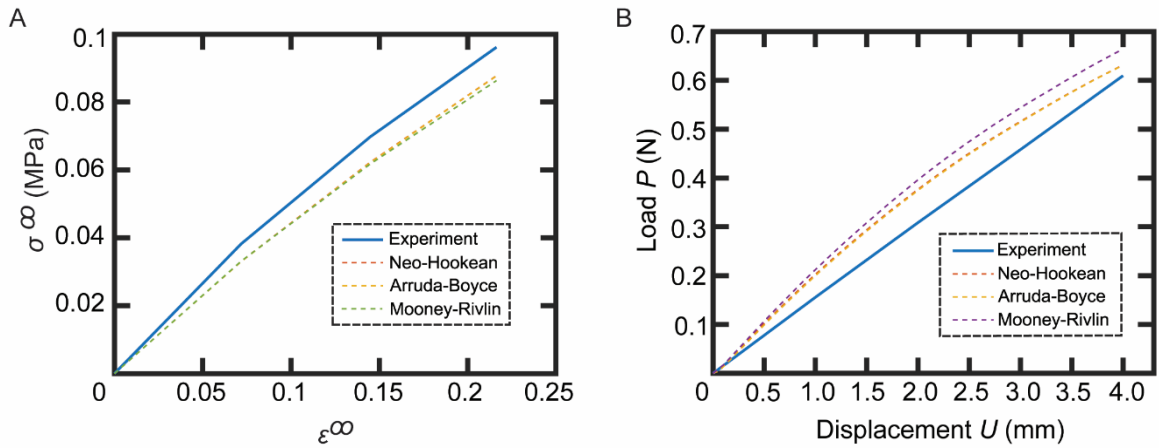

**Figure S14:** Comparison between measurements and hyperelastic model predictions using the parameters inferred by FEMU from the $H/D = 2$ tensile test. The (a) tensile nominal stress $\sigma^\infty$ versus nominal strain $\varepsilon^\infty$ of the $H/D = 1$ specimen and (b) 3-point bending response of the Silicone rubber beam from Fig. 4.

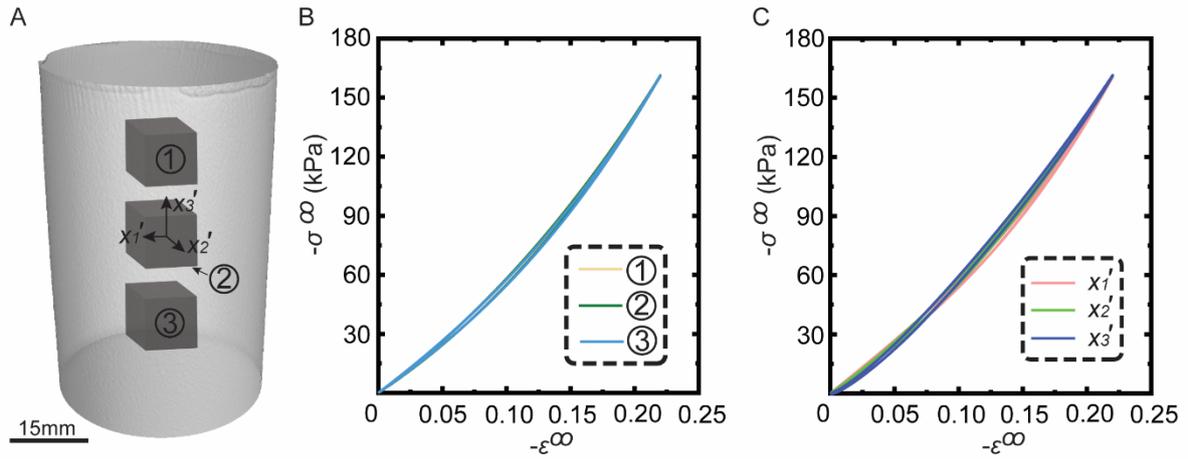

**Figure S15:** (a) Sketch of the cylindrical Silicone rubber specimen indicating locations where cubes were cut from. A local co-ordinate system $x'_i$ is also indicated on the cubes. (b) The measured compressive nominal stress versus nominal strain responses of these cubes cut from the three locations and tested in the $x'_1-$direction. (c) The compressive response of the cube from location 2 in the 3 orthogonal directions. The compressive responses were measured with the specimens glued to platens.

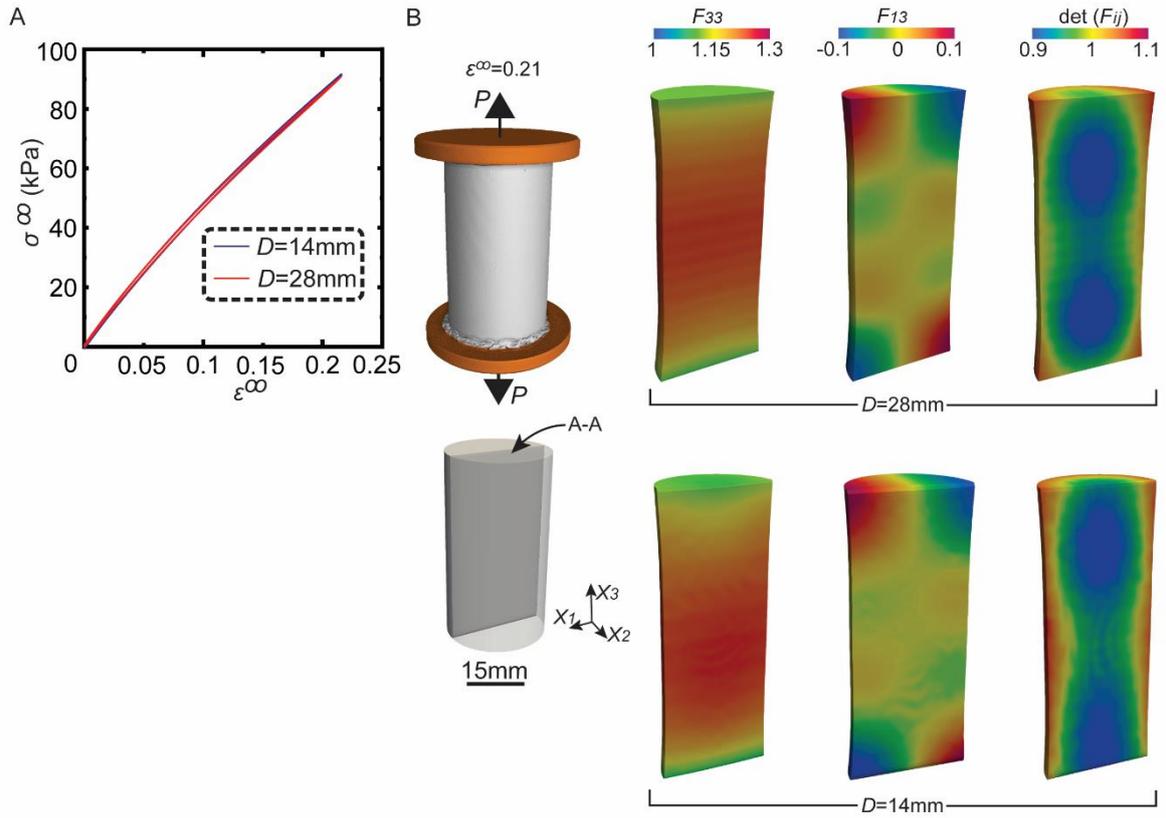

**Figure S16:** (a) Comparison between the measured nominal tensile stress $\sigma^\infty$ versus nominal strain $\varepsilon^\infty$ response of geometrically self-similar $H/D = 2$ Silicone rubber specimens with $D = 28$ mm and 14 mm. (b) The corresponding FETC measurements of spatial distributions of $\det(F_{ij})$, $F_{33}$ and $F_{23}$ on a diametrical plane through the specimens at an applied strain $\varepsilon^\infty = 0.21$. The specimen images have been scaled so that images for both specimen sizes are of the same size. The measured deformation gradient distributions are plotted on the deformed configuration.

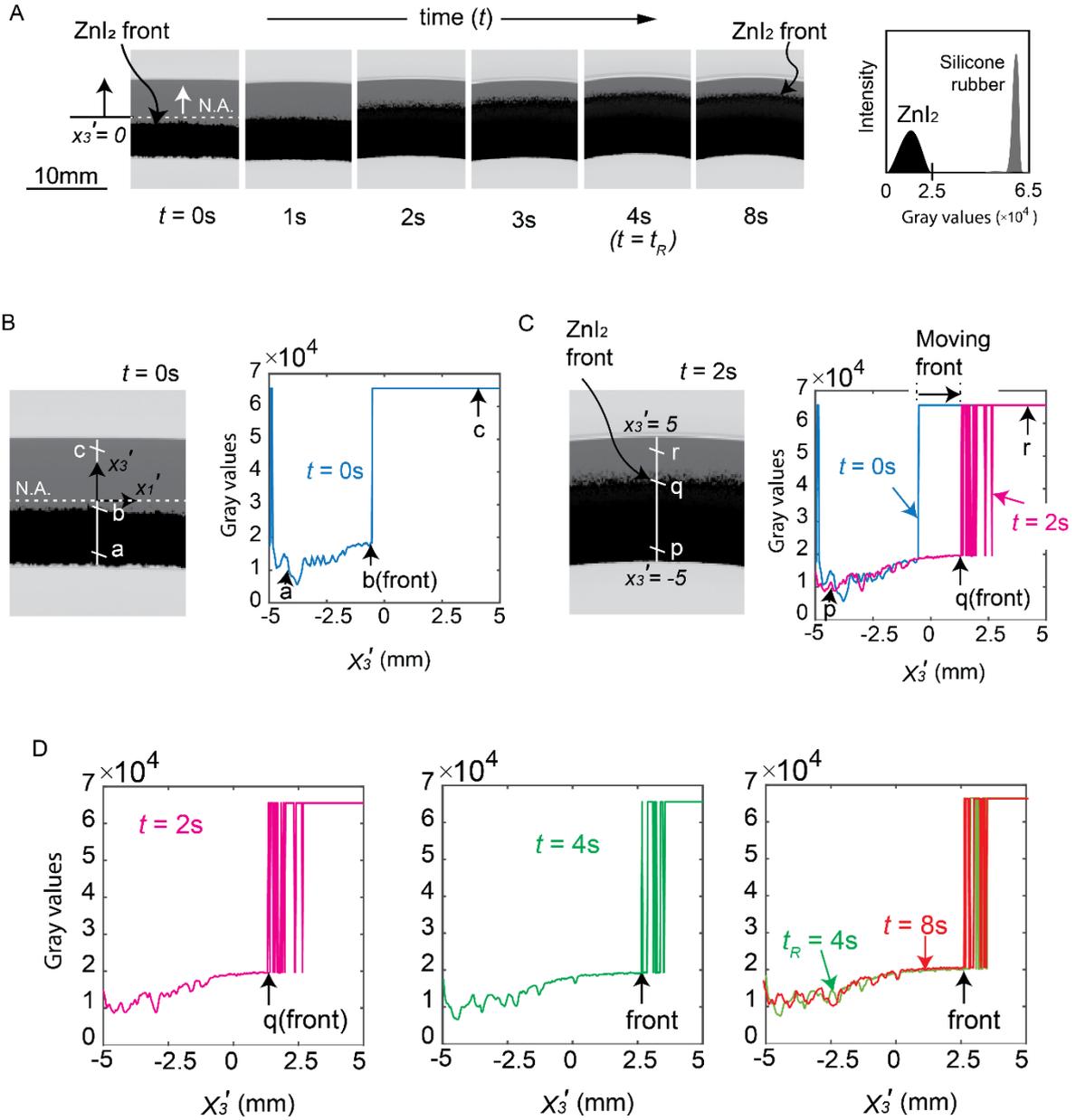

**Figure S17:** (a) The radiographs showing the motion of the front of the $ZnI_2$ tracers in the 4-point bend experiment reproduced from Fig. 3d without the colour tone. The histogram of gray values at $t = 0$ s is also shown along with a local co-ordinate system $x'_1 - x'_3$ with origin located at the mid-point (neutral axis) of the beam. The fixed local co-ordinate system is fixed to that material point so that it translates with the beam. The variation of the gray values with $x'_3$ at $x'_1 = 0$ at (b) $t = 0$ s and (c) $t = 2$ s. In (b) and (c) we show locations "a, b, c" and "p, q, r" respectively on the beams and these are marked in the plotted variation of the gray values. In (c) the gray value variation for both $t = 0$ s and $t = 2$ s is included to indicate the movement of the tracer front. (d) The variation of the gray values with $x'_3$ at $x'_1 = 0$ are plotted for times $t = 2$ s, $4$ s and $8$ s. The front remains stationary for $t > t_R$, where $t_R = 4$ s. This is clearly seen in the final graph in (d) where we include the gray value variations for both $t = 4$ s and $8$ s.

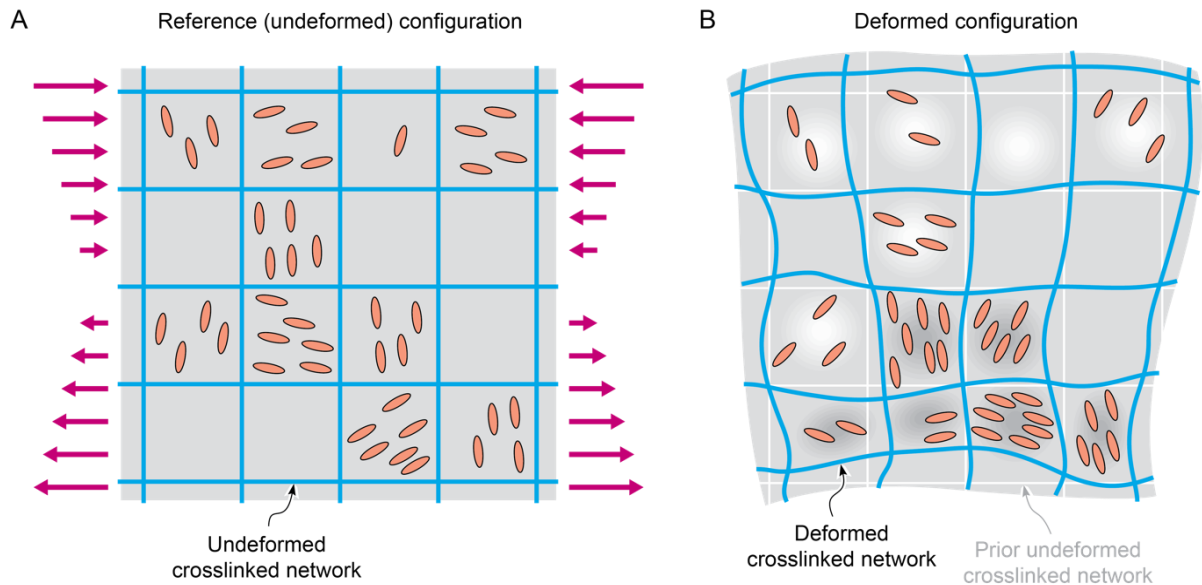

**Figure S18:** Sketch of the envisioned microstructure of the Silicone rubber comprising a network of the crosslinked polymeric chains with the un-crosslinked non-spherical molecules fully or partially filling the pores. (a) The undeformed or reference configuration subjected to a spatial gradient of loading similar to bending and (b) the deformed configuration where the un-crosslinked molecules have migrated from the region of compressive stresses to the region of tensile stresses. This migration is accompanied by dilation of the pores on the compressive side (and vice versa) as well as changes in the shapes of the pores associated with the nematic ordering (alignment) of the non-spherical molecules. To better illustrate the deformation of the crosslinked network, the undeformed network is simplified as a square grid in (a) and shown as white lines in (b) along with the deformed crosslinked network. The dilation due to reduction in attractive forces between the un-crosslinked molecules and the network is illustrated by the pale background while the darkened background denotes the increased attractive forces causing the contraction of the pores.

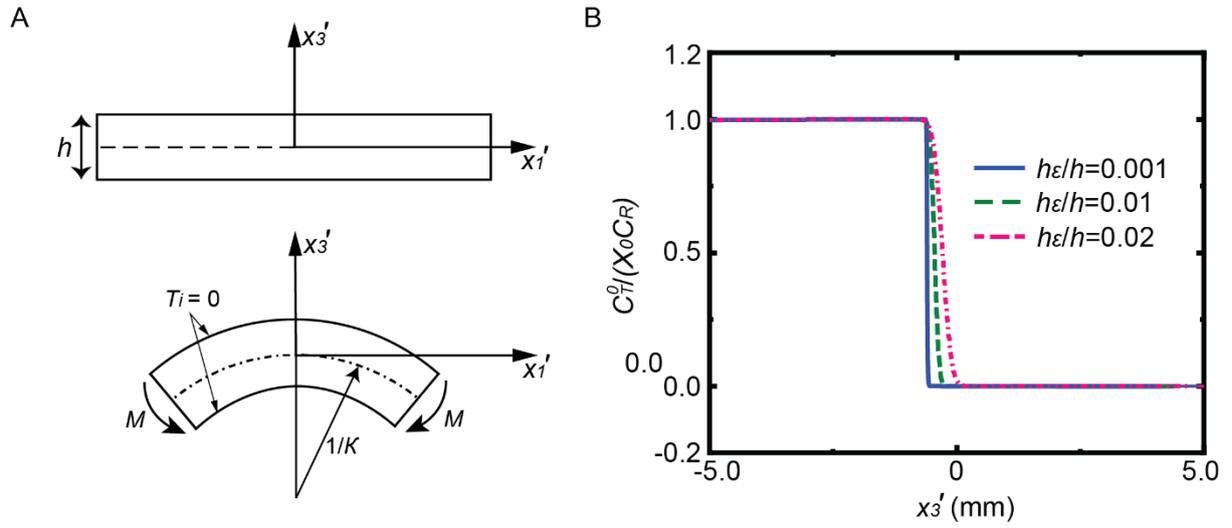

**Figure S19:** (a) Sketch of the pure bending problem analysed along with the co-ordinate system $x_1' - x_3'$. (b) Plot of the normalized initial ($t = 0$) tracer concentration distribution $c_T^0/(\chi_0 c_R)$ given by (4.37) for selected choices of the regularising parameter $h_\varepsilon/h$. The distribution is shown for the beam size in Fig. 3d and the measured value of $h_T^0 = -0.6$ mm.

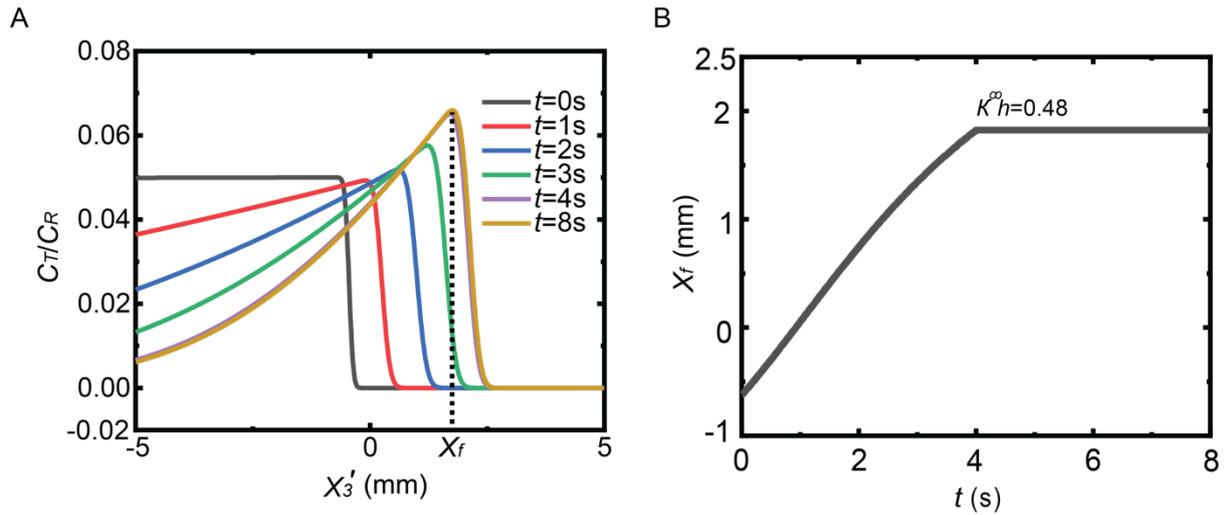

**Figure S20:** Predictions of the evolution of the tracer concentration $c_T$ for the experimentally applied loading protocol, viz. $\dot{\kappa}h = 0.12\ \text{s}^{-1}$ over time period $0 \leq t \leq 4$ s and $\dot{\kappa}h = 0$ for $t > 4$ s and the measured value of $h_T^0 = -0.6$ mm. (a) Spatio-temporal evolution of $c_T/c_R$. The definition of the tracer front position $x_f$ is indicated. (b) Corresponding prediction of the temporal evolution of $x_f$. All predictions are shown for the choice $\chi_0 = 0.05$.

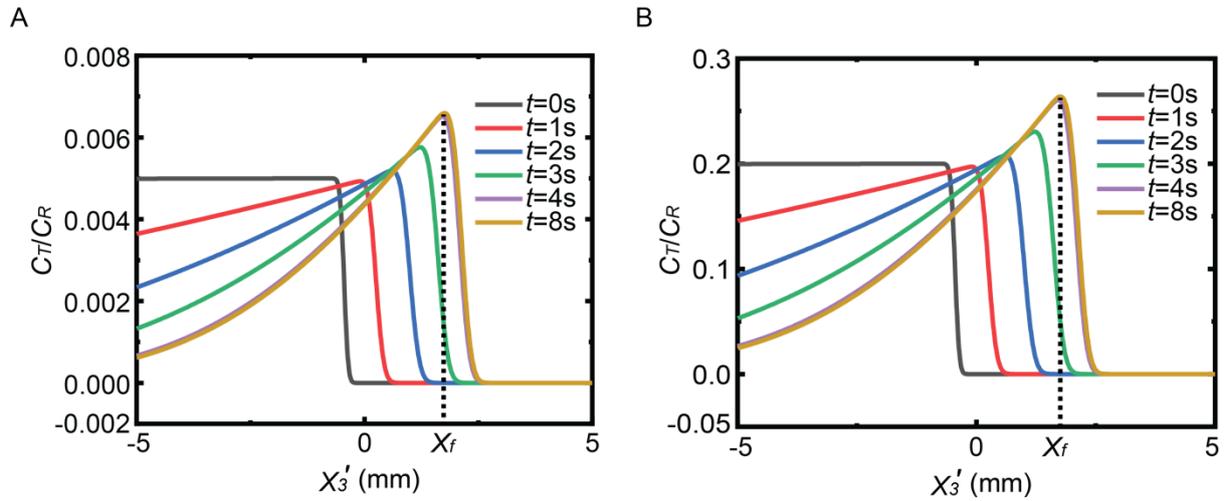

**Figure S21:** Predictions corresponding to Fig. S20 but with different choices of $\chi_0$. Spatio-temporal evolution of $c_T/c_R$ with choices (a) $\chi_0 = 0.005$ and (b) $\chi_0 = 0.2$. The definition of the tracer front position $x_f$ is indicated.

**Supplementary Tables**

| | **Specimens** | **Anode material** | **Thickness of Cu filter** | **Source voltage (kVp)** |
|---|---|---|---|---|
| Cylinder | $H = 28$ mm, $D = 28$ mm | Tungsten | 0.25 mm | 155 |
| | $H = 56$ mm, $D = 28$ mm | Tungsten | 0.25 mm | 155 |
| (Silicone rubber) | $H = 28$ mm, $D = 14$ mm | Tungsten | 0.35 mm | 150 |
| | $H = 112$ mm, $D = 28$ mm | Tungsten | 0.25 mm | 155 |
| | $H = 120$ mm, $D = 12$ mm | Tungsten | 0.50 mm | 165 |
| Beam (Silicone rubber) | $h = B = 10$ mm, $L = 70$ mm<br>$s = 50$ mm<br>(3-point bend)<br>$h = B = 10$ mm, $L = 70$ mm<br>$s_1 = 30$ mm, $s_2 = 60$ mm<br>(4-point bend) | Tungsten | 0.25 mm | 180 |
| | $h = B = 10$ mm, $L = 70$ mm<br>$s_1 = 30$ mm, $s_2 = 60$ mm<br>(4-point bending of composite beam with $ZnI_2$ tracers) | Tungsten | 0.25 mm | 135 |
| | $h = B = 10$ mm, $L = 70$ mm<br>$s_1 = 30$ mm, $s_2 = 60$ mm<br>(Fast radiography 4-point bending of composite beam with $ZnI_2$ tracers) | Molybdenum | 0.10 mm | 50 |
| Beam (HDPE) | $h = B = 10$ mm, $L = 70$ mm<br>(3-point bend) | Tungsten | 0.10 mm | 120 |
| Beam (Nylon6) | $h = B = 10$ mm, $L = 70$ mm<br>(3-point bend) | Tungsten | 0.10 mm | 125 |

**Table S1:** Dimensions and materials of all specimens investigated here. The definitions of the geometric parameters are in Fig. 1 for the cylindrical specimens and in Figs. 2 and 4 for the 4-point and 3-point bending specimens, respectively. The table also provides the radiography parameters for all test configurations considered in this study. All tests using the W target were conducted at a power of 30 W while the Mo target tests used a power of 10 W.

(a)

| Model | Model parameters | $\mathcal{C}_u$ | $\mathcal{C}_P$ | $\mathcal{C}$ |
|---|---|---|---|---|
| Neo-Hookean | $C_{10} = 0.083$ MPa<br>$D_1 = 0.768$ MPa$^{-1}$ | 0.244 | 0.030 | 0.274 |
| Arruda-Boyce | $C_1 = 0.154$ MPa<br>$\lambda_m = 5.275$<br>$D = 0.397$ MPa$^{-1}$ | 0.228 | 0.016 | 0.244 |
| Mooney-Rivlin | $C_{10} = 0.072$ MPa<br>$C_{01} = 0.023$ MPa<br>$D_1 = 1.977$ MPa$^{-1}$ | 0.296 | 0.029 | 0.324 |

(b)

| Model | Model parameters | $\mathcal{C}_u$ | $\mathcal{C}_P$ | $\mathcal{C}$ |
|---|---|---|---|---|
| Neo-Hookean | $C_{10} = 0.07$ MPa<br>$D_1 = 0.179$ MPa$^{-1}$ | 0.131 | 0.013 | 0.143 |
| Arruda-Boyce | $C_1 = 0.135$ MPa<br>$\lambda_m = 3.872$<br>$D = 0.286$ MPa$^{-1}$ | 0.132 | 0.013 | 0.145 |
| Mooney-Rivlin | $C_{10} = 0.065$ MPa<br>$C_{01} = 0.011$ MPa<br>$D_1 = 1.413$ MPa$^{-1}$ | 0.154 | 0.013 | 0.167 |

**Table S2:** The optimal parameters for the hyperelastic modules as deduced by the FEMU analysis along with the corresponding values of the optimised cost functions $\mathcal{C}_u$, $\mathcal{C}_P$ and $\mathcal{C}$. This data is shown for properties inferred from the tensile tests on the (a) $H/D = 1$ and (b) $H/D = 2$ specimens with $D = 28$ mm.

| Material parameter | |
|---|---|
| $G$ | 0.42 MPa |
| $\nu$ | 0.45 |
| $k$ | 2 |
| $\theta_0$ | 0.05 |
| $R\mathcal{T}c_R$ | 0.2 MPa |

**Table S3:** The material parameters for the thermodynamically consistent model for Silicone rubber. These parameters are used for all the predictions shown in Figs. 2 and 3.

**Supplementary Movies**

**Movie S1:** A walkthrough into the volume of the $H/D = 1$ Silicone rubber specimen (Fig. 1e) subjected to a nominal strain $\varepsilon^\infty = 0.21$ showing the spatial variation of $\Delta V/V_0$.

**Movie S2:** A walkthrough into the volume of the 4-point bend specimen Silicone rubber specimen (Fig. 2d) showing the spatial variation of $\Delta V/V_0$ at nominal strain $\kappa h = 0.24$.

**Movie S3:** A walkthrough into the volume of the bent composite beams (Fig. 3b) to show transport of the $ZnI_2$ tracers when the tracers were present on the (a) on the compressive side and (b) on the tensile side.

**Movie S4:** Video stitched together from the radiographs of Fig. 3d of the composite beam subjected to a step loading 4-point bend experiment with $\dot\kappa h = 0.12 \text{ s}^{-1}$ and $t_R = 4$ s (i.e., maximum $\kappa h = 0.48$). The total time of the experiment is $t = 10$ s and shows there is no motion of the tracer beyond $t_R = 4$ s. The video was constructed from radiographs with an interframe time of 0.5 s.

**Movie S5:** A walkthrough into the volume of the 3-point bend specimen of the commercial Nylon6 (Fig. 4c) showing the spatial variation of $\Delta V/V_0$ at a displacement $U = 4$ mm.